\documentclass[12pt]{article}

\usepackage{amsfonts,amsmath}
\usepackage{amssymb}
\usepackage{ytableau}
\usepackage{mathtools} 
\usepackage{quiver} 
\usepackage{tikz-cd} 

\usepackage{pgfplots}
\pgfplotsset{compat=1.5}
\usepgfplotslibrary{fillbetween} 
\usepackage{tikz}
\usetikzlibrary{arrows.meta,bending,patterns,angles,calc}
\usetikzlibrary{patterns.meta}
\usepackage{siunitx}
\usetikzlibrary{decorations.shapes}
\tikzset{decorate sep/.style 2 args=
{decorate,decoration={shape backgrounds,shape=circle,shape size=#1,shape sep=#2}}}

\pgfplotsset{ignore zero/.style={%
  #1ticklabel={\ifdim\tick pt=0pt \else\pgfmathprintnumber{\tick}\fi}
}} 

\newcommand{\oast}{\circledast}

\newcommand{\dr}{{{\rm d}}}

\makeatletter \@addtoreset{equation}{section} \makeatother

{\vspace{3mm} }

\def\al{\alpha}

\def\*{\star}

\def\E2{\mathbf{E}}

\def\y{\mathbf{y}}
\def\v{\mathbf{v}}
\def\u{\mathbf{u}}

\newcommand{\be}{\begin{equation}}
\newcommand{\ee}{\end{equation}}
\newcommand{\bee}{\begin{eqnarray}}
\newcommand{\beee}{\begin{array}}
\newcommand{\eee}{\end{eqnarray}}
\newcommand{\eeee}{\end{array}}
\newcommand{\gb}{\beta}
\newcommand{\gga}{\gamma}

\newcommand{\gd}{\delta}

\newcommand{\gk}{\varkappa}
\newcommand{\gep}{\epsilon}

\newcommand{\gs}{\sigma}
\newcommand{\go}{\omega}

\newcommand{\dal}{\dot \alpha}
\newcommand{\dgb}{\dot \beta}

\newcommand{\p}{\partial}

\newcommand{\ff}{\frac}      

\begin{document}
    
    \begin{flushright}
FIAN/TD/13-2024\\
\end{flushright}

\vspace{0.5cm}
\begin{center}
{\large\bf All vertices for unconstrained symmetric gauge fields}

\vspace{1 cm}

\textbf{V.E.~Didenko$^{1}$ and M.A.~Povarnin$^{2}$}\\

\vspace{1 cm}

\textbf{}\textbf{}\\
 \vspace{0.5cm}
 \textit{$^{1}$I.E. Tamm Department of Theoretical Physics,
Lebedev Physical Institute,}\\
 \textit{ Leninsky prospect 53, 119991, Moscow, Russia }\\
 \vspace{0.5cm}
 {\it
			$^2$Moscow Institute of Physics and Technology,\\
			Institutsky lane 9, 141700, Dolgoprudny, Moscow region, Russia}

\par\end{center}

\begin{center}
\vspace{0.6cm}
e-mails: didenko@lpi.ru, povarnin.ma@gmail.com \\
\par\end{center}

\vspace{0.4cm}

\begin{abstract}
\noindent Recently, the generating system that describes
interacting symmetric higher-spin gauge fields at the level of equations of motion was proposed. The interaction vertices it offers are 'off the mass shell' unless constrained by the prescribed factorization condition that properly removes traceful components. In this paper we detail the structure of the unconstrained, i.e., traceful vertices. We derive their manifest form to all orders along with a net of the associated dualities, thus providing the complete higher-spin vertex analysis at the unconstrained level for the bosonic theory in any dimension. These vertices are shown to be the minimal space-time local and  have a form of the peculiar integrals over a space of closed polygons, which we scrutinize in the paper. The obtained results directly apply to the holomorphic sector of the four-dimensional theory, where the interaction is on shell, producing the all-order chiral higher-spin vertices.
\end{abstract}
\newpage

\tableofcontents
\newpage

\section{Introduction}
The higher-spin (HS) gauge theories are quite well studied and understood up to the lowest cubic interaction order \cite{Bengtsson:1983pd}-\cite{Tatarenko:2024csa}. The
legacy of these studies include appreciation that the principles of
gauge invariance are so compelling that they force one to abandon the
habitual minimal gravitational coupling as inconsistent for the fields
of higher spins $(s>2)$,  \cite{Aragone:1979hx}. Yet, the interaction of symmetric gauge fields
with each other is governed by the Eastwood-Vasiliev algebra, known as
the HS symmetry \cite{Eastwood:2002su, Vasiliev:2003ev}, the transformation laws of which exhibit
higher derivatives growing with spin. Its effect is an infinite
spectrum of HS fields $(d>3)$ and a certain degree of nonlocality.
The latter, however, is under full control at the cubic level
meaning that the number of derivatives that the interaction vertex $s_1-s_2-s_3$
may have is still finite for the fixed three spins.

A nontrivial implication of the HS symmetry is also the vacuum
invoking the nonzero cosmological constant \cite{Fradkin:1987ks}.  It lends a potentially powerful tool in the form of the Klebanov-Polyakov
AdS/CFT conjecture \cite{Klebanov:2002ja} (see also \cite{Leigh:2003gk, Sezgin:2003pt}), which in the
simplest case, relates the free boundary $O(N)$ model to an
interacting theory of HS gauge fields. Perhaps, this may open a
way for reconstructing the latter from the free boundary theory
directly. This plan has come to a successful conclusion at the
level of the three-point functions justifying the original
expectation due to \cite{Klebanov:2002ja} and \cite{Giombi:2009wh} and led to a partially
gauged fixed cubic HS action \cite{Sleight:2016dba}. Moreover, the obtained
couplings indeed encode the structure constants of the HS algebra \cite{Sleight:2016xqq}.
However, as one proceeds to the four-point function level aiming
at the quartic HS interactions \cite{Bekaert:2015tva}, the resulting vertex gets
too nonlocal to be dealt with using the standard field theory
tools \cite{Sleight:2017pcz}. Thus, the holographic reconstruction, as it seems,
has reached the point where it loses predictability once the
freedom in nonlocal field redefinitions affecting the interaction
couplings \cite{Barnich:1993vg} naturally shows up.

Present difficulty with the holographic approach has led to at least two
parallel views of the problem. According to one, which assumes the
HS theories are essentially nonlocal, they should be taken with
an eye to the boundary. Roughly speaking, their definition should include the
proper correlation functions of the boundary currents made of, e.g.,
free fields. This 'boundary first' perspective does not directly
suggest an independent bulk definition of the HS interactions. For
example, one may attempt to qualify the bulk nonlocalities via the
Mellin amplitude singularities as it has been undertaken in
\cite{Ponomarev:2017qab}. There is also a collective dipole approach of \cite{deMelloKoch:2010wdf} and
its further implementation of \cite{Aharony:2020omh}, which, however, departs
from the conventional Fronsdal formulation in the bulk. An extreme
version of the boundary based reasoning is a narrative asserting
the conventional HS theories do not exist in AdS at all, much as
they seem unlikely to exist in the Minkowski space-time.

A parallel approach to the locality problem rests on the first
principle analysis of the degree of the HS nonlocality, if
any,\footnote{The quartic nonlocality relies on the HS duality
conjecture, which despite the three-point agreement might not hold
beyond \cite{Vasiliev:2012vf, Diaz:2024kpr, Diaz:2024iuz}, but even if it does, there are still potential
loopholes in the original nonlocality argument of \cite{Sleight:2017pcz}. In particular, in \cite{Neiman:2023orj}, the nonlocality argument was argued to be space-time signature dependent. This
gives the holographic HS locality a slim chance (see also \cite{Ponomarev:2017qab} for other options).}
rather, than on tethering it to the Klebanov-Polyakov conjecture.
Having the access to all-order HS interactions along with the manifest
gauge invariance, this approach is known due to Vasiliev, who
proposed generating equations in four \cite{Vasiliev:1992av} and any \cite{Vasiliev:2003ev}
dimensions. Designed for making up interaction vertices, these
equations can be used in analysis of (non)locality and holographic duality.

An important remark in this regard is in order. Vasiliev's
strategy in formulating dynamical problem is very close in spirit
to the earlier ideas of Penrose and his twistor space \cite{Penrose}.
Namely, HS interactions may acquire a simpler form once pushed out
from space-time into a certain correspondence space. In practice,
the correspondence space can be identified using the so-called
unfolded formalism \cite{Vasiliev:1988sa} (see also \cite{Misuna:2022cma, Misuna:2024ccj, Misuna:2024dlx} for recent advances). As a result, the natural notion of
locality is prescribed by the locality in the fiber space, rather
than in space-time. We call the former fiber {\it spin locality},
see \cite{Gelfond:2019tac}, in order to distinguish it from the usual space-time
locality. The relation between the two is not straightforward
beyond cubic order, while a condition that brings space-time locality from the fiber spin locality has been recently identified in \cite{Vasiliev:2022med}.

Spin locality of the Vasiliev theory is not manifest even to the
lowest interaction order. This fact has already caused a big deal
of confusion in the literature for it seemingly suggests a certain extent of
arbitrariness of the Vasiliev's theory prediction. We find it useful to clear up this situation here.

Inherent freedom in the choice of physical
variables\footnote{These appear in the form of
$\dr_z$ cohomologies. For reviews on the Vasiliev theory see \cite{Vasiliev:1999ba}-\cite{Didenko:2014dwa}.} is a salient feature of Vasiliev's
formulation. The generating equations alone do not give the clue
to what these variables are to meet the (non)locality criterion
(presently missing). For example, a seemingly natural choice of
these variables used by Giombi and Yin in \cite{Giombi:2009wh} produced
infinities in the process of taking the boundary limit $z\to 0$ at
the cubic order. These divergencies are not expected at this
order, since the cubic approximation is perfectly local for fixed
spins. Thus, the local vertex differs from that of Giombi and Yin
by a nonlocal field redefinition. One may argue that ambiguity in
nonlocal field redefinitions can modify the interaction coupling
constants, and, therefore, the theory loses its predictive
performance. This is indeed what happens to the standard Noether
procedure, as follows from \cite{Barnich:1993vg}, once nonlocalities are
admitted. However, any feasible field redefinition has a very
peculiar form of a polynomial of the HS modules\footnote{In general, such a form of field redefinition is not meant to be local in the usual space-time sense. However, organized in
terms of the HS modules that carry representation of the HS algebra, it
differs from the generic field redefinition available within the
Noether approach.} in the Vasiliev case. In
particular, there is no way of getting rid of the lowest HS vertex,
which stores information on the free HS equations along with a part
of the interaction. In other words, the field redefinitions, being
local or not, have no bearing on the HS vertices, which are fixed in
accordance with the HS symmetry. Their explicit form, of course, is
field dependent. Instead, what may happen in the case of unlucky
choice of field variables is the relevant quantities, such as
vertices or the boundary correlators, leave their convergence domains
in some neighborhood of, or even in the whole space-time, causing
divergences or ill-defined expressions. Therefore, the minimally
nonlocal framework is required. To recapitulate, the setup based
on the HS modules and the infinite HS symmetry underlying
Vasiliev's equations preclude fine-tuning of the interaction
constants. The crucial difference of Vasiliev's approach from the standard Noether procedure applied to higher spins is the latter is defined perturbatively about the AdS (or Minkowski) vacuum, while the former is formulated on {\it arbitrary} HS background. In \cite{Vasiliev:2016xui}, for example, the local variables have been identified at the lowest interaction order by field redefining the
Giombi-Yin vertex. This allowed one to check the cubic HS vertices
against the holography expectations. The result is a perfect match
\cite{Sezgin:2017jgm, Didenko:2017lsn, Misuna:2017bjb}, including the parity broken case.

The all-order (non)locality structure of the HS theory is a challenging and still unsolved problem. Its solution is necessary to define the proper free from divergences setup. The problem can
be posed as follows. Is Vasiliev theory spin local, and if not what
is the minimum degree of nonlocality? In a series of papers
\cite{Gelfond:2019tac}, \cite{Vasiliev:2017cae}-\cite{Vasiliev:2023yzx}, the locality issue was analyzed mostly in
four dimensions. While the full answer is not known yet, the
obtained partial results uncover a highly nontrivial structure of
the HS interactions constrained by locality. These findings can be
briefly summarized as follows.
\begin{itemize}
\item {\bf Ultralocality.} The four-dimensional HS theory in its spinor
formulation contains three sectors: the holomorhic, antiholomorphic
and the mixed one. The vertices from the (anti)holomorhic gauge sector
were shown to be spin local up to few orders in \cite{Didenko:2018fgx, Didenko:2019xzz, Didenko:2020bxd}.
However, the observed locality turned out to be stronger than
expected. Namely, the vertices from the gauge sector went maximally local in the following sense: The number of derivatives entering them do not grow with the spins
of fields that couple via the infinite-dimensional Weyl module. This
phenomenon was called {\it ultralocality}\footnote{The simplest
manifestation of ultralocality is the vertex from the so-called
central on mass shell theorem \cite{Vasiliev:1988sa, Bychkov:2021zvd} that contains the primary field contribution from the Weyl module only. As a matter of principle, it could have contained a tail of descendants violating
ultralocality.} and led to a conjecture \cite{Gelfond:2018vmi} that the whole
(anti)holomorhic sector is all-order spin local. The conjecture was
manifestly proven, while ultralocality of the gauge sector
was established in \cite{Didenko:2022qga}. That being said, ultralocality
automatically implies space-time locality \cite{Gelfond:2019tac}.

\item {\bf Shift symmetry.} There is a remarkable interplay between
spin locality and a specific shift symmetry of the HS vertices. The
symmetry acts as shifts by a constant in momenta in the Fourier
vertices and leads to their simple transformation. Observed in
\cite{Didenko:2022qga}, as a development of the earlier ideas from \cite{Gelfond:2018vmi}, it was
then shown in \cite{Didenko:2022eso} that under some mild assumptions on the HS vertices, the symmetry implies spin locality. Moreover, its action can be extended to define field redefinitions that leave interaction spin
local. On a practical side, shift symmetry implies the vertices
acquire a form of the integrals over closed polygons.

\item {\bf Holomorphic sector.} This sector of the four-dimensional
theory, also known as chiral, can be considered independently from
the rest. Using observation from \cite{Didenko:2019xzz} (based on \cite{DeFilippi:2019jqq}, see also, \cite{Iazeolla:2020jee, DeFilippi:2021xon}), it was shown that spin locality effectively results in a different large star product than the original Vasiliev one. While this star product, called the limiting one, cannot be seemingly used in the full $4d$ theory, it leads to a certain generating equations for the (anti)holomorphic sector \cite{Didenko:2022qga}. Among the advantages of the proposed equations are
(i) manifest ultralocality, (ii) manifest shift symmetry, (iii) a
simple form of the corresponding all-order interaction vertices. The limiting star product was also used in \cite{Sharapov:2022awp} for the analysis of the HS vertices from the holomorphic sector.

\item{\bf Quartic interaction in the mixed sector.} Recall that the
holographic reconstruction somehow came to a halt at the quartic level leaving the status of the Klebanov-Polyakov
conjecture in a suspense. The verification remains one of the
urgent problems in the field. Recently, Gelfond managed to
calculate the contribution to this vertex \cite{Gelfond:2023fwe} using the formalism of \cite{Vasiliev:2023yzx}. Impressively, even though the result obtained is not spin local, it falls into the same category as a purely
(anti)holomorphic vertex. Namely, the two carry the same degree of
the exponential star-product contractions\footnote{These contractions receive two contributions in the holomorphic sector: the local holomorphic and another (irrelevant) nonlocal antiholomorphic,
which has degree one. The mixed sector contains both nonlocal
contributions, but the overall degree is still equal to one.}
that measure the degree of (non)locality . Given that, it is not
unlikely that the vertex admits a well-defined boundary limit. It would
be very useful to test the resulting tree level boundary correlators using this vertex.

\item{\bf Generating equations in $d$ dimensions.} The important
result that came out from the ongoing locality quest is the
Vasiliev-like system of the interacting symmetric bosonic gauge fields
\cite{Didenko:2023vna}. It literally stems from the holomorphic equations of
\cite{Didenko:2022qga}, but unlike the latter, it describes the full fledged interaction in any dimensions. The system is based on the so-called off-shell HS algebra \cite{Vasiliev:2003ev}. Thus, similarly to the system from \cite{Vasiliev:2003ev}, it does not describe propagating HS fields; rather, it describes a collection of the generalized HS Bianchi identities. There is a mechanism that sets the system on shell in the original Vasiliev
case. It is based on the deformed oscillator algebra, which is
built in the equations and yields the required ideal that has to
be factored out order by order. There is no deformed oscillators
in the system of \cite{Didenko:2023vna}, however. Interestingly, the system still
admits the ideal responsible for the on-shell reduction. This ideal was manifestly identified in \cite{Didenko:2023vna} and shown to
receive no higher-order field corrections, which hopefully makes it
useful in practice. At the off-shell level, the vertices were
shown to be all order (ultra)local and featured the $d$-dimensional version of the shift symmetry. While the relation
between the equations from \cite{Didenko:2023vna} and the original Vasiliev ones from \cite{Vasiliev:2003ev} is
not known, we tend to think the former might follow from the
latter in the star-product reordering limit restricted to the proper
functional class, along the lines of \cite{Didenko:2019xzz}.
\end{itemize}

\paragraph{Goals and results.} Present paper focuses on the last item
above. As the equations of \cite{Didenko:2023vna} give straightforward access to the
off-shell HS vertices in their local form, one wishes to have the
list of all of them. This would bring the HS (non)locality problem
to the next level being the analysis of the factorization
condition with respect to the known traceful ideal. An example of
the vertex that corresponds to the most convenient field ordering has
already been given in \cite{Didenko:2023vna} to any order of perturbations. This
vertex has a form of an integral over two simplices supplemented
with a shoelace condition that admits a nice geometric interpretation. Here, we extend these results to the vertices of all possible orderings, thus providing the full list of them. We show that the vertices have a form of integrals, the configuration space of which is represented by certain closed polygons. These polygons range from the purely concave to the purely convex ones via intermediate in accordance with {\it impurities} of the gauge field insertions. 

The important result of our study leads to a somewhat unexpected structure of the obtained unconstrained unfolded equations. Let us briefly summarize it here.    

The HS unfolded equations have two sectors: the gauge sector of the 1-forms $(\go)$ and the Weyl sector of the 0-forms $(C)$. These will be specified below. Schematically, each sector starts with a primary field of, e.g., spin $s$ that generates its descendants,
\begin{subequations}
\begin{align}
&C_s:=C_s^0\to C_s^1\to C_s^2\to\dots\to C_s^{\infty}=C_{s}^{k},\quad k\geq 0\,,\label{chain_C} \\
&\go_s:=\go_s^0\to\go_s^1\to\go_s^2\to\dots\to\go_s^{s-1}=\go_s^l\,,\quad 0\leq l\leq s-1\,.\label{chain_omega}
\end{align}
\end{subequations}
Notice, that unlike $\omega_s$, there are infinitely many descendants for $C_s$ that effectively capture all space-time derivatives of a given HS field. This fact plays crucial role in the locality issue. In addition, the $C$ equations essentially control the HS nonlinearities. The latter turn out to have the following proliferated form:
\be\label{dC=U}
\dr_x C_s^k=\sum_{i=0}^{k}\sum_{\vec{n}, \vec{s}}\Upsilon_s^i(\go_{s'}^{n'}, \overbrace{C_{s_0}^{n_0},\dots ,C_{s_i}^{n_i}}^{i+1})\,,
\ee
where $\vec n=(n',n_0\dots, n_i)$, $\vec s=(s',s_0\dots s_i)$, and $\Upsilon_s^i$ are the monomials of their arguments of the degree $i+2$, which we manifestly found. Its main features are as follows:
\begin{itemize}
    \item For any fixed $\vec s$, the sum over $\vec n$ contains only finite number of terms. This fact implies spin locality and has already been  proven in \cite{Didenko:2023vna}. In addition, the vertices are space-time local and carry the minimal number of derivatives, while the structure of $\Upsilon$'s contains peculiar integrals over a space of certain polygons with the steady convexity properties.  
    \item Less expected is that the right-hand side of \eqref{dC=U} terminates at $i=k$ despite being all-order exact. Indeed, as a matter of principle, once \eqref{dC=U} is all-order exact, one could have expected contribution of ever growing powers of 0-forms on the right for fixed $k$. However, this does not happen. Rather we see how the nonlinearities are bounded by the depth of the descendant $C_s^k$ as these appear in a controlled fashion; they grow intact with the depth $k$ of the descendant $C_s^k$. On top of that, the right-hand side vanishes for sufficiently  distant spins. As will be also explained later, such a form fits the notion of the projectively compact spin locality introduced in  \cite{Vasiliev:2022med} that in the on-shell situation implies space-time locality of interaction.
\item Another consequence of the proliferated nonlinearities of \eqref{dC=U} is the highly restrictive integrability constraints that follow from the inspection of $\dr_x^2=0$. These allow one to fix the right-hand side of 
\be
\dr_x\go=\mathcal{V}(\go, \go, C\dots C)
\ee
in terms of $\Upsilon$ from \eqref{dC=U} unambiguously in a simple algebraic way using no generating equations for this sector. Moreover, this way, we also derive various algebraic (horizontal) relations between vertices from the 1-form sector. So arising {\it vertex dualities} are systematized in the paper.
\end{itemize}
Finally, as a byproduct of our analysis, we provide the full list of the holomorphic on-shell vertices of the four
dimensional HS theory. This problem is straightforward to solve because the off-shell system turns into the chiral one on shell in four dimensions upon properly adjusting the undeformed part of the corresponding HS algebra.

The paper is structured as follows. In Section \ref{sec:Section1}, we briefly provide the standard unfolded setup for the HS problem of symmetric gauge fields. We also address the locality issue there with the  emphasis  on  the conditions in the fiber that result in space-time locality. We then proceed with the HS generating system that we will work with throughout the paper. In Section \ref{Sec: Pert}, we investigate perturbations, as we lay out the manifest expressions for the generating master field and HS vertices, as well as all-order HS gauge transformations. In Section \ref{Sec:Geometry} the geometric interpretation of the obtained vertices is elaborated. In Section \ref{Sec:lowest}, we focus on the first lower-order examples to demonstrate general features of the off-shell vertices. In Section \ref{section:DualRel}, a set of the vertex dualities is derived. Straightforward implication of the obtained results for the four-dimensional holomorphic sector is given in Section \ref{Sec:holomor}. A brief leading-order comparison of our vertices with those that can be extracted from the Vasiliev generating equations is carried out in Section \ref{Sec:Vasiliev}. We conclude in Section \ref{Sec:Conclusion}. The paper is supplemented with seven technical appendices.

\section{Higher spin generating equations}
\label{sec:Section1} \index{Section1}
\subsection{Unfolded form}
The unfolded approach \cite{Vasiliev:1988sa} supplemented with the HS algebra concept suggests the following schematic form of the HS equations:
\begin{subequations}
    \label{nonlinear}
    \begin{align}
        \label{nonlinear:1-sector}& \dr_x \omega + \omega \ast \omega = \mathcal{V}(\omega, \omega, C) + \mathcal{V}(\omega, \omega, C, C) + ...\, \\
        \label{nonlinear:0-sector}& \dr_x C + \omega \ast C - C \ast \pi(\omega) = \Upsilon(\omega, C, C) + \Upsilon(\omega, C, C, C) +
        ...
    \end{align}
\end{subequations}
Their content is as follows. Field spectrum is packed into 1-form
$\go(Y|x)$ and 0-form $C(Y|x)$. These are the collection of the
two-row Young diagrams labeled by integer spin $s$ 
\be\label{w}
\go^{a(s-1),\, b(n)}=\dr x^{\mu}\go^{a(s-1),\,
b(n)}_{\mu}=\overbrace{\ytableausetup{mathmode, boxsize=1.25em}
\begin{ytableau}
{} & {\bullet} & {\bullet}
& {\bullet} & {} \\
{} &  {\bullet} & {\bullet} & {}
\end{ytableau}}^{s-1}\,\,,\qquad 0\leq n\leq
s-1
\ee
and
\be\label{C}
C^{a(m),\, b(s)}=\overbrace{\ytableausetup{mathmode, boxsize=1.25em}
\begin{ytableau}
{} & {\bullet} & {\bullet}
& {\bullet} & {} \\
{} &  {\bullet} & {\bullet} & {}
\end{ytableau}}^{m}\,\,,\qquad m\geq s\,,
\ee
where the Lorentz $o(d-1,1)$ indices range $a,b=0,\dots, d-1$,
while the Young condition implies that symmetrization of an index
from the second row of the Young diagram with all the indices from
the first row is zero\footnote{We use symmetric basis for Young
diagrams, which assumes the Lorentz indices are symmetric along the
rows. We also use the symmetrization convention, e.g.,
$\phi^{a(k)}:=\phi^{(a_1\dots a_k)}$ }
\be\label{Young}
\go^{a(s-1),\, ab(n-1)}=C^{a(m),\, ab(s-1)}=0\,.
\ee
The way fields \eqref{w} and \eqref{C} are packed into master
fields $\go(Y|x)$ and $C(Y|x)$ is via the generating variables
\be\label{Y}
Y=(y_{\al}, \y^{a}_{\gb})\,,
\ee
where $\al, \gb=1,2$ counts rows of the Young diagram. To guarantee the Young condition for coefficients of polynomials in
$Y$, the Greek indices need to be contracted in the $sp(2)$ invariant
fashion, e.g., $y\y^a:=y_{\al}\y_{\gb}^{a}\gep^{\al\gb}$, where
$\gep_{\al\gb}=-\gep_{\gb\al}$ is a canonical $sp(2)$ form with
its inverse $\gep^{\al\gb}\gep_{\al\gga}=\gd_{\gga}{}^{\gb}$. By
adopting convention for raising and lowering indices
\be
y^{\al}=\gep^{\al\gb}y_{\gb}\,,\qquad
y_{\al}=y^{\gb}\gep_{\gb\al}\,,
\ee
one makes the formalism $sp(2)$ covariant allowing us to state
\cite{Vasiliev:2003ev} that coefficients in the Taylor expansion of any function
$f=f(y, \vec\y)$ are the two-row Young diagrams provided the $sp(2)$
singlet condition is imposed
\be\label{spop}
\left(\vec\y_{\al}\cdot\ff{\p}{\p \vec\y^{\gb}}+y_{\al}\ff{\p}{\p
y^{\gb}}+\al\leftrightarrow\gb\right)f=0\,.
\ee
Practically, the latter implies the fields \eqref{w} and \eqref{C} are
packed by means of the following $sp(2)$ invariant combinations:
\begin{equation}
    \label{generators}
    P_a = \frac{1}{2} \, y_\alpha \y^\alpha_a, \quad M_{ab} = \frac{1}{2} \, \epsilon_{\alpha\beta} \y^\alpha_a \y^\beta_b \, ,
\end{equation}
as 
\begin{subequations}
    \label{MFP}
    \begin{align}
        &\omega(Y|x) = i \cdot \sum_{s \geqslant n+1} \frac{1}{(s-1+n)!} \,  \omega^{a_1 \dots a_{s-1} \, , \, b_1 \dots b_{n}}  M_{a_1 b_1} \dots  M_{a_n b_n} P_{a_{n+1}} \dots  P_{a_{s-1}} \,,\label{wman} \\
        &C(Y|x) = i \cdot \sum_{m\geqslant s} \, (-i)^{m-s} \, C^{a_1 \dots a_m \, ,  \, b_1 \dots b_s}  M_{a_1 b_1}  \dots  M_{a_s b_s}  P_{a_{s+1}} \dots P_{a_m} \,.\label{Cman}
    \end{align}
\end{subequations}
Presence of the imaginary unit in \eqref{MFP} is driven by the reality conditions for the fields $\omega(Y|x)$ and $C(Y|x)$ that will be explored later in the text. The choice of the normalization coefficients is ambiguous, but the one used is convenient in practical calculations.

Star product on the left-hand sides of \eqref{nonlinear:1-sector}
and \eqref{nonlinear:0-sector} is the usual Moyal product
\be\label{moyal}
(f*g)(y, \vec\y)= \int f\left(y+u, \vec\y+\vec\u\right)
g\left(y+v, \vec\y+\vec\v\right)
e^{iu_{\al}v^{\al}+i\vec\u_{\al}\cdot\vec\v^{\al}}\,,
\ee
which defines the following commutation relations:
\be
[y_{\al}, y_{\gb}]_*=2i\gep_{\al\gb}\,,\qquad [\y^{a}_{\al},
\y^{b}_{\gb}]_*=2i\gep_{\al\gb}\eta_{ab}\,,\qquad [\y^{a}_{\al},
y_{\gb}]_*=0\,.
\ee
The integration measure of the introduced star product is chosen such that $1 \ast 1 = 1$.

The operation $\pi$  in \eqref{nonlinear:0-sector} is defined as a star-product automorphism via the following reflection:
\be
\pi f(y, \vec\y):=f(-y, \vec\y)\,.
\ee
In terms of the star-product commutators, the $sp(2)$-invariant $P_a$ and $M_{ab}$ form representation of $so(d-1,2)$: 
\begin{subequations}
    \begin{align}
        [P_a, P_b]_{\ast} &= i M_{ab} \, , \\
        [M_{ab}, P_c]_{\ast} &= i (\eta_{ac} P_b - \eta_{bc} P_a) \, , \\
        [M_{ab}, M_{cd}]_{\ast} &= i (\eta_{ac}M_{bd} - \eta_{ad}M_{bc} - \eta_{bc}M_{ad}+\eta_{bd}M_{ac}) \, .
    \end{align}
\end{subequations}
Equation \eqref{spop} acquires a neat form of a singlet condition in terms of star product
\be\label{singlet} 
[t_{\al\gb}^0,f(y, \vec\y)]_*=0\,,
\ee
where
\be\label{t0}
t_{\al\gb}^0=\ff{1}{4i}(\y^{a}_{\al}\y_{a\gb}+y_{\al}y_{\gb})
\ee
are the generators of the $sp(2)$ algebra
\be\label{sp20}
[t^0_{\al\gb},
t^{0}_{\gga\gd}]_*=\gep_{\al\gga}t_{\gb\gd}^0+\gep_{\gb\gga}t_{\al\gd}^0+
\gep_{\al\gd}t_{\gb\gga}^0+ \gep_{\gb\gd}t_{\al\gga}^0\,.
\ee
Thus, we arrive at the concept of the off-shell HS algebra as an
associative algebra with the product given by \eqref{moyal}, while
its elements are the $sp(2)$ singlets \eqref{singlet}. It is clear
that star product of any two such elements $f_1*f_2$ is again
an $sp(2)$ singlet
\be
[t^0_{\al\gb}, f_1*f_2]_*=[t^0_{\al\gb},
f_1]_**f_2+f_1*[t^0_{\al\gb}, f_2]_*=0\,.
\ee
To describe the HS dynamics, the fields \eqref{w} and \eqref{C} have to be
properly traceless (for example, Lorentz traceless).
Unfortunately, the tracelessness  condition is not protected by the star
product \eqref{moyal}. As a result, Eqs.
\eqref{nonlinear} do not
describe propagation of fields for the traceful master fields $\go$
and $C$; rather, they provide a set of the generalized Bianchi
identities for the HS Weyl tensors and  a set of conditions
that relate certain field components to derivatives of other fields.
This explains the name of the corresponding off-shell HS algebra.
Similarly, the vertices on the right-hand sides of
\eqref{nonlinear} are called
off-shell or, equivalently, unconstrained.

The on-shell HS algebra responsible for real field dynamics
arises from the off-shell one as a factor algebra upon stripping
off the two-sided ideal spanned by elements of the form
\be
A^{\al\gb}*t^0_{\al\gb}=t^0_{\al\gb}*A^{\al\gb}\,,
\ee
where $A^{\al\gb}(y,\vec\y)$ transforms as a symmetric $sp(2)$
tensor under the adjoint action of \eqref{t0}. The problem here is
once the vertices on the right of
\eqref{nonlinear} result from
a certain deformation of the off-shell algebra, its ideal deforms
too. Thus, one needs the manifestly field deformed ideal in order to
set the HS equations on shell
\be
t^{0}_{\al\gb}\to t_{\al\gb}[C]\,.
\ee
The required deformation is described using the generating system, which will be given below.

\subsection{HS (non)locality}

As mentioned, the unfolded equations \eqref{nonlinear} contain huge spectrum of fields. The physical ones are the spin $s$ gauge potential \eqref{w} with $n=0$
\begin{equation}\label{w:phys}
  \go^{a(s-1)}=\overbrace{\ytableausetup{mathmode, boxsize=1.25em}
\begin{ytableau}
{} & {\bullet} & {\bullet}  & {\bullet} & {}
\end{ytableau}}^{s-1}\,,
\end{equation}
and their gauge invariant (linearized) field strengths stored in \eqref{C} at $m=s$ 
\begin{equation}\label{C:phys}
    C^{a(s),\, b(s)}=\overbrace{\ytableausetup{mathmode, boxsize=1.25em}
\begin{ytableau}
{} & {\bullet} & {\bullet}
& {\bullet} & {} \\
{} & {\bullet} & {\bullet}
& {\bullet} & {}
\end{ytableau}}^{s}\,.
\end{equation}
These are supposed to be properly traceless in order to describe on-shell dynamics. For example, the $s=2$ case identifies \eqref{C:phys} with the standard Weyl tensor, while \eqref{w:phys} with the gravitational frame field (vielbein). The rest components packed in \eqref{w} and \eqref{C} are expressed in terms of  physical fields through equations of motion. At the free level on AdS background, for example, such a relation for the Weyl module is organized via derivatives 
\begin{equation}\label{C:nphys}
    C^{a(s+n),\, b(s)}=\overbrace{D^{a}\dots D^a}^{n}C^{a(s),\, b(s)}+\dots\,,
\end{equation}
where $D^a$ is the AdS-covariant derivative, and we omitted the additional lower derivative contribution, as well as terms that ensure the above expression is traceless and properly Young-symmetric. Here, it is important that the number of derivatives is finite. At higher orders  Eq. \eqref{C:nphys} generally receives nonlinear corrections, the structure of which might contain the whole infinite Weyl module of a given spin, e.g., 
\begin{equation}\label{CC:contr}
    \sum_n C^{..b(n)\,, \dots}C_{..b(n),}{}^{\dots}\,.
\end{equation}
Above we omitted the index structure except for contractions over first rows of the Young diagrams \eqref{C}. Taking into account that the depth of the first row is proportional to the number of derivatives relating the physical field to its descendant \eqref{C:nphys}, the unlimited amount of contractions along the first rows carries over into the unlimited amount of space-time derivatives. The latter implies nonlocality. Summarizing at this stage, it is convenient to introduce the following definition: Whenever vertices on the right of \eqref{nonlinear} contain unbounded contractions of the Weyl modules $C$'s of fixed spins $s_i$, one calls such vertices {\it spin nonlocal}. Otherwise the vertices are {\it spin local}; see \cite{Gelfond:2018vmi}. The distinguished role of the Weyl module $C$ in context of locality is related to the fact that it contains infinitely many descendants for a given physical field, as opposed to the gauge module $\omega$, which is finite dimensional once its spin is fixed. 

Recent progress in understanding the HS (non)locality has led to a certain refinement of the notion of spin locality \cite{Didenko:2018fgx, Vasiliev:2022med}. Let us give the necessary definitions and some consequences thereof.
\begin{itemize}
    \item Vertex $\mathcal{V}(\omega_{s'}, \omega_{s''}, C_{s_1}\dots C_{s_n})$ is said to be {\it ultralocal} if it is spin local and the amount of index contractions between various $C_i$ is bounded by a number $N(s', s'')$ independent of $s_i$. Ultralocality of the $\mathcal{V}$ sector plays a crucial role in the HS locality problem. Namely, it turns out that there would be no spin locality at order $C^n$ once $\mathcal{V}$ was not ultralocal at order $C^{n-1}$, \cite{Didenko:2022qga}.  Ultralocal HS vertices were first identified in \cite{Didenko:2018fgx}. Ultralocality was then shown \cite{Didenko:2022qga} to be the characteristic property of the whole HS holomorphic gauge sector in four dimensions. Moreover, as follows from \cite{Gelfond:2019tac}, ultralocality implies the usual space-time locality.

    \item Another important concept introduced in \cite{Vasiliev:2022med} is {\it projectively compact spin locality}. A HS vertex is said to be projectively compact spin local if (i) it is spin local and (ii) its projection on the first descendant is nonzero only for those spins, which are not 'too far away from each other'. Specifically, consider the contribution to the physical sector
    \begin{equation}
        \dr_x C_{a(s_0), b(s_0)}=\Big[\Upsilon(s_1, \dots, s_n)\Big]_{a(s_0), b(s_0)}\,,
    \end{equation}
    which can be decomposed as follows:
    \begin{equation}\label{projection}
        \dr_x C_{a(s_0), b(s_0)}=e^c \Upsilon_{a(s_0)c, b(s_0)}(s_1, \dots, s_n)+\dots\,,
    \end{equation}
    where $e^a$ stands for the gravitational frame field\footnote{The original definition of \cite{Vasiliev:2022med} is confined to the AdS background.}, while $\Upsilon_{a(s_0)c, b(s_0)}$ denotes projection on the first descendant specified by a diagram with one extra box in the first row. Other contributions are omitted. Then, it is required that  
    \begin{equation}\label{proj-comp}
        \Upsilon_{a(s_0+1), b(s_0)}(s_1, s_2,\dots, s_k+t_k, \dots, s_n)=0\quad\text{for}\quad t_k>t^0_k
    \end{equation}
   with some constant $t^0_k$ and any $k=0\dots n$. The relevance of this notion has to do with the fact that the projectively compact spin-local vertices \eqref{proj-comp} are proved to be local in the usual space-time sense (not only in the fiber), \cite{Vasiliev:2022med}. For example, HS vertices of the holomorphic theory in four dimensions analyzed in \cite{Didenko:2022qga} belong to this class and, therefore, are space-time local to all orders. An intuition behind condition \eqref{proj-comp} is as follows. Suppose a given vertex is spin local. This implies that descendants should be expressed via physical fields in a way that contains finite number of derivatives. This does not exclude a situation when a descendant of {\it any} spin contains derivatives of a certain fixed spin $s_0$. Whenever this happens the sum over all spins yields an infinite derivative tail associated with the spin $s_0$. The condition \eqref{proj-comp} prevents this from happening and implies spin locality is equivalent to the space-time locality \cite{Vasiliev:2022med}.            
\end{itemize}

In the event the fields \eqref{w:phys} and \eqref{C:phys} are unconstrained, i.e., no conditions on traces imposed, one has no field dynamics,\footnote{As an illustration, take the pure spin two case. Traceless \eqref{C:phys} is the 'unfolded' way of saying the linearized Einstein equations imposed. Indeed,  differential constraints that follow from \eqref{nonlinear} amount to the linearized Bianchi identities in this approximation, meaning that the $s=2$ diagram \eqref{C:phys} is the linearized Riemann tensor, while its tracelessness is equivalent to Ricci=0 i.e., the vacuum Einstein equations. } while the fields themselves should be called primaries, rather than physical. In this case, all the above notions of the on-shell locality literally apply. The meaning of the locality is different, however. It tells one whether the nonlinear HS Bianchi identities for  primaries are local or not. One of the results of our study and \cite{Didenko:2023vna} is the unconstrained vertices from \eqref{nonlinear:1-sector} are ultralocal, while those from \eqref{nonlinear:0-sector} are projectively compact spin local. Moreover, these vertices contain the {\it minimal} number of derivatives. In the on-shell situation this would imply space-time locality of the HS theory, while in the unconstrained case, one can only claim that the nonlocality, if present, is associated with the ideal of the off-shell HS algebra \cite{Didenko:2023vna}.

\subsection{Generating equations}
As shown in \cite{Didenko:2023vna}, the unconstrained (off-shell) vertices on the
right-hand sides of
\eqref{nonlinear:1-sector}-\eqref{nonlinear:0-sector} can be
systematically recovered from the following Vasiliev-like system\footnote{The crucial difference from the original Vasiliev equations is the choice of the large star-product algebra and the lack of the $z$-dependent $B$ module (see \cite{Didenko:2022qga} for more detail.)}
\begin{subequations}
    \label{generating}
    \begin{align}
        \label{generating:1-sector}& \dr_x W + W \ast W = 0\,, \\
        \label{generating:W}& \dr_z W + \{W, \Lambda\}_{\ast} + \dr_x \Lambda = 0\,, \\
        \label{generating:0-sector}& \dr_x C + \big(W(z'; y, \Vec{\bf{y}}) \ast C - C \ast W(z'; -y, \Vec{\bf{y}}) \big)\Big|_{z'=-y} =
        0\,.
    \end{align}
\end{subequations}
Here, $C = C(y, \Vec{\bf{y}}| x)$ is the same $z$-independent field
\eqref{Cman}, while $\Lambda$ is the following $\dr z$ form:
\begin{equation}
    \label{LambdaDef}
    \Lambda[C] = \dr z^\alpha \, z_\alpha \int_0^1 \dr \tau \, \tau C(-\tau z, \Vec{\bf{y}}) \, e^{i \tau z_\alpha
    y^\alpha}\,,
\end{equation}
which satisfies
\begin{equation}
    \label{LambdaProp}
    \dr_z \Lambda = C \ast \gamma, \quad \gamma = \frac{1}{2} \, e^{izy} \dr z^\beta \dr z_\beta, \quad \dr_z :=
    \dr z^\alpha \frac{\partial}{\partial z^\alpha}\,,
\end{equation}
where we use the following short-hand notation for the index contraction:
\begin{equation}\label{indxconv}
    z_\alpha y^\alpha := zy = - yz\,.
\end{equation}
The evolution along $z$ governed by \eqref{generating:W} suggests
that $W$ must have the following form:
\begin{equation}
    \label{PowerExpansion}
    W(z; Y) = W^{(0)}(z; Y) + W^{(1)}(z; Y) + ...
\end{equation}
where $k$ in $W^{(k)}(z; Y)$ means the order of the perturbative expansion in powers of $C$. So, for the 0th power, we have
\begin{equation}\label{W0}
    \dr_z W^{(0)}(z; Y) = 0 \longrightarrow W^{(0)}(z; Y) = \omega(Y)
\end{equation}
with $\omega(Y)$ being a $z$-independent $\dr x$ form that enters
\eqref{nonlinear}. One then reconstructs the equations \eqref{nonlinear} in the following way. First, one determines $W(z; Y)$ in terms of $\omega$ and $C$ from \eqref{generating:W} and then
substitutes the result into \eqref{generating:0-sector},
\eqref{generating:1-sector}.
The necessary details on the system \eqref{generating} are available
in \cite{Didenko:2022qga, Didenko:2023vna}. Here, we briefly point out its most important features.

\begin{itemize}
    \item As typical of the Vasiliev system, while $W$ is
$z$ dependent, \eqref{PowerExpansion}, the left-hand side of
\eqref{generating:1-sector} is not. Its $z$ independence follows
from the very special evolution of $W$ along $z$ driven by
\eqref{generating:W}.

\item Equation \eqref{generating:0-sector} is not independent. It
follows from \eqref{generating:W} as a consequence of the
so-called projective identities on the functions $\phi(z, y, \vec\y)$
from a special class that $W$ as a solution of \eqref{generating:W} belongs; see \cite{Didenko:2022qga},
\be\label{iden}
\dr_z (\phi*\Lambda)=(\phi(z',y;
\vec\y)*C)\Big|_{z'=-y}*\gga\,,\qquad \dr_z
(\Lambda*\phi)=(C*\phi(z',-y; \vec\y))\Big|_{z'=-y}*\gga\,.
\ee
Their validity critically relies on the concise form of $\Lambda$
from \eqref{LambdaDef} and the specific star product in
\eqref{generating} defined as
\be\label{limst}
(f*g)(z; Y)= \int f\left(z+u', y+u; \y \right)\star
g\left(z-v,y+v+v'; \y \right)
\exp({iu_{\al}v^{\al}+iu'_{\al}v'^{\al}})\,,
\ee
where $\star$ is the Moyal part of the star product \eqref{moyal}, which
acts on $\vec\y$ only
\be\label{starvec}
(f\star g)(\vec\y)= \int f\left(\vec\y+\vec{\mathbf{u}}\right)
g\left(\vec\y+\vec{\mathbf{v}} \right)
\exp({i\vec{\mathbf{u}}_{\al}\cdot\vec{\mathbf{v}}^{\al}})\,.
\ee
The integration measures of the introduced star-product definitions are chosen such, that $1*1=1\star 1=1$.
The manifest action of the generating variables then reads
\begin{subequations}\label{star:gen}
\begin{align}
&y* =y+i\ff{\p}{\p y}-i\ff{\p}{\p z}\,,\qquad z* =z+i\ff{\p}{\p
y}\,,\\
&* y=y-i\ff{\p}{\p y}-i\ff{\p}{\p z}\,,\qquad * z=z+i\ff{\p}{\p
y}\,,\\
&\vec\y* =\vec\y+i\ff{\p}{\p \vec\y}\,,\qquad *
\vec\y=\vec\y-i\ff{\p}{\p \vec\y}\,.
\end{align}
\end{subequations}
Notice, while the product of $z$ is not pointwise, it commutes with any function. 

The origin of \eqref{iden} is little understood, except that its
direct verification can be carried out, \cite{Didenko:2022qga}. Perhaps, the
observed projection has something to do with the fact that the element
\be
\gd=e^{izy}
\ee
manifesting in \eqref{LambdaProp} behaves like a $\gd(z)$ function
under star product \eqref{limst},
\be\label{delta}
f(z)*\gd=\gd*f(z)=\gd\cdot f(0)\,.
\ee

\item In the process of the order by order reconstruction of $W$ from
\eqref{generating:W} it is convenient to fix ambiguity of
the $z$-independent part called $\go(Y)$ at the first step \eqref{W0} in
such a way that
\be\label{can}
W=\go(Y|x)+W_1(z, Y|x)+W_2(z, Y|x)+\dots\,,\quad W_n(0,
Y|x)=0\,,\quad\forall n\geq 1\,.
\ee
This way one introduces the so-called {\it canonical embedding}.
Its great advantage is the resulting HS vertices are all-order
spin (ultra)local, \cite{Didenko:2022qga, Didenko:2023vna}.

\item The equations \eqref{generating} are invariant under the local gauge
transformations parametrized by $\epsilon$
\begin{subequations}
    \label{gauge}
    \begin{align}
        \label{gauge:W}& \delta_{\epsilon} W = \dr_x \epsilon + [W, \epsilon]_{\ast} \\
        \label{gauge:Lambda}& \delta_{\epsilon} \Lambda = \dr_z \epsilon + [\Lambda, \epsilon]_{\ast} \\
        \label{gauge:C}& \delta_{\epsilon} C = -\big(\epsilon(z'; y, \Vec{\bf{y}}) \ast C - C \ast \epsilon(z'; -y, \Vec{\bf{y}}) \big)\Big|_{z'=-y}\,.
    \end{align}
\end{subequations}
The above transformations leave \eqref{generating} and
\eqref{LambdaProp} invariant, but not yet \eqref{LambdaDef}. In
order to keep \eqref{LambdaDef} intact one has to require
\begin{equation}
    \delta_{\epsilon} \Lambda[C] = \Lambda[\delta_\epsilon C]\,,
\end{equation}
which is equivalent to
\begin{equation}\label{epseq}
    \dr_z \epsilon + \big[\Lambda[C], \epsilon\big]_{\ast} - \Lambda[\delta_\epsilon C] = 0\,.
\end{equation}

\item The equations \eqref{generating}
possess the $sp(2)$ global symmetry spanned by the generators $t_{\al\gb}$,
\begin{align}
&[t_{\al\gb},
t_{\gga\gd}]_*=\gep_{\al\gga}t_{\gb\gd}+\gep_{\gb\gga}t_{\al\gd}+
\gep_{\al\gd}t_{\gb\gga}+ \gep_{\gb\gd}t_{\al\gga}\,,\label{1}\\
&\gd_{t_{\al\gb}}W=\dr_x t_{\al\gb}+[W, t_{\al\gb}]_*=0\,,\label{2}\\
&\gd_{t_{\al\gb}}\Lambda=\dr_z t_{\al\gb}+[\Lambda,
t_{\al\gb}]_*=0\,.\label{3}
\end{align}
This symmetry defines deformation of the traceful ideal from the
off-shell HS algebra discussed in the previous section. Once the
generators $t_{\al\gb}$ are known, the equations \eqref{nonlinear} can
be set on shell. The form of these generators is remarkably simple
for the canonical embedding \eqref{can}, \cite{Didenko:2023vna}:
\be
t_{\al\gb}=t^0_{\al\gb}-z_{\al}z_{\gb}\int_{0}^{1}\dr\tau\,\tau(1-\tau)\,C(-\tau
z, \vec\y)\,e^{i\tau zy}\,,
\ee
where $t^0$ is defined in \eqref{t0}.
\end{itemize}

\subsection{Reality conditions}
Let us specify the reality conditions that have been briefly mentioned. These are driven by the two requirements: 
\begin{enumerate}
    \item The Hermitian conjugation must be an involution of the star product \eqref{limst}
    \begin{equation}
        (f \ast g)^\dag = g^\dag \ast f^\dag \, ,
    \end{equation}
    from which the following conditions on $Y$ and $z$ arise:
    \begin{equation}
        \label{RealVar}
        (y_\alpha)^\dag = y_\alpha, \quad (\y_\alpha)^\dag = \y_\alpha, \quad (z_\alpha)^\dag = -z_\alpha \, .
    \end{equation}
    \item The generating equations \eqref{generating} must be consistent with the reality conditions on $W(z; Y)$ and $\Lambda[C]$ via involution. Regarding \eqref{RealVar}, one can see that 
    \begin{equation}
        \label{RealMaster}
        \Big(W(z; Y) \Big)^\dag = -W(z; Y), \quad \Big( \Lambda(z; Y)\Big)^\dag = -\Lambda(z; Y)
    \end{equation}
is consistent with the generating equations.
\end{enumerate}

Necessity of some reality conditions is driven by the free level analysis. At the free level the fields $\omega^{a(s-1), b(n)}$ and $ C^{a(s+k), b(s)}$ originate from covariant derivatives of \textit{real} Fronsdal fields $\phi_{\mu(s)}$. Therefore, descendant fields must be real. Due to $W(z; Y)$ and $\Lambda(z; Y)$ containing $\omega(Y)$ and $C(Y)$, some reality conditions on $z$-dependent fields must be imposed. However, $W(z; Y)$ and $\Lambda(z;Y)$ obey \eqref{generating}, and so the reality conditions should be consistent with these equations. For example, one may try to set
\begin{equation}
     \Big(W(z; Y) \Big)^\dag = +W(z; Y)\,.
\end{equation}
However, it comes in conflict with \eqref{generating:1-sector}
\begin{equation}
    0 = (\dr_x W + W \ast W)^\dag = \dr_x W - W \ast W\,.
\end{equation}

Due to the fact that $\omega(Y | x)$ is defined as the 0th power of the perturbative expansion in $W(z; Y)$, Eq. \eqref{RealMaster} provides the following reality condition for $\omega(Y | x)$:
\begin{equation}
    \label{RealCondW}
    \big(\omega(Y|x) \big)^\dag = - \omega(Y|x) \, .
\end{equation}
Since $\Lambda[C]$ is defined via $C(Y | x)$, the reality condition for $\Lambda[C]$ translates into the reality condition for $C(Y | x)$
\begin{equation}
    \label{RealCondC}
    \big(C(Y|x) \big)^\dag = -\pi\big( C(Y|x) \big) \, .
\end{equation}
For the above reality conditions, the imaginary unit in \eqref{MFP} ensures the fields $\omega^{a(n), b(m)}, C^{a(n), b(m)}$ are real 
\begin{equation}
    \big(\omega^{a(n), b(m)} \big)^\dag = \omega^{a(n), b(m)}, \, \quad \big( C^{a(n), b(m)} \big)^\dag =  C^{a(n), b(m)} \, .
\end{equation}

\section{Perturbative analysis}\label{Sec: Pert}
The perturbation theory for \eqref{generating} amounts to the calculation of $W$ from \eqref{generating:1-sector} order by order. Having $W$
at a given order in the $C$ expansion, one recovers the corresponding
vertices as follows:
\begin{subequations}
    \begin{align}
        \label{Ups1}&\mathcal{V}(\omega, \omega, C^n) = -\Big(\sum_{k+m=n}W^{(k)} \ast W^{(m)} +(\dr_x W)^{(n)}\Big)\,,\\
        \label{Ups0}&\Upsilon(\omega, C^n) = -\Big(W^{(n)}(z'; y, \Vec{\bf{y}}) \ast C - C \ast W^{(n)}(z'; -y, \Vec{\bf{y}}) \Big)\Big|_{z'=-y}\,.
    \end{align}
\end{subequations}
Due to the $z$ independence of \eqref{generating:1-sector}, the rhs of \eqref{Ups1} is actually $z$ independent, so we conveniently set $z=0$, which, thanks to the canonical embedding \eqref{can}, gives
\begin{equation}
    \label{V1}
    \mathcal{V}(\omega, \omega, C^n) = -\Big(\sum_{k+m=n}W^{(k)} \ast W^{(m)} \Big)\Big|_{z=0}\,.
\end{equation}
More details, as well as the explicit form of the particular
$n$-order vertex, are provided in \cite{Didenko:2023vna}. Here we complete the initiated earlier analysis for all vertices \eqref{nonlinear} throughout.

\label{sec:Section2} \index{Section2}
\subsection{Source prescription}
In what follows, we use the source prescription of \cite{Didenko:2018fgx} that facilitates much the star-product calculation and improves readability of
the final expressions. A simple idea behind is the star product of two
exponentials is easy to calculate.  So, it is convenient to use
the following \textit{source prescription}:
\begin{subequations}
    \begin{align}
    & \omega(Y) = \omega(y_\alpha, \Vec{\bf{y}}) = \exp\bigg(y_\alpha \frac{\partial}{\partial y^t_\alpha}\bigg)\omega(y^t_\alpha,\Vec{\bf{y}})\Big|_{y^t_\alpha=0} = e^{-iy_\alpha t^\alpha}\omega(y^t_\alpha,\Vec{\bf{y}})\Big|_{y^t_\alpha=0}\,,\label{w:source} \\
    &C(Y) = C(y_\alpha, \Vec{\bf{y}}) = \exp\bigg( y_\alpha \frac{\partial}{\partial y^p_\alpha} \bigg) C(y^p_\alpha,\Vec{\bf{y}})\Big|_{y^p_\alpha=0} = e^{-iy_\alpha p^\alpha}C(y^p_\alpha,\Vec{\bf{y}})\Big|_{y^p_\alpha=0}\,,\label{C:source}
    \end{align}
\end{subequations}
where we introduced
\begin{equation}
    \label{source}
    t^\alpha := i \frac{\partial}{\partial y^t_\alpha}  \,, \quad  p^\alpha := i \frac{\partial}{\partial y^p_\alpha}\, .
\end{equation}
Although $t^\alpha$ and $p^\alpha$ are differential operators, we will treat them just as formal commuting $sp(2)$ variables in our calculations because such representation is equivalent to the Fourier integral representation.

The source prescription implies that one operates with the exponentials
leaving fields $\go$ and $C$ aside. Notice also that we do not
introduce source for the dependence on $\vec\y$. The reason is the
off-shell HS algebra remains undeformed in its $\vec\y$ part.
Therefore, the dependence on $\vec\y$ in \eqref{nonlinear} is
organized plainly via the star product \eqref{starvec}. Since
$\Vec{\bf{y}}$ is ``passive'' in all calculations, $W^{(n)}$ has the
following structure:
\begin{equation}
    \label{W-Sources}
    W^{(n)}(z; Y) = \sum_{k=0}^{n} \mathcal{W}^{(k|n)}(z; y | t; p)\big(\overbrace{C...C}^{k} \, \omega \,\overbrace{C...C}^{n-k}\big) \, \, 
    \Big|_{y^t,y^p=0}\,,
\end{equation}
where $\mathcal{W}^{(k|n)}$ collects the \textit{source terms}, while
the lines composed out of $\omega$ and $C$ are
\begin{equation}
    \big(\overbrace{C...C}^{k} \, \omega \,\overbrace{C...C}^{n-k}\big) = \dots \star C\Big(y^{p_k}, \Vec{\bf{y}}  \Big) \star \omega \Big( y^t, \Vec{\bf{y}}  \Big) \star  C\Big(  y^{p_{k+1}}, \Vec{\bf{y}}  \Big) \star \dots
\end{equation}
with the convention that $y^{p_1}$ corresponds to the leftmost $C$, while $y^{p_n}$ to the rightmost one.
We also introduce sources for $\Upsilon(\omega, C^n)$ and
$\mathcal{V}(\omega, \omega, C^n)$ in the following manner:
\begin{subequations}
    \label{VertexSourcesDef}
    \begin{align}
        \label{VertexSourcesDef:0-form}&\Upsilon(\omega, C^n) = \sum_{k=0}^{n}\Phi^{[k]}_n(y|t; p) \, \big(\overbrace{C...C}^{k} \,
         \omega \,\overbrace{C...C}^{n-k}\big)  \Big|_{y^t,y^p=0}\,,\\
        \label{VertexSourcesDef:1-form}&\mathcal{V}(\omega, \omega, C^n) = \, \,
        \smashoperator{\sum_{0\leqslant k_1 \leqslant k_2 \leqslant n} } \, \Psi^{[k_1, k_2]}_{n}(y|t_1, t_2; p) \, \big(\overbrace{C...C}^{k_1} \,
        \omega \, \overbrace{C...C}^{k_2-k_1} \, \omega \, \overbrace{C...C}^{n-k_2}\big)  \Big|_{y^t,  y^p=0}\,.
    \end{align}
\end{subequations}
Whenever $k\neq 0, n$ we refer of expressions like \eqref{W-Sources} or \eqref{VertexSourcesDef:0-form} as having $\go$ {\it impurity} at the $k$th position. Similarly, for \eqref{VertexSourcesDef:1-form} if the line of $C$'s has a spacer of one or two $\go$'s in between, we call it as having $\go$ impurities at the respective positions, say $k_1$ and $k_2$. The line is called {\it pure} otherwise. An example of a pure line is $\go C\dots C$ or $C\dots C\go\go$.   

Among other things, the advantage of using the source prescription is the
analysis of spin locality becomes transparent this way. Namely, the absence of nonpolynomial contractions $(p_ip_j)$ within sources $\Phi$ and $\Psi$ guarantees their spin locality; see e.g., \cite{Gelfond:2018vmi}. Such contractions in terms of Young diagrams correspond to the index contractions over first rows between descendants \eqref{C}. Besides, the absence of the contractions $(yp_i)$ implies ultralocality \cite{Didenko:2018fgx} already discussed in Section \ref{sec:Section1}.

\subsection{Master field $W$}
Using \eqref{PowerExpansion}, from \eqref{generating:W}, one
arrives at the following iterative system of differential (with
respect to $z$) equations:
\begin{equation}
    \dr_z W^{(n+1)} + \{ W^{(n)}, \Lambda \}_{\ast} + (\dr_x \Lambda)^{(n+1)} =
    0\,,
\end{equation}
which can be solved via the standard contracting homotopy operator $\Delta$
at any given order\footnote{The standard homotopy operator \eqref{homotopy} annihilates $(d_x \Lambda)^{(n+1)}$ due to $\Lambda_\alpha \thicksim z_\alpha$ and $z^{\alpha}z_{\alpha}\equiv 0$.}
\begin{equation}
    \label{reccurent}
    \begin{cases}
        W^{(n+1)}_{\mu} = \Delta \big[ W^{(n)}_\mu \ast \Lambda - \Lambda \ast W^{(n)}_\mu \big] \\
        W^{(0)}_\mu = \omega_\mu\,,
    \end{cases}
\end{equation}
where we recall that $W$ is a space-time 1-form $W=\dr x^{\mu}
W_{\mu}$, while the standard homotopy operator is given by
\be
\label{homotopy}
\Delta\big(f_\alpha(z) \, \dr z^\alpha\big) = z^\alpha \int_0^1 d\tau f_\alpha(\tau z)\,.
\ee
In terms of sources, Eq.  \eqref{reccurent} becomes the
equation\footnote{With the convention $\mathcal{W}^{(-1|n-1)}
 = 0$ and $\mathcal{W}^{(n|n-1)} = 0$.} for sources
$\mathcal{W}^{(k|n)}$, where we recall $n$ is the order of perturbation, while $k$ counts the amount of $C$'s before $\go$ in the line of $C$'s; see \eqref{W-Sources}
\begin{equation}
    \label{reccurentMAIN}
    \begin{cases}
        \mathcal{W}^{(k|n)} = \Delta \big[\mathcal{W}^{(k|n-1)} \ast \Lambda' - \Lambda' \ast \mathcal{W}^{(k-1|n-1)}  \big], \, \, k \in [0, n] \\
        \mathcal{W}^{(0|0)} = e^{-iyt}
    \end{cases}
\end{equation}
with $\Lambda'$ being the source for $\Lambda$
\begin{equation}
    \Lambda'(z; y | p) = \dr z^\alpha \, z_\alpha \, \smashoperator{ \int_{0}^{1} } d\sigma \, \sigma \, e^{i\sigma z(y+p)} \longrightarrow \Lambda(z; Y) = \Lambda'(z; y | p) C\Big( y^p, \Vec{\bf{y}}  \Big)  \Big|_{y^p=0}\,.
\end{equation}
In solving \eqref{reccurentMAIN}, one should keep track of
the $p$ ordering. More specifically, in $\mathcal{W}^{(k|n-1)}
\ast \Lambda'$ the source term $\mathcal{W}^{(k|n-1)}$ has
arguments $p_1, ..., p_{n-1}$, and $\Lambda'$ has $p_{n}$, while
in $\Lambda' \ast \mathcal{W}^{(k-1|n-1)}$ the source term
$\Lambda'$ has argument $p_1$, and $\mathcal{W}^{(k|n-1)}$ has
arguments $p_2, ..., p_{n}$.

The full solution of system \eqref{reccurentMAIN} was not known, although the lowest order representatives $\mathcal{W}^{(0|1)}$ and $\mathcal{W}^{(1|1)}$ as well as the all-order source-term $\mathcal{W}^{(0|n)}$ corresponding to the leftmost ordering have been calculated in \cite{Didenko:2023vna}.

Our study generalizes these results to all orders $n$ and any $k
\in [0,n]$. Namely, we show that
\begin{equation}
    \label{W_k|n}
    \begin{split}
    \mathcal{W}^{(k|n)}&(z; y| t; p) = \\
    &= (-)^{k} \, (zt)^n \, \, \smashoperator{\int_{\Delta(n-1)}} d \lambda \, \, \,
    \smashoperator{ \int_{\Delta^{*}_{k,n}(\lambda) } } d\nu\smashoperator{ \int_{0}^{1} } d\tau \, (1-\tau) \tau^{n-1} \, e^{i\tau z(y+B_{k,n}) + i (1-\tau) y A_n - i\tau B_{k,n}A_n + i C_n} \, ,
    \end{split}
\end{equation}
where (mind the convention \eqref{indxconv}) 
\begin{subequations}
    \label{ABC}
    \begin{align}
        &A_n = -\Big(1 - \sum_{s=1}^{n}\nu_s \Big)t\,, \quad  C_n = \Big(\sum_{s=1}^{n} \nu_s p_s \Big)t\,, \\
        B_{k,n} =\sum_{s=1}^{n}&\lambda_s p_s + \Big[ -\sum_{s=1}^{k} \lambda_s + \sum_{s=k+1}^{n} \lambda_s + \sum_{i<j}^{n}(\lambda_i \nu_j  - \lambda_j \nu_i ) \Big]t \,,
    \end{align}
\end{subequations}
and the integration domains are
\begin{subequations}
    \begin{align}
        & \Delta(n-1) = \big\{\lambda_1 + ... +\lambda_n = 1 \,,\quad \lambda_i\geq 0\,\big\}\,,\label{lmbddmn} \\
        & \Delta^{*}_{k,n}(\lambda) = \begin{cases}
        \nu_1 + ... +\nu_n\leqslant 1 \,,\quad \nu_i\geq 0\,, \\
        \nu_i \lambda_{i+1} - \nu_{i+1} \lambda_{i} \leqslant 0\,,\quad i \in [1\,,\,k-1] \\
        \nu_i \lambda_{i+1} - \nu_{i+1} \lambda_{i} \geqslant 0\,,\quad i \in [k+1\,,\,n-1]\,. \label{eq:W-domain}
    \end{cases}
    \end{align}
\end{subequations}
The range of $i$ for the $\lambda$-dependent constraints in the definition of $\Delta^{*}_{k,n}(\lambda)$ should be understood in the following way:
\begin{itemize}
    \item If $k-1 < 1$, then there is not the associated $\nu_i \lambda_{i+1} - \nu_{i+1} \lambda_{i} \leqslant 0$ condition.
    \item Analogously, if $k+1 > n-1$, the associated constraint  $\nu_i \lambda_{i+1} - \nu_{i+1} \lambda_{i} \geqslant 0$ is also lacking.
\end{itemize}
To make it clear, we provide the following examples: 
\begin{equation*}
    \Delta^{*}_{0,2}(\lambda) = \begin{Bmatrix}
        \nu_1 + \nu_2 \leqslant 1\,, \quad \nu_i \geqslant 0\,, \\
        \nu_1 \lambda_2 - \nu_2 \lambda_1 \geqslant 0
    \end{Bmatrix}\,, \quad \Delta^{*}_{2,2}(\lambda) = \begin{Bmatrix}
        \nu_1 + \nu_2 \leqslant 1\,, \quad \nu_i \geqslant 0\,, \\
        \nu_1 \lambda_2 - \nu_2 \lambda_1 \leqslant 0
    \end{Bmatrix}\,,
\end{equation*}
\begin{equation*}
    \Delta^{*}_{1,2}(\lambda) = \begin{Bmatrix}
        \nu_1 + \nu_2 \leqslant 1\,, \quad \nu_i \geqslant 0
    \end{Bmatrix}\,, \quad \Delta^{*}_{2,4}(\lambda) = \begin{Bmatrix}
        \nu_1 + ... +\nu_4 \leqslant 1\,, \quad \nu_i \geqslant 0\,, \\
        \nu_1 \lambda_2 - \nu_2 \lambda_1 \leqslant 0 \\
        \nu_3 \lambda_4 - \nu_4 \lambda_3 \geqslant 0
    \end{Bmatrix}\,.
\end{equation*}
Notice that the integration domain corresponding to the pure line has always $n-1$ shoelace constraints at order $n$, while the one containing the $\go$ impurity, -- $n-2$. We leave derivation of the main result \eqref{W_k|n} to the Appendix \ref{W-derivation}.

\subsection{Vertices $\Upsilon(\omega, C^n)$}
Computing the simple star product in \eqref{Ups0}, we find that
\begin{equation}
    \label{eq:0-VertexFromSources}
    \begin{split}
        \Phi^{[k]}_n(y|t; p_1, & ..., p_n)
        = \mathcal{W}^{(k-1|n-1)}(-y; -y - p_1 | t; p_2, ..., p_n)e^{-iyp_1} - \\ &- \mathcal{W}^{(k|n-1)}(-y; y - p_n | t; p_1, ...,
        p_{n-1})e^{-iyp_n}\,,
    \end{split}
\end{equation}
which upon change of the integration variables takes the following
neat structure:
\begin{equation}
    \label{eq:0-VertexSource}
\Phi^{[k]}_n(y|t; p_1, ..., p_n)=(-)^{k+1}(ty)^{n-1}
\smashoperator{\int_{\mathcal{D}^{[k]}_n} } d\xi d\eta \,
e^{-iy P_n(\xi) - it P_n(\eta) +  i (ty) \cdot S^{[k]}_{n} }\,,
\end{equation}
where
\begin{equation}\label{PS}
    P_n(\zeta) = \sum_{s=1}^{n} \zeta_s p_s\,, \quad S^{[k]}_n = -\sum_{s=1}^{k} \xi_s + \sum_{s=k+1}^{n} \xi_s + \sum_{i<j}^{n}(\xi_i \eta_j - \xi_j \eta_i)\,,
\end{equation}
while the integration domain $\mathcal{D}^{[k]}_n$ is
\begin{equation}
    \label{eq:0-VertexDomain}
    \mathcal{D}^{[k]}_n =
    \begin{cases}
        \eta_1 + ... + \eta_{n} = 1\,,\quad\eta_i\geq 0\,, \\
        \xi_1 + ... + \xi_{n} = 1\,,\quad\xi_i\geq 0\,, \\
        \eta_i \xi_{i+1} - \eta_{i+1} \xi_{i} \leqslant 0 \,, \quad i \in [1 \,,\, k-1]\,, \\
        \eta_i \xi_{i+1} - \eta_{i+1} \xi_{i} \geqslant 0\,, \quad i \in [k+1\,,\,n-1]\,.
    \end{cases}
\end{equation}
and we used the same convention for the range of $i$ as in \eqref{eq:W-domain}.

Using the definitions \eqref{VertexSourcesDef:0-form} and
\eqref{source} we can complete differentiation with respect to
sources by separating the dependence on $t$ and $p_i$ using the following integration formula:
\be
e^{-iyP_n(\xi) - itP_n(\eta) + i(ty) \cdot S^{[k]}_n} =  \int \frac{d^2u d^2 v}{(2\pi)^2} \, e^{iuv -iut-iyP_n(\xi) - i v P_n(\eta) - i (yt) S^{[k]}_n}\,,
\ee
which features the translation operators associated with the exponentials
of $t$ and $p$. The final result is then readily available
\begin{equation}
    \label{verC}
    \begin{split}
        &\Upsilon(\omega, C^n) =
    (-i)^{n-1}\sum_{k=0}^{n}(-)^{k+1}\smashoperator{\int_{\mathcal{D}_{n}^{[k]}} } d\xi d\eta \, 
    \int \frac{d^2 u d^2 v}{(2\pi)^2} \,e^{iuv}\times\\
    &\times\left({\prod_{i=1}^{k}}\star C(\xi_i y+\eta_i v,
    \vec\y)\right)\star (y^{\al}\p_{\al})^{n-1}\go (S_{n}^{[k]}y+u,
    \vec\y)\star \left({\prod_{j=k+1}^{n}}\star C(\xi_j y+\eta_j v,
\vec\y)\right)\,.
    \end{split}
\end{equation}
Here $\star$ defined in \eqref{starvec} does not act on $y$. For
$k=0$ or $k=n$ the corresponding left or right product $\prod$ is
missing. For example, for $n=1$ one has
\begin{align}\label{Upsilon:wC}
&\Upsilon(\omega, C)=-\int e^{iuv}\left(\go(y-u, \vec\y)\star
C(y-v, \vec\y)-C(y-v,\vec\y)\star\go(-y-v,\vec\y)\right)=\nonumber\\
&=-\go*C+C*\pi(\go)\,,
\end{align}
which is the standard twisted-adjoint vertex on the left of
\eqref{nonlinear:0-sector}. Interestingly, even though \eqref{verC} was derived for $n\geqslant 2$, it can be applied for $n=1$.  Notice also that $\p_{\al}$
differentiates with respect to the first argument of $\go$.

\paragraph{Locality} Let us show that the vertex \eqref{verC} is projectively compact spin local and, therefore, space-time local. Recall, that for this to happen, the two conditions should be satisfied: (i) The vertex should be spin local, and (ii) its contribution to the first descendant must vanish for sufficiently distant spins. Vertices \eqref{verC} are clearly spin local. Indeed, for polynomial $\omega$, the integration over $u$ and $v$ yields finite derivatives with respect to first arguments of $C$'s and, therefore, finite number of contractions over first rows of the respective Young diagrams. Another way seeing this is Eq. \eqref{eq:0-VertexSource} contains no $p_ip_j$ contribution. 

Now let us look into the projection to primaries. Recall that a spin $s$ primary is contained in the rectangular diagram \eqref{C:phys}, while descendants have longer first rows. For example, the first descendant is given by 
\begin{equation}\label{C:1descendent}
    C^{a(s+1),\, b(s)}=\overbrace{\ytableausetup{mathmode, boxsize=1.25em}
\begin{ytableau}
{} & {\bullet} & {\bullet}
& {\bullet} & {} & {} \\
{} & {\bullet} & {\bullet}
& {\bullet} & {}
\end{ytableau}}^{s+1}\,.
\end{equation}
In terms of variables $Y=(y, \vec\y)$ indices from the first row are contracted with\footnote{This and $\y^a_{\al}\y^{b\al}$ are the only $sp(2)$-invariant combinations made of $Y$'s. The latter carrying vector indices $a$ and $b$ involves two boxes of Young diagram from both rows.} $\y_{\al}^{a}y^{\al}$ once they have no pair below on the second row. So, the first descendant is proportional to the first power of $y$: $C_s^{1}(Y|x)\sim y$ and so on. Therefore, vertices \eqref{verC} being proportional to $y^{n-1}$ do not contribute to equation for the primaries for $n> 1$, as we have \eqref{Upsilon:wC}
\begin{equation}\label{Eq:primary}
    \dr_x C_{a(s), b(s)}=\Big[\Upsilon(\omega, C) \Big]_{a(s), b(s)}\,.
\end{equation}
Now, since $\omega$ is a polynomial in $Y$ for a given fixed spins, it contains no more than finite number of rectangular Young diagrams. Hence,  large spins in $C$ with sufficiently large second rows cannot contribute to \eqref{Eq:primary} because they lead to the excess of the length $s$ of the second row. The result is the projection \eqref{projection} is zero for distant spins, while vertices \eqref{verC} are projectively compact spin local. In addition, as explained in \cite{Vasiliev:2023yzx}, the presence of the factor $(ty)^{n-1}$ in \eqref{eq:0-VertexSource} makes these vertices the {\it minimal} ones, i.e., containing the minimal number of derivatives.

Let us point out few other properties of \eqref{verC}. As mentioned, a peculiar property of \eqref{verC} is that $\Upsilon(\omega, C^n)$ as a function of $y_\alpha$ starts with $(n-1)$th power of $y_\alpha$, which implies that the vertices with $n \leqslant k+1$ only contribute to the equation for $C_{a(s+k), b(s)}$, 
\begin{subequations}\label{dC}
    \label{3.24}
    \begin{align}
        \dr_x C_{a(s+k), b(s)} = \sum_{n=1}^{k+1} \Big[\Upsilon(\omega, C^n) \Big]_{a(s+k), b(s)} \, , \\
        \forall n > k+1: \Big[\Upsilon(\omega, C^n) \Big]_{a(s+k), b(s)} = 0 \, . 
    \end{align}
\end{subequations}
This property simply follows from the fact that  $C_{a(s+k), b(s)}$ enters $C(y, \vec\y)$ accompanied with the $k$th power of $y_\alpha$ (recall the definition \eqref{Cman}). As a consequence of \eqref{3.24}, combined with spin locality of $\Upsilon(\omega, C^n)$, the equations for $ C_{a(s+k), b(s)}$ become polynomial in $ C_{a(m), b(n)}$. Let us stress once again that the form Eq. \eqref{dC} is a manifestation of the projectively compact spin locality.

Another interesting property follows from the fact that $\omega$ entering \eqref{verC} is supplied with $(n-1)$ derivatives. Since $\omega_{a(s-1), b(k-1)}$ appears in $\omega(y,\vec\y)$ accompanied with $(s-k)$th power of $y_\alpha$ (recall the definition \eqref{wman}), the following properties hold:
\begin{enumerate}
    \item The spin $s$ gauge field $\omega_s$ does not contribute to the vertices $\Upsilon(\omega, C^n)$ with $n> s$.
    \item The generalized spin connections $\omega_{a(s-1) \, , \, b(s-1)}$ do not contribute to the rhs of \eqref{nonlinear:0-sector}. 
\end{enumerate}

\subsection{Vertices $\mathcal{V}(\omega, \omega, C^n)$}
Having obtained master field $W$ in the form \eqref{W_k|n}, one can
straightforwardly compute the 1-form vertices using \eqref{V1}. In terms of the sources the result amounts to the following star products:
\begin{equation}
\label{PsiFromGen}
    \begin{split}
        &\Psi^{[k_1, k_2]}_{n}(y|t_1, t_2; p_1, . .., p_n) = \\
        &=-\Big( \sum_{m_1 + m_2 = n}\mathcal{W}^{(k_1|m_1)}(z; y| t_1; p_1, ..., p_{m_1}) \ast \mathcal{W}^{(k_2 - m_1|m_2)}(z; y|t_2; p_{m_1+1}, ..., p_{n}) \Big)\Big|_{z=0} \, .
    \end{split}
\end{equation}
Recall that $n$ stands here for the perturbation order (number of $C$'s) and $k_{1,2}$ point at the two $\omega$ impurities. 
While it is possible to complete the above gaussian
integration, it is interesting that the final result can be
extracted from the already found 0-form vertices
\eqref{eq:0-VertexSource}. Various relations between vertices
arise from the integrability constraints of \eqref{nonlinear} and the distinctive properties of \eqref{eq:0-VertexSource}. We call the aforementioned relations the \textit{vertex dualities} and leave their analysis to the Section \ref{section:DualRel}. Technical details on the derivation of $\Psi$ via the duality can be found in the Appendix \ref{sec:Appendix4}. Here we provide the following final form:
\begin{equation}
    \label{eq:1-VertexSource}
    \Psi_n^{[k_1, k_2]}(y|t_1, t_2; p) = (-)^{k_2 - k_1 + 1} (t_2 t_1)^{n} \smashoperator{\int\limits_{\mathcal{D}^{[k_1, k_2]}_{n+1} }  }
    d\xi d\eta  \, e^{-iy (\xi_{n+1}t_1 + \eta_{n+1}t_2) - it_1 P_n(\xi) - it_2 P_n(\eta) +  i (t_2 t_1) \cdot S^{[k_1, k_2]}_{n}} \,,
\end{equation}
where
\begin{equation}
    S^{[k_1, k_2]}_n = 1 - \sum_{s=1}^{k_2} \xi_s + \sum_{s=k_2+1}^{n}\xi_s +
    \sum_{s=1}^{k_1}\eta_s -
    \sum_{s=k_1 + 1}^{n}\eta_s + \sum_{i<j}^{n}(\xi_i \eta_j - \xi_j
    \eta_i)\,,
\end{equation}
and $P_n$ is defined in \eqref{PS}. The integration domain reads
\begin{equation}
    \label{eq:1-VertexDomain}
    \mathcal{D}^{[k_1, k_2]}_{n+1} =
    \begin{cases}
        \eta_1 + ... + \eta_{n+1} = 1, \quad \eta_i \geq 0 \, , \\
        \xi_1 + ... + \xi_{n+1} = 1, \quad \xi_i \geq 0 \, , \\
        \eta_i \xi_{i+1} - \eta_{i+1}\xi_i \geqslant 0, \quad i \in [0, k_1-1] \, ,  \\
        \eta_i \xi_{i+1} - \eta_{i+1}\xi_i \,  \boldsymbol{\leqslant }0, \quad i \in [k_1 + 1, k_2-1] \, , \\
        \eta_i \xi_{i+1} - \eta_{i+1}\xi_i \geqslant 0, \quad i \in [k_2 + 1, n]\,,
    \end{cases}
\end{equation}
and $\eta_0$ and $\xi_0$ should be identified with $\eta_{n+1}$ and $\xi_{n+1}$ correspondingly.

Similarly to the 0-form vertex \eqref{verC}, we can separate
variables $t$ and $p$ in \eqref{eq:1-VertexSource} by means of
extra integration. Specifically, recalling that $t$ and $p$ are just $-i\ff{\p}{\p y}$ acting on
the associated fields ($\omega$ and $C$), it is convenient to perform the following chain of identical transformations
\begin{align*}
    &e^{-iy(\xi_{n+1}t_1 + \eta_{n+1}t_2) - i t_1 P_n(\xi) - it_2 P_n(\eta) + i (t_2 t_1) \cdot S^{[k_1, k_2]}_{n}} = \\
    &=\int d^2 u d^2 v \, \delta(v-t_1) \delta(u-t_2) \, e^{-iy(\xi_{n+1}t_1 + \eta_{n+1}t_2) - i v P_n(\xi) - i u P_n(\eta) + i (u t_1) \frac{S^{[k_1, k_2]}_{n} }{2} - i (vt_2) \frac{S^{[k_1, k_2]}_{n} }{2} } = \\
    & = \int \frac{d^2 u d^2 u' d^2 v d^2 v'}{(2\pi)^4} \, e^{iv'(v-t)+iu'(u-t_2)-iy(\xi_{n+1}t_1 + \eta_{n+1}t_2) - i v P_n(\xi) - i u P_n(\eta) + i (u t_1) \frac{S^{[k_1, k_2]}_{n} }{2} - i (vt_2) \frac{S^{[k_1, k_2]}_{n} }{2} } \, .
\end{align*}
Relabeling now $v' \leftrightarrow u$, one arrives at 
\begin{equation}
    \label{verw}
    \begin{split}
        \mathcal{V}(&\omega, \omega, C^n) = \smashoperator{ \sum_{0\leqslant k_1 \leqslant k_2 \leqslant n} } \, (-)^{k_2 - k_1 + 1} \smashoperator{\int_{\mathcal{D}^{[k_1, k_2]}_{n+1} } } d\xi d\eta\, \times   \\
        & \times \int \frac{d^2 u d^2 u' d^2 v d^2 v'}{(2\pi)^4} \, e^{iuv + iu'v'}\Bigg( \prod_{i=1}^{k_1} \star C\Big(\xi_i v + \eta_i v', \vec\y \Big) \Bigg) \star  \\
        &  \star \partial^n \omega\Big(\xi_{n+1} y + u - \frac{S^{[k_1, k_2]}_{n} }{2} \, v', \vec\y \Big) \star \Bigg( \prod_{j=k_1+1}^{k_2} \star C\Big(\xi_j v + \eta_j v', \vec\y \Big) \Bigg) \star\\
        & \star \partial^n \omega\Big(\eta_{n+1} y + u' + \frac{S^{[k_1, k_2]}_{n} }{2} \, v, \vec\y \Big) \star \Bigg( \prod_{j=k_2+1}^{n} \star C\Big(\xi_l v + \eta_l v', \vec\y \Big) \Bigg)\,,
    \end{split}
\end{equation}
where $\p^n\go\dots\p^n\go$ should be understood as
$(\gep^{\al\gb}\p_{1\al}\p_{2\gb})^n\go\dots\go$ with $\p_1$
differentiating the first argument of $\go$ from the second line
in \eqref{verw}, while $\p_2$ does the same with $\go$ from the
third line. Let us stress that the dependence on $y$ drops off
the arguments of each $C$. This is a manifestation of ultralocality (see \cite{Didenko:2023vna}).
Let us also note, that \eqref{verw} captures
\be
\mathcal{V}(\go,\go)=-\go*\go
\ee
for $n=0$.

Although splitting $S^{[k_1, k_2]}_n$ in \eqref{verw} is ambiguous, it seems reasonable to divide it into two equal parts, because $S^{[k_1, k_2]}_n$ originates from the \textit{twice} of difference of areas of some polygons, which will be specified later in the text.  This splitting leads to a natural generalization of the geometric form of the Moyal star product which features the area of triangle; see, e.g., \cite{Zachos:1999mp}.  

An important feature of \eqref{verw} is that $\omega$ enters it via the $n$th derivative with respect to its first argument that leads to a conclusion similar to $\Upsilon(\omega, C^n)$: 
\begin{enumerate}
    \item Gauge part of spin-$s$ field ($\omega$) does not contribute to vertices $\mathcal{V}(\omega, \omega, C^n)$ with $n \geqslant s$.
    \item The generalized spin-connections $\omega_{a(s-1) \, , \, b(s-1)}$ do not contribute to the rhs of \eqref{nonlinear:1-sector}.
    \item Vertices are ultralocal (no $yp$ contractions in \eqref{eq:1-VertexSource}), implying space-time locality.
\end{enumerate}

\subsection{Gauge transformations}

In this section, we obtain manifest form of the gauge
transformations \eqref{gauge:W}, \eqref{gauge:C} that leave
Eqs. \eqref{nonlinear} invariant. Interestingly, while it
makes sense addressing the locality concern at the level of the HS
vertices, it turns out that the gauge transformations reveal locality
structure at the level of fields, too.

Let us remind that the generating system \eqref{generating} is
invariant under local gauge symmetry governed by a parameter
$\epsilon(z; Y)$, which is subject to condition \eqref{epseq}. The
latter determines the $z$ dependence of the gauge transformations in
perturbations. Equation \eqref{epseq} can be naturally solved using the
standard homotopy operator \eqref{homotopy}. Since
\begin{equation*}
     \Delta \Bigg( \Lambda(\delta_\epsilon C) \Bigg) \equiv 0\,,
\end{equation*}
the further order by order analysis literally reproduces the calculation of $W^{n}$ and allows us to write down the solution of \eqref{epseq} right away,
\begin{equation}
    \epsilon(z; Y) = \varepsilon(Y) + \epsilon^{(1)}(z; Y) + \epsilon^{(2)}(z; Y) + ...
\end{equation}
with $\epsilon^{(n)}(z; Y)$ being the expansion in powers of $C$
of the following form:
\begin{equation}\label{epsexpl}
    \epsilon^{(n)}(z; Y) = \sum_{k=0}^{n} 
\mathcal{W}^{(k|n)}(z; y | q, p) \, \, \big(\overbrace{C...C}^{k} \,
\varepsilon \,\overbrace{C...C}^{n-k}\big)
\end{equation}
where $\mathcal{W}^{(k|n)}$ is given by \eqref{W_k|n} and we also
introduce $q$ in a sourcelike way as
\begin{equation}
    \varepsilon(Y) =  \varepsilon(y_\alpha, \Vec{\bf{y}}) = \exp\bigg(y_\alpha \frac{\partial}{\partial y^q_\alpha}\bigg) \varepsilon(y^q_\alpha,\Vec{\bf{y}})\Big|_{y^q_\alpha=0} = e^{-iy_\alpha q^\alpha} \varepsilon(y^q_\alpha,\Vec{\bf{y}})\Big|_{y^q_\alpha=0}\,,\,.
\end{equation}
with 
\begin{equation}
    q^\alpha := i \frac{\partial}{\partial y^q_\alpha} \, .
\end{equation}
Unsurprisingly, all gauge parameters are actually contained in
$\dr_z$ cohomology $\varepsilon(Y)$.

It is straightforward now to arrive at the manifest expressions
for gauge transformation of the Weyl module $C$. For that matter
we use \eqref{gauge:C}. Analogously to the derivation of
\eqref{epsexpl}, the computation of $\gd C$ repeats the one of
\eqref{Ups0} via \eqref{eq:0-VertexFromSources}. The final result
then can be expressed in terms of the 0-form vertices with
$\varepsilon(Y)$ in place of $\go$
\begin{equation}
    \label{GaugeTransform:C}
    \delta_{\varepsilon} C = - \varepsilon \ast C + C \ast \pi(\varepsilon) + \Upsilon(\varepsilon, C, C) + \Upsilon(\varepsilon, C, C, C) + ...\,,
\end{equation}
where $\Upsilon(\varepsilon, C^n)$ is just $\Upsilon(\omega, C^n)$
with $\omega$ replaced by $\varepsilon$. Interestingly, given
vertices $\Upsilon$ are spin local, $\delta_{\varepsilon} C$ is
spin local too for the polynomial in $y$ parameter $\varepsilon$.

Obtaining explicit expression for gauge transformation of the gauge sector $\omega$ is a bit trickier. Recalling the transformation law of $W$,  \eqref{gauge:W} and using the manifest form of $W$, one concludes that 
\begin{equation}
    \delta_{\varepsilon} \omega = \delta_{\epsilon} W\Big|_{z=0} \, .
\end{equation}
Leaving technical details to the Appendix \ref{App:gauge}, we provide the following result for this transformation:
\begin{equation}
    \label{GaugeTransformation:omega}
    \delta_{\varepsilon} \omega = \dr_x\varepsilon - \varepsilon \ast \omega + \omega \ast \varepsilon + \mathcal{V}(\varepsilon, \omega, C) - \mathcal{V}(\omega, \varepsilon, C) + \mathcal{V}(\varepsilon, \omega, C, C)  - \mathcal{V}(\omega, \varepsilon, C, C) -... \, .
\end{equation}
Although we arrived at \eqref{GaugeTransform:C}, \eqref{GaugeTransformation:omega} through a straightforward calculation, the final result was to some extent expected. Indeed, the general  consistent unfolded system
\begin{equation}
    \dr_x W^{A} = F^{A}(W^B) \, ,
\end{equation}
admits gauge transformation of the formal form 
\begin{equation}
    \delta W^{A} = \dr_x\varepsilon^{A} + \varepsilon^B\frac{\delta F^A}{\delta W^B} \,.
\end{equation}

\section{Geometric interpretation: convex/concave polygons}\label{Sec:Geometry}
An interesting feature of the integration domains $\mathcal{D}^{[k]}_n$ \eqref{eq:0-VertexDomain} is that each of these with fixed $n$ and $k$ corresponds to a set of closed polygons with special convexity properties. On top of that, the peculiar $y$ dependence of $\omega$ in \eqref{verC} that manifests itself weighted by the integration of $S_n^{[k]}$ gains some clear geometric interpretation. 

To show this, let us take $(\xi_1, ... \,, \xi_n ; \eta_1, ... \,, \eta_n) \in \mathcal{D}^{[k]}_n$ from \eqref{eq:0-VertexDomain} and let us introduce a set of $(n+1)$-points on the 2d plane $(\mathtt{x}\,, \mathtt{y})$ with coordinates\footnote{The introduced coordinates $(\mathtt{x}_i\,, \mathtt{y}_i)$ should not be confused with the space-time $x$ and the generating oscillators $y_{\al}$.} $(\mathtt{x}_i\,, \mathtt{y}_i)$, $i=0\dots n$ as follows: 
    \begin{subequations}\label{xy-plane}
        \begin{align}
            &\mathtt{x}_0 = 0 \,, \quad \mathtt{x}_i = \sum_{s=1}^i \xi_s \, , \\
            &\mathtt{y}_0 = 0, \quad \mathtt{y}_i = \sum_{s=1}^i \eta_s \, .
        \end{align}
    \end{subequations}
Notice that these coordinates are given by construction in the ascending order:
\be
\mathtt{x}_i\geq\mathtt{x}_j\,,\quad \mathtt{y}_i\geq \mathtt{y}_j\,,\quad i>j
\ee 
and due to \eqref{eq:0-VertexDomain}, 
\be
\mathtt{x}_i\leq 1\,,\quad \mathtt{y}_i\leq 1\,,\quad\mathtt{x}_n=\mathtt{y}_n=1\,.
\ee
Let us associate coordinates $(\mathtt{x}_i, \mathtt{y}_i)$ on the two-plane with insertions of the fields $\omega$ and $C$ from the line \eqref{VertexSourcesDef:0-form}. Then the point $(\mathtt{x}_k, \mathtt{y}_k)$ corresponds to the field $\omega$, while the rest $i\neq k$ with various $C$'s.   
Inequalities \eqref{eq:0-VertexDomain} lead to a geometric picture, which is easy to capture for $k=0$ and $k=n$ first. These are the no impurity cases corresponding to the leftmost vertex $\Upsilon_{\go C\dots C}$ for $k=0$ and the rightmost $\Upsilon_{C\dots C \go}$ for $k=n$, respectively. In the former case $(k=0)$, we have from \eqref{eq:0-VertexDomain}
\be\label{D0}
\mathcal{D}_n^{[0]}:\qquad \eta_i\xi_{i+1}-\eta_{i+1}\xi_i\geq 0\,,\quad i=1,\dots, n-1\,,
\ee
which entails
\be
\ff{\eta_1}{\xi_1}\geq\ff{\eta_2}{\xi_2}\geq\dots\geq\ff{\eta_n}{\xi_n}\,.
\ee
The latter condition just tells us that the slope of the line connecting two points on the plane $(\mathtt{x}_j, \mathtt{y}_j)$ and $(\mathtt{x}_{j+1}, \mathtt{y}_{j+1})$ decreases as $j$ grows from 0 to $n-1$. Thus, the corresponding sections form a convex polygon that starts at the origin $(0,0)$ and ends at $(1,1)$ as shown in Fig. 1.
\begin{figure}[h!]
    \centering
    \begin{tikzpicture}
    \begin{axis}[
    name=ax1,
    axis lines = left,
    xlabel = $\mathtt{x}$,
    ylabel = {$\mathtt{y}$},
    xmin = 0,
    xmax = 1.1,
    ymin = 0, 
    ymax = 1.1,
    xtick={0, 1},
    ytick={0, 1},
    x label style={at={(axis description cs:0.975, 0.125)},anchor=north},
    y label style={at={(axis description cs:-0.05, 1.06)}, rotate=270, anchor=west},
    every axis plot/.append style={ultra thick},
    ignore zero=x
]

    \addplot [
          only marks,
          mark=*,
          mark size=2pt,
          point meta=explicit symbolic,
          visualization depends on=\thisrow{alignment} \as \alignment,
          nodes near coords,
          every node near coord/.style={anchor=\alignment}
        ]
        table[meta index = 2]
        {
        x      y         label      alignment
        0      0         $\omega$   120
        0.1   0.4        $C$        130
        0.3   0.7        $C$        155
        0.55    0.9       $C$        110
        1      1         $C$        130
        };

    \draw[color=black, dashed]
        (axis cs:0, 0) -- (axis cs: 1, 1);
    \draw[color = black]
        (axis cs: 1, 0) -- (axis cs: 1, 1);
    \draw[color = black, name path = U]
        (axis cs: 0, 1) -- (axis cs: 1, 1);
    \path[name path = R]
        (axis cs: 0, 0) -- (axis cs: 0.1, 0.4) -- (axis cs: 0.3, 0.7) -- (axis cs: 0.55, 0.9) --  (axis cs: 1, 1);
    
    \draw[color = black, thick]
        (axis cs: 0, 0) -- (axis cs: 0.1, 0.4) -- (axis cs: 0.3, 0.7);
    \draw[decorate sep={1mm}{5mm}, fill]
        (axis cs: 0.3, 0.7) -- (axis cs: 0.55, 0.9);
    \draw[color = black, thick]
        (axis cs: 0.55, 0.9) -- (axis cs: 1, 1);

    \addplot[draw, pattern = {Lines[angle=135,distance=5pt, line width=1pt]}, pattern color = red, draw = none] 
    fill between[of=R and U, soft clip={domain=0:1}];  
    
\end{axis}

\begin{axis}[
    at={(ax1.outer north east)},anchor=outer north west,
    axis lines = left,
    xlabel = $\mathtt{x}$,
    ylabel = {$\mathtt{y}$},
    xmin = 0,
    xmax = 1.1,
    ymin = 0, 
    ymax = 1.1,
    xtick={0, 1},
    ytick={0, 1},
    x label style={at={(axis description cs:0.975, 0.125)},anchor=north},
    y label style={at={(axis description cs:-0.05, 1.06)}, rotate=270, anchor=west},
    every axis plot/.append style={ultra thick},
    ignore zero=x
]

    \addplot [
          only marks,
          mark=*,
          mark size=2pt,
          point meta=explicit symbolic,
          visualization depends on=\thisrow{alignment} \as \alignment,
          nodes near coords,
          every node near coord/.style={anchor=\alignment}
        ]
        table[meta index = 2]
        {
        x      y         label      alignment
        0      0         $C$        120
        0.4   0.1        $C$        -90
        0.7   0.3        $C$        -70
        0.9   0.6        $C$        -30
        1      1         $\omega$        130
        };

    \draw[color=black, dashed]
        (axis cs:0, 0) -- (axis cs: 1, 1);
    \draw[color = black]
        (axis cs: 1, 0) -- (axis cs: 1, 1);
    \draw[color = black]
        (axis cs: 0, 1) -- (axis cs: 1, 1);
    \path[name path = B]
        (axis cs: 0, 0) -- (axis cs: 1, 0);
    \path[name path = L]
        (axis cs: 0, 0) -- (axis cs: 0.4, 0.1) -- (axis cs: 0.7, 0.3) -- (axis cs: 0.9, 0.6) --  (axis cs: 1, 1);

    \draw[color = black, thick]
        (axis cs: 0, 0) -- (axis cs: 0.4, 0.1) -- (axis cs: 0.7, 0.3);
    \draw[decorate sep={1mm}{5mm}, fill]
        (axis cs: 0.7, 0.3) -- (axis cs: 0.9, 0.6);
    \draw[color = black, thick]
        (axis cs: 0.9, 0.6) -- (axis cs: 1, 1);

    \addplot[draw, pattern = {Lines[angle=135,distance=5pt, line width=1pt]}, pattern color = blue, draw = none] 
    fill between[of=L and B, soft clip={domain=0:1}];  
    
\end{axis}

\end{tikzpicture}
    \label{figure:1}
    \caption{Representations of $\mathcal{D}^{[0]}_n$(left) and $\mathcal{D}^{[n]}_n$(right) as a set of convex/concave polygons. Dashed areas are half of $S^{[0]}_n$ and $-S^{[n]}_n$, correspondingly.}
\end{figure}
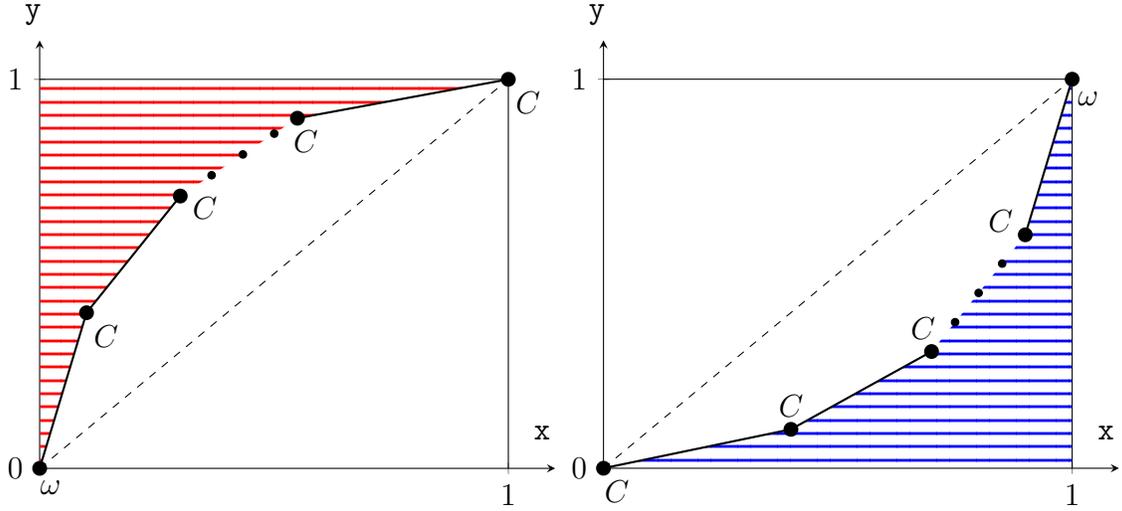

Similarly, for the rightmost $(k=n)$ vertex $\Upsilon_{C\dots C \go}$ the integration domain is   
\be
\mathcal{D}_n^{[n]}:\qquad \eta_i\xi_{i+1}-\eta_{i+1}\xi_i\leq 0\,,\quad i=1,\dots, n-1
\ee
which differs from \eqref{D0} by the signs of inequalities,  
\be
\ff{\eta_1}{\xi_1}\leq\ff{\eta_2}{\xi_2}\leq\dots\leq\ff{\eta_n}{\xi_n}\,.
\ee
The legs between points $(\mathtt{x}_j, \mathtt{y}_j)$ and $(\mathtt{x}_{j+1}, \mathtt{y}_{j+1})$ form a concave polygon bridging points $(0,0)$ and $(1,1)$.

Now, the general $k$ case of $\mathcal{D}_n^{[k]}$ brings us to a combination of the concave condition arising below the $\go$ impurity placed at the $k$th position  
\be\label{inequality-concave}
\eta_i\xi_{i+1}-\eta_{i+1}\xi_i\leq 0\,,\qquad i<k
\ee
and the convex one 
\be\label{inequality-convex}
\eta_i\xi_{i+1}-\eta_{i+1}\xi_i\geq 0\,,\qquad i>k
\ee
which holds beyond the $\go$ impurity. There is no condition at $i=k$:  therefore, the resulting picture is a junction of the two types of polygons. The concave one starts at $(0,0)$ and ends at the $\go$ impurity $(\mathtt{x}_k, \mathtt{y}_k)$, while the convex starts at $(\mathtt{x}_k, \mathtt{y}_k)$ ending at $(1, 1)$.

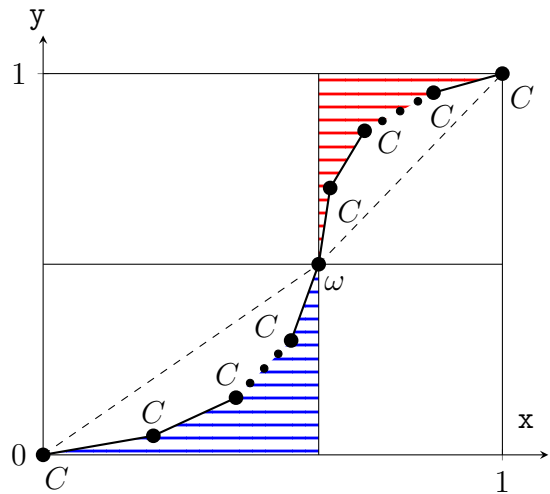
\begin{figure}[h]
    \centering
    \begin{tikzpicture}[scale=0.98]
    \begin{axis}[
    axis lines = left,
    xlabel = $\mathtt{x}$,
    ylabel = {$\mathtt{y}$},
    xmin = 0,
    xmax = 1.1,
    ymin = 0, 
    ymax = 1.1,
    xtick={0, 1},
    ytick={0, 1},
    x label style={at={(axis description cs:0.975, 0.125)},anchor=north},
    y label style={at={(axis description cs:-0.05, 1.06)}, rotate=270, anchor=west},
    every axis plot/.append style={ultra thick},
    ignore zero=x
]

    \addplot [
          only marks,
          mark=*,
          mark size=2pt,
          point meta=explicit symbolic,
          visualization depends on=\thisrow{alignment} \as \alignment,
          nodes near coords,
          every node near coord/.style={anchor=\alignment}
        ]
        table[meta index = 2]
        {
        x      y         label      alignment
        0      0         $C$        120
        0.24   0.05      $C$        -90
        0.42   0.15      $C$        -70
        0.54   0.3       $C$        -30
        0.6    0.5       $\omega$   130
        0.625   0.7       $C$        130
        0.7   0.85      $C$        155
        0.85  0.95      $C$        110
        1      1         $C$        130
        };

    \draw[color=black, dashed]
        (axis cs:0, 0) -- (axis cs: 0.6, 0.5);
    \draw[color = black, dashed]
        (axis cs: 0.6, 0.5) -- (axis cs: 1, 1);
    \draw[color = black]
        (axis cs: 1, 0) -- (axis cs: 1, 1);
    \draw[color = black]
        (axis cs: 0, 1) -- (axis cs: 1, 1);
    \path[name path = B]
        (axis cs: 0, 0) -- (axis cs: 0.6, 0);
    \path[name path = L]
        (axis cs: 0, 0) -- (axis cs: 0.24, 0.05) -- (axis cs: 0.42, 0.15) -- (axis cs: 0.54, 0.3) --  (axis cs: 0.6, 0.5);
    \path[name path = R]
        (axis cs: 0.6, 0.5) -- (axis cs:  0.625, 0.7) -- (axis cs:  0.7, 0.85) -- (axis cs: 0.85, 0.95) -- (axis cs: 1, 1);
    \path[name path = U]
        (axis cs: 0.6, 1) -- (axis cs: 1, 1);
    \draw[color = black, thin]
        (axis cs: 0.6, 0) -- (axis cs: 0.6, 1);
    \draw[color = black, thin]
        (axis cs: 0, 0.5) -- (axis cs: 1, 0.5);

    \draw[color = black, thick]
        (axis cs: 0, 0) -- (axis cs: 0.24, 0.05) -- (axis cs: 0.42, 0.15);
    \draw[decorate sep={1mm}{2.7mm}, fill]
        (axis cs: 0.42, 0.15) -- (axis cs: 0.54, 0.3);
    \draw[color = black, thick]
        (axis cs: 0.54, 0.3) -- (axis cs: 0.6, 0.5);
    \draw[color = black, thick]
        (axis cs: 0.6, 0.5) -- (axis cs:  0.625, 0.7) -- (axis cs:  0.7, 0.85);
    \draw[decorate sep={1mm}{2.7mm}, fill]
        (axis cs: 0.7, 0.85) -- (axis cs: 0.85, 0.95);
    \draw[color = black, thick]
        (axis cs: 0.85, 0.95) -- (axis cs: 1, 1);

    \addplot[draw, pattern = {Lines[angle=135,distance=5pt, line width=1pt]}, pattern color = blue, draw = none] 
    fill between[of=L and B, soft clip={domain=0:1}];  
    \addplot[draw, pattern = {Lines[angle=135,distance=5pt, line width=1pt]}, pattern color = red, draw = none] 
    fill between[of=R and U, soft clip={domain=0:1}]; 
    
\end{axis}
\end{tikzpicture}
    \label{figure:2}
    \caption{Representation of $\mathcal{D}^{k}_n$. Note the change of the area that contributes to $S^{[k]}_n$ in the $\omega$ impurity:  It is given by the area under the polygonal chain with the negative sign before $\omega$, while after the impurity, it is the area above the polygonal chain with the positive sign.}
\end{figure}
Let us specify various properties that the observed geometric picture suggests 

\begin{itemize}
    \item The two-polygon chain junction is always placed inside unit square $(0\, , \,0) - (1\, , \,0) - (1\, , \,1) - (0\, , \,1)$. Its concave part lies below line $(0,0) - (\mathtt{x}_k, \mathtt{y}_k)$, while the convex part lies above line $(\mathtt{x}_k, \mathtt{y}_k) -(1,1)$. This simple fact leads to consequences that might not be immediately visible from inequalities \eqref{eq:0-VertexDomain}. For example, for $k=0$ it is clear from Fig.1 that $\eta_1\geq\xi_1$. Likewise, $\eta_1\leq\xi_1$ for $k=n$ as is seen from Fig.2. 
    
    \item There is a somewhat degenerate case corresponding to domain $\mathcal{D}_2^{[1]}$. For these particular values of $n=2$ and $k=1$ inequalities \eqref{inequality-concave} and \eqref{inequality-convex} do not apply.       
   
    \item As mentioned, the point $(\mathtt{x}_k\,, \mathtt{y}_k)$ is associated with $\omega$, while all other points with $C$'s. Now, $\go$ depends on $y$ via the combination $S^{[k]}_n$ as a function of $\xi_i$ and $\eta_i$ defined in \eqref{PS}. In terms of the introduced variables \eqref{xy-plane} it boils down to
    \begin{align}
        S^{[k]}_n = -\sum_{s=1}^{k} \xi_s + \sum_{s=k+1}^{n} \xi_s + \sum_{i<j}^{n}(\xi_i \eta_j - \xi_j \eta_i)=1-2\mathtt{x}_k+\sum_{i=1}^{n-1}(\mathtt{x}_i\mathtt{y}_{i+1}-\mathtt{x}_{i+1}\mathtt{y}_{i})\,.
    \end{align}
By means of the Gauss shoelace formula  for the oriented closed polygon area, $S^{[k]}_n$ can be shown to acquire the following form: 
\be
\ff12S^{[k]}_n=A^+-A^-\,,
\ee
where $A^-$ and $A^+$ are the areas of the dashed concave and convex parts of the polygons, correspondingly. Therefore, $S^{[k]}_n$ is just twice the difference between the area enclosed by the polygon chain $(\mathtt{x}_{k}\,, \mathtt{y}_k) - (\mathtt{x}_{k}\,, 1) - (1 \,, 1) - (\mathtt{x}_{n-1}\,, \mathtt{y}_{n-1}) - \dots - (\mathtt{x}_{k}\,, \mathtt{y}_k)$ and area enclosed by the chain $(\mathtt{x}_{k}\,, \mathtt{y}_k) - (\mathtt{x}_{k}\,, 0) - (0\,,  0) - (\mathtt{x}_{1}\,, \mathtt{y}_{1}) - \dots - (\mathtt{x}_{k}\,, \mathtt{y}_k)$.
\end{itemize}
Let us note that while the concave part of the polygon chain below the impurity point $i\leq k$ gives non-negative angles between vectors $\vec{r}_i = (\mathtt{x}_{i} - \mathtt{x}_{i-1}\,, \mathtt{y}_{i} - \mathtt{y}_{i-1})$ and $\vec{r}_{i-1}$, and, similarly, the convex part $i\geq k$ corresponds to non-positive such angles, there are no constraints on the sign of angle between $\vec{r}_{k+1}$ and $\vec{r}_{k}$ at the impurity point $k$ itself. This is because \eqref{eq:0-VertexDomain} lacks inequalities at $i=k$ There are obviously some constraints on this angle itself, but its sign is not fixed by inequalities \eqref{eq:0-VertexDomain}.

As an example, let us consider domain $\mathcal{D}^{[2]}_4$, corresponding to the vertex $\Upsilon_{CC\omega CC}$:
\begin{equation}
    \label{example}
    \mathcal{D}^{[2]}_4 = \begin{cases}
        \eta_1 + \eta_2 + \eta_3 +  \eta_{4} = 1\,,\quad \eta_i \geqslant  0\,, \\
        \xi_1 + \xi_2 + \xi_3 +  \xi_{4} = 1\,,\quad \xi_i \geqslant 0\,, \\
        \eta_1 \xi_2 - \eta_2 \xi_1 \leqslant 0 \,, \\
        \eta_3 \xi_4 - \eta_4 \xi_3 \geqslant 0\, .
    \end{cases}
\end{equation}

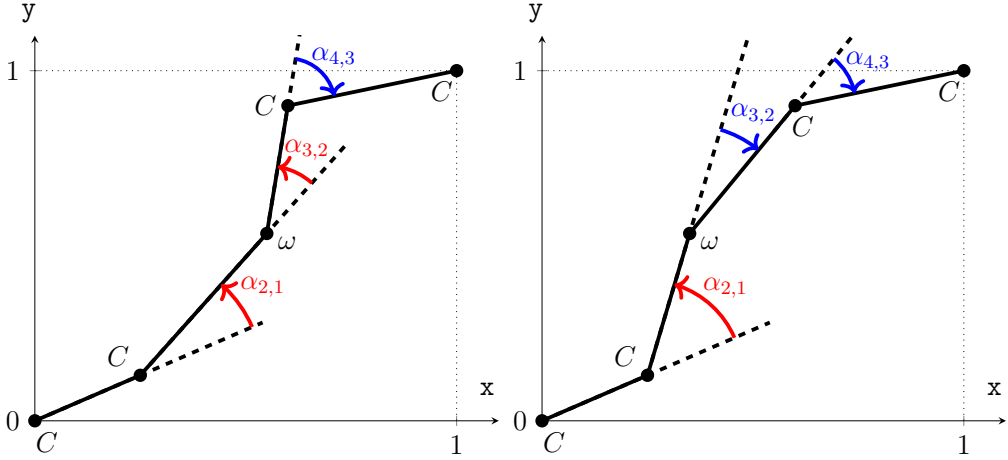
\begin{figure}[h]
    \centering
    \begin{tikzpicture}[scale=0.9]
    \begin{axis}[
    name=ax1,
    axis lines = left,
    xlabel = $\mathtt{x}$,
    ylabel = {$\mathtt{y}$},
    xmin = 0,
    xmax = 1.1,
    ymin = 0, 
    ymax = 1.1,
    xtick={0, 1},
    ytick={0, 1},
    x label style={at={(axis description cs:0.975, 0.125)},anchor=north},
    y label style={at={(axis description cs:-0.05, 1.06)}, rotate=270, anchor=west},
    every axis plot/.append style={ultra thick},
    ignore zero=x
]

\addplot[domain=0:0.25, samples=100, color=black, ultra thick]
    {(0.13/0.25)*x};
\addplot[domain=0.25:0.55, samples=100, color=black, ultra thick]
    {0.13+(x-0.25)*(0.4-0.13)/(0.45-0.25)};
\addplot[domain=0.55:0.6, samples=100, color=black, ultra thick]
    {0.535+(x-0.55)*(0.9-0.535)/(0.6-0.55)};
\addplot[domain=0.6:1, samples=100, color=black, ultra thick]
    {0.9+(x-0.6)*(1-0.9)/(1-0.6)};

\addplot [
          only marks,
          mark=*,
          mark size=2pt,
          point meta=explicit symbolic,
          visualization depends on=\thisrow{alignment} \as \alignment,
          nodes near coords,
          every node near coord/.style={anchor=\alignment}
        ]
        table[meta index = 2]
        {
        x      y         label      alignment
        0      0         $C$        120
        0.25   0.13      $C$        -40
        0.55   0.535     $\omega$   155
        0.6    0.9       $C$        0
        1      1         $C$        55
        };

\addplot[domain=0:0.54, samples=100, color=black, ultra thick, dashed]
    {(0.13/0.25)*x};
\addplot[domain=0.25:0.74, samples=100, color=black, ultra thick, dashed]
    {0.13+(x-0.25)*(0.4-0.13)/(0.45-0.25)};
\addplot[domain=0.55:0.9, samples=100, color=black, ultra thick, dashed]
    {0.535+(x-0.55)*(0.9-0.535)/(0.6-0.55)};

\coordinate (A1) at (axis cs: 0.4, 0.208);
\coordinate (O1) at (axis cs: 0.25, 0.13);
\coordinate (B1) at (axis cs: 0.4, 0.3325);
\pic  [pic text=$\alpha_{2,1}$,red,draw, ->, angle eccentricity=1.24,angle radius=1.8cm, ultra thick]{angle=A1--O1--B1};

\coordinate (A2) at (axis cs: 0.74, 0.7915);
\coordinate (O2) at (axis cs: 0.55, 0.535);
\coordinate (B2) at (axis cs: 0.6, 0.9);
\pic  [pic text=$\alpha_{3,2}$,red,draw, ->, angle eccentricity=1.37,angle radius=1cm, ultra thick]{angle=A2--O2--B2};

\coordinate (A3) at (axis cs: 1, 1); 
\coordinate (O3) at (axis cs: 0.6, 0.9);
\coordinate (B3) at (axis cs: 0.7, 1.63);
\pic  [pic text=$\alpha_{4,3}$,blue,draw, <-, angle eccentricity=1.4,angle radius=0.7cm, ultra thick]{angle=A3--O3--B3};

\draw[color = black, thin, dotted]
    (axis cs: 0, 1) -- (axis cs: 1 , 1) -- (axis cs: 1, 0); 

\end{axis}

\begin{axis}[
    at={(ax1.outer north east)},anchor=outer north west,
    axis lines = left,
    xlabel = $\mathtt{x}$,
    ylabel = {$\mathtt{y}$},
    xmin = 0,
    xmax = 1.1,
    ymin = 0, 
    ymax = 1.1,
    xtick={0, 1},
    ytick={0, 1},
    x label style={at={(axis description cs:0.975, 0.125)},anchor=north},
    y label style={at={(axis description cs:-0.05, 1.06)}, rotate=270, anchor=west},
    every axis plot/.append style={ultra thick},
    ignore zero=x
]

\addplot[domain=0:0.25, samples=100, color=black, ultra thick]
    {(0.13/0.25)*x};
\addplot[domain=0.25:0.35, samples=100, color=black, ultra thick]
    {0.13+(x-0.25)*(0.535-0.13)/(0.35-0.25)};
\addplot[domain=0.35:0.6, samples=100, color=black, ultra thick]
    {0.535+(x-0.35)*(0.9-0.535)/(0.6-0.35)};
\addplot[domain=0.6:1, samples=100, color=black, ultra thick]
    {0.9+(x-0.6)*(1-0.9)/(1-0.6)};

\addplot [
          only marks,
          mark=*,
          mark size=2pt,
          point meta=explicit symbolic,
          visualization depends on=\thisrow{alignment} \as \alignment,
          nodes near coords,
          every node near coord/.style={anchor=\alignment}
        ]
        table[meta index = 2]
        {
        x      y         label      alignment
        0      0         $C$        120
        0.25   0.13      $C$        -40
        0.35   0.535     $\omega$   155
        0.6    0.9       $C$        110
        1      1         $C$        55
        };

\addplot[domain=0:0.54, samples=100, color=black, ultra thick, dashed]
    {(0.13/0.25)*x};
\addplot[domain=0.25:0.7, samples=100, color=black, ultra thick, dashed]
    {0.13+(x-0.25)*(0.535-0.13)/(0.35-0.25)};
\addplot[domain=0.55:0.9, samples=100, color=black, ultra thick, dashed]
    {0.535+(x-0.35)*(0.9-0.535)/(0.6-0.35)};

\coordinate (A1) at (axis cs: 0.4, 0.208);
\coordinate (O1) at (axis cs: 0.25, 0.13);
\coordinate (B1) at (axis cs: 0.4, 0.7375);
\pic  [pic text=$\alpha_{2,1}$,red,draw, ->, angle eccentricity=1.24,angle radius=1.4cm, ultra thick]{angle=A1--O1--B1};

\coordinate (B2) at (axis cs: 0.5, 1.1425);
\coordinate (O2) at (axis cs: 0.35, 0.535);
\coordinate (A2) at (axis cs: 0.6, 0.9);
\pic  [pic text=$\alpha_{3,2}$,blue,draw, <-, angle eccentricity=1.25,angle radius=1.6cm, ultra thick]{angle=A2--O2--B2};

\coordinate (A3) at (axis cs: 1, 1); 
\coordinate (O3) at (axis cs: 0.6, 0.9);
\coordinate (B3) at (axis cs: 0.7, 1.046);
\pic  [pic text=$\alpha_{4,3}$,blue,draw, <-, angle eccentricity=1.4,angle radius=0.9cm, ultra thick]{angle=A3--O3--B3};

\draw[color = black, thin, dotted]
    (axis cs: 0, 1) -- (axis cs: 1 , 1) -- (axis cs: 1, 0); 
    
\end{axis}

\end{tikzpicture}
    \label{figure:3}
    \caption{Schematic form of the two possible cases for the sign of angle $\alpha_{3,2}$ for the polygon, corresponding to $\mathcal{D}^{[2]}_4$. Each joint associated with either $\omega$ or $C$ placed in the order, which is the same as in the vertex.}
\end{figure}
The angle $\alpha_{2,1}$ being positive and $\alpha_{4,3}$ being negative can be derived from inequalities \eqref{example}. However, there is no sign constraint on the angle $\alpha_{3,4}$, see Fig.3.

According to \eqref{PS},
\begin{equation}
    S^{[2]}_4 = -\xi_1 - \xi_2 + \xi_3 + \xi_4 + \sum_{i<j}^{4} \big(\xi_i \eta_j  - \xi_j \eta_i \big)=1-2\mathtt{x}_2+\sum_{i=1}^{3}(\mathtt{x}_i\mathtt{y}_{i+1}-\mathtt{x}_{i+1}\mathtt{y}_i)
\end{equation}
equals twice the difference between areas as illustrated in Fig.4. 
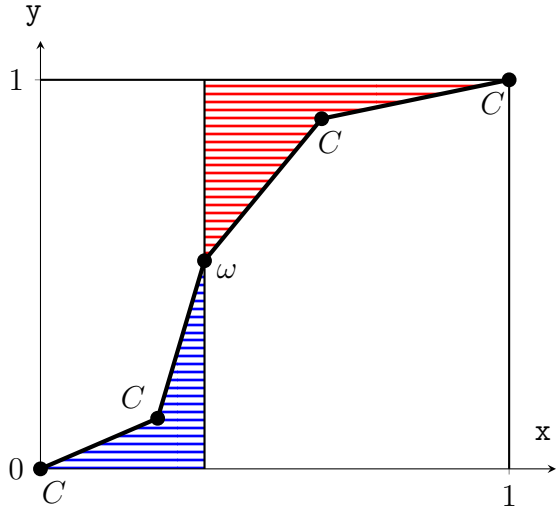
\begin{figure}[h!]
    \centering
    \begin{tikzpicture}
    \begin{axis}[
    axis lines = left,
    xlabel = $\mathtt{x}$,
    ylabel = {$\mathtt{y}$},
    xmin = 0,
    xmax = 1.1,
    ymin = 0, 
    ymax = 1.1,
    xtick={0, 1},
    ytick={0, 1},
    x label style={at={(axis description cs:0.975, 0.125)},anchor=north},
    y label style={at={(axis description cs:-0.05, 1.06)}, rotate=270, anchor=west},
    every axis plot/.append style={ultra thick},
    ignore zero=x
]

\addplot[domain=0:0.25, samples=100, color=black, ultra thick]
    {(0.13/0.25)*x};
\addplot[ domain=0.25:0.35, samples=100, color=black, ultra thick]
    {0.13+(x-0.25)*(0.535-0.13)/(0.35-0.25)};
\addplot[domain=0.35:0.6, samples=100, color=black, ultra thick]
    {0.535+(x-0.35)*(0.9-0.535)/(0.6-0.35)};
\addplot[domain=0.6:1, samples=100, color=black, ultra thick]
    {0.9+(x-0.6)*(1-0.9)/(1-0.6)};

\addplot [
          only marks,
          mark=*,
          mark size=2pt,
          point meta=explicit symbolic,
          visualization depends on=\thisrow{alignment} \as \alignment,
          nodes near coords,
          every node near coord/.style={anchor=\alignment}
        ]
        table[meta index = 2]
        {
        x      y         label      alignment
        0      0         $C$        120
        0.25   0.13      $C$        -40
        0.35   0.535     $\omega$   155
        0.6    0.9       $C$        110
        1      1         $C$        55
        };

    \draw[color=black, thick]
        (axis cs:0, 1) -- (axis cs:1, 1);
    \draw[color=black, thick]
        (axis cs:1, 0) -- (axis cs:1, 1);
    \draw[color=black, thick]
        (axis cs:0.35, 0) -- (axis cs:0.35, 1);

    \path[name path = B]
        (axis cs: 0, 0) -- (axis cs:0.35, 0);
    \path[name path = L]
        (axis cs: 0, 0) -- (axis cs:0.25, 0.13) -- (axis cs:0.35, 0.535);
    \path[name path = U]
        (axis cs: 0.35, 1) -- (axis cs: 1, 1);
    \path[name path = R]
        (axis cs: 0.35, 0.535) -- (axis cs: 0.6, 0.9) -- (axis cs: 1, 1);
    
    \addplot[draw, pattern ={Lines[angle=135,distance=3pt, line width=1pt]}, pattern color = blue] 
    fill between[of=L and B, soft clip={domain=0:0.35}];
    
    \addplot[draw, pattern ={Lines[angle=135,distance=3pt, line width=1pt]}, pattern color = red] 
    fill between[of=R and U, soft clip={domain=0.35:1}];
\end{axis}
\end{tikzpicture}
    \label{figure:4}
    \caption{Schematic representation of $\mathcal{D}^{[2]}_4$. Coefficient $S^{[2]}_4$ is just twice the difference between red and blue areas}
\end{figure}
Let us note, that for the domains $\mathcal{D}^{[0]}_n$ every angle $\alpha_{i+1, i}$ is positive, while for the domains  $\mathcal{D}^{[n]}_n$ all angles $\alpha_{i+1, i}$ are negative.

\section{Lowest orders}\label{Sec:lowest}
Before proceeding further, we provide the lowest order vertices of both sectors, namely, $\mathcal{V}(\omega, \omega, C)$ and $\Upsilon(\omega, C, C)$. These were already derived\footnote{The structure of vertex $\mathcal{V}(\omega, \omega, C)$ in $d$-dimensional theory with respect to sp(2)-variables is identical to that of $\mathcal{V}_{\eta}(\omega, \omega, C)$ or $\mathcal{V}_{\bar{\eta}}(\omega, \omega, C)$ of the four-dimensional theory with respect to the spinoral variables $y_\alpha$ or $\bar{y}_{\dot{\alpha}}$ derived in \cite{Didenko:2018fgx} .} in \cite{Didenko:2023vna}, but the integration domain parametrization used in \cite{Didenko:2023vna} differs from that of our study. We provide the necessary technical details of the transition from one form to another in Appendix \ref{App:Match}.

\subsection{Vertex $\mathcal{V}(\omega, \omega, C)$}

Vertex $\mathcal{V}(\omega, \omega, C)$ is the most known and well studied for its relation to the so-called central on mass shell theorem  \cite{Vasiliev:1986td}. Also, $\mathcal{V}(\omega, \omega, C)$ being invariant under field redefinition $\omega \longrightarrow \omega + f(\omega, C)$, is unique in that sense, unlike all other vertices from both sectors. In the form \eqref{verw} this vertex reads 
\begin{equation}
    \label{wwC}
    \begin{split}
        &\mathcal{V}(\omega, \omega, C) =\epsilon^{\alpha\beta}\int \frac{d^2 u d^2 v d^2 u' d^2 v'}{(2\pi)^4} \, e^{iuv + iu'v'} \, \times\\
        &\bigg[-\smashoperator{ \int_{\mathcal{D}^{[0,0]}_2} } d\xi d\eta \,  \partial_\alpha \omega\Big( \xi_2 y + u - \frac{S^{[0,0]}_1}{2} v' , \vec\y \Big) \star \partial_\beta \omega\Big( \eta_2 y + u' + \frac{S^{[0,0]}_1}{2} v , \vec\y \Big) \star C\Big(\xi_1 v + \eta_1 v', \vec\y \Big) + \\
        &+ \smashoperator{ \int_{\mathcal{D}^{[0,1]}_2} } d\xi d\eta \,  \partial_\alpha \omega\Big( \xi_2 y + u - \frac{S^{[0,1]}_1}{2} v' , \vec\y \Big) \star C\Big(\xi_1 v + \eta_1 v', \vec\y \Big)  \star \partial_\beta \omega\Big( \eta_2 y + u' + \frac{S^{[0,1]}_1}{2} v , \vec\y \Big) - \\
        & - \smashoperator{ \int_{\mathcal{D}^{[1,1]}_2} } d\xi d\eta \,  C\Big(\xi_1 v + \eta_1 v', \vec\y \Big) \star \partial_\alpha \omega\Big( \xi_2 y + u - \frac{S^{[1,1]}_1}{2} v' , \vec\y \Big)   \star \partial_\beta \omega\Big( \eta_2 y + u' + \frac{S^{[1,1]}_1}{2} v , \vec\y \Big) \bigg]
    \end{split}
\end{equation}
Due to the presence of derivatives with respect to the first argument of $\omega(y, \vec\y)$ in \eqref{wwC}, the vertex $\mathcal{V}(\omega, \omega, C)$ treats different parts of the gauge module $\omega(y, \vec\y)$ unequally even for fixed spins. To show this, we consider the lower spin case for $\omega(y, \vec\y)$, i.e., we truncate it up to $s\leqslant 2$
\begin{equation}
    \omega_{\text{low}}(Y | x) = i \big( A(x) + e^a(x) P_a + \frac{1}{2} \omega^{a,b}(x) M_{ab} \big)\, .
\end{equation}
The star product on the lhs of \eqref{nonlinear:1-sector} does not contain field $A$ but contains both the vielbein $e^a$ and the spin connection $\omega^{a,b}$
\begin{equation}
    \label{ww:s=2}
    \omega_{\text{low}} \ast \omega_{\text{low}} = - i\bigg( \frac{1}{2} \, e^a e^b \, M_{ab} + \omega^{a,b} e^c \eta_{ac} \, P_b + \frac{1}{2} \, \omega^{a,b}\omega^{c,d} \eta_{ac} \, M_{bd} \bigg)\,,
\end{equation}
while the vertex \eqref{wwC} depends only on the vielbein $e^a$ and the Weyl module $C(y, \vec\y)$
\begin{equation}
    \label{wwC:grav}
    \mathcal{V}(\omega_{\text{low}}, \omega_{\text{low}}, C) = \frac{\epsilon^{\alpha\beta}}{4} \, e_a e_b \, \frac{\partial^2 C}{\partial y^\alpha_a \partial y^\beta_b}(0, \vec\y)\,.
\end{equation}
All higher-order vertices of this sector vanish
\begin{equation}
    \mathcal{V}(\omega_{\text{low}}, \omega_{\text{low}}, C, C, ...) = 0\,.
\end{equation}
The formula \eqref{wwC:grav} is very important in the context of linearization over the AdS solution and is tightly related to the central on-mass shell theorem.

For $s \leqslant 2$ the Weyl module $C(y, \vec\y)$ has the following structure:
\begin{equation}
    \begin{split}
       C_{\text{low}}(Y | x) = i \, \Big(& \underbrace{\phi - i \, \phi^a P_a -  \, \phi^{ab}P_{a}P_{b} + \dots }_{\text{spin-0}}  \Big) + i \, \Big(\underbrace{ F^{a,b}M_{ab} - i F^{ab,c}P_a M_{bc}  + \dots }_{ \text{spin-1} } \Big) +  \\
       &+i \, \Big( \underbrace{ R^{ab,cd}M_{ac}M_{bd} - i \, R^{abc,de}P_a M_{bd}M_{ce} + \dots}_{\text{spin-2}} \Big)\,,
    \end{split}
\end{equation}
where we used the new notation for the lower spin fields instead of the conventional one $C^{a(m), b(n)}$ for the components of the Weyl module. Substituting this into \eqref{wwC:grav}, one finds that upon considering only $s\leqslant 2$, the equation \eqref{nonlinear:1-sector} decomposes into 
\begin{subequations}
    \label{low:example}
    \begin{align}
        \label{low:example1}&\dr_x A = \frac{1}{2} \, e_a e_b \, F^{a,b} \, , \\
        \label{low:example2}&\dr_x e^a + \omega^{a}\,_{b} \, e^b = 0 \, , \\
        \label{low:example3}&\dr_x \omega^{a,b} + \omega^{a}\,_{c}\, \omega^{c,b}  - e^a e^b = R^{ac, db} e_c e_d \, .
    \end{align}
\end{subequations}
The obtained equations are not dynamical but rather the definitions of one set of fields as derivatives of the others. Indeed, \eqref{low:example1} is the standard definition of the Faraday tensor $F_{a,b}(x)$; \eqref{low:example2} tells how $\omega^{a,b}(x)$ is expressed via derivatives of the veilbein $e^a(x)$, while \eqref{low:example3} defines (up to the AdS contribution $e^a e^b$) the Riemann tensor  $R^{ad,cb}(x)$  through derivatives of $\omega^{a,b}(x)$ and veilbein $e^a(x)$.

One important thing to keep in mind is the equations \eqref{low:example} were derived for artificially truncated set of fields $s \leqslant 2$. Taking into consideration the tower of all spins would drastically modify equations \eqref{low:example}. For example, adding spin-3 gauge field only would lead to a nontrivial contribution from the quadratic vertex in the form of $\mathcal{V}(\omega_{s=3}, \omega_{s=3}, C, C)$. There would be an additional contribution to the linear in $C$ terms from $\mathcal{V}(\omega_{s\leqslant 2}, \omega_{s=3}, C), \, \mathcal{V}(\omega_{s=3}, \omega_{s\leqslant 3}, C), \, \mathcal{V}(\omega_{s=3}, \omega_{s=3}, C)$. The contribution to the lhs of \eqref{low:example2} would acquire the term proportional to $\omega_{ab}(x) \omega^{ab,c}(x)$. Those higher-spin interaction would manifest in the \textit{change} of definitions for $F^{a,b}, R^{ad,cb}$ and $\omega^{a,b}$.

The analogy of such a deformation of definitions is well known within the standard field theory, e.g., the transition from noninteracting photons to non-Abelian Yang-Mills theory. In such a transition, the definition of the Faraday tensors gets deformed by the quadratic in gauge-potential terms.

\subsection{Vertex $\Upsilon(\omega, C, C)$}

In the form \eqref{verC} the vertex amounts to 
\begin{equation}\label{wCC:expl}
    \begin{split}
        &\Upsilon(\omega, C, C) = i y^\alpha \int \frac{d^2 u d^2 v}{(2\pi)^2} \, e^{iuv} \times \\
        &\bigg[\smashoperator{ \int_{\mathcal{D}^{[0]}_2} } d\xi d\eta \, \partial_\alpha  \omega\big( S^{[0]}_2 y+u , \vec\y\big) \star C\big( \xi_1 y + \eta_1 v, \vec\y \big) \star C\big( \xi_2 y + \eta_2 v, \vec\y \big) - \\ 
        &- \smashoperator{ \int_{\mathcal{D}^{[1]}_2} } d\xi d\eta \, C\big( \xi_1 y + \eta_1 v, \vec\y \big) \star \partial_\alpha \omega\big( S^{[1]}_2 y+u , \vec\y\big) \star C\big( \xi_2 y + \eta_2 v, \vec\y \big) + \\
        &+\smashoperator{ \int_{\mathcal{D}^{[2]}_2} } d\xi d\eta \,   C\big( \xi_1 y + \eta_1 v, \vec\y \big) \star C\big( \xi_2 y + \eta_2 v, \vec\y \big) \star \partial_\alpha  \omega\big( S^{[2]}_2 y+u , \vec\y\big) \bigg] \, .
    \end{split}
\end{equation}
As it was done for $\mathcal{V}(\omega, \omega, C)$, let us confine ourselves to the lower-spin case. The twisted commutator in this case is
\begin{equation}
    \begin{split}
        &\omega_{\text{low}} \ast C - \pi(\omega_{\text{low}}) \ast C = i \, e^a \, \{P_a, C \}_{\ast} + \frac{i}{2} \, \omega^{a,b} \big[M_{ab}, C \big]_{\ast} = \\
        & = i \, e^a\Big(2P_a \cdot C(y, \vec\y) + \frac{\partial^2 C}{\partial y^\alpha \partial y^a_\alpha}(y, \vec\y) \Big) +  \omega^{a,b} \, \epsilon_{\alpha\beta} \, y^\beta_b \frac{\partial C}{\partial y^a_\alpha}(y, \vec\y) \, .
    \end{split}
\end{equation}
Meanwhile, $\Upsilon(\omega_{\text{low}}, C, C)$ contains no spin connection $\omega^{a,b}$ and can be set into the following form: 
\begin{equation}
    \Upsilon(\omega_{\text{low}}, C, C) = i \, e^a \, y_\alpha \cdot \smashoperator{ \int_{\mathcal{D}^{[0]}_2} } d\xi d\eta \, \bigg[\frac{\partial C}{\partial y^a_\alpha}(\xi_1 y, \vec\y), C(\xi_2 y, \vec\y)  \bigg]_{\star} \, .
\end{equation}
All higher-order vertices vanish identically 
\begin{equation}
    \Upsilon(\omega_{\text{low}}, C, C, C, ...) = 0 \, .
\end{equation}
To see how nonlinearities affect definitions of field descendants, let us once again truncate the spectrum of $C(y, \vec\y)$ to the scalar, vector and spin two field. Given the general field component formula for $\Upsilon(\omega_{\text{low}}, C_{\text{low}}, C_{\text{low}})$ is quite cumbersome, we consider the equations for the lowest components only. Let us  start with the scalar field,
\begin{equation}
    \label{spin-0:example1}
    \dr_x \phi - e^a \phi_a = 0 \, .
\end{equation}
Let us stress once again that this is not a dynamical equation, but just the definition of $\phi^a$. It follows from \eqref{spin-0:example1} that 
\begin{equation}
    \phi_\mu(x) := \phi_a(x) e^a_\mu(x) =  \frac{\partial \phi}{\partial x^\mu}(x) \, .
\end{equation}
However, with the next descendant the relations get somewhat more complicated:
\begin{equation}
    \label{spin-0:example2}
    \dr_x \phi^a + \omega^{a,b} \phi_{b} - 2e^a \, \phi - 3 \phi^{ab} \,  e_b = -2 F^{a,b} F_{b,c} \, e^c - 6 R^{ab,cd}(R_{cd,bf} - R_{cb,df}) \, e^f 
\end{equation}
Indeed, Eq. \eqref{spin-0:example2} not only contains $\phi(x)$ and $ \phi^a(x)$, but also the squares of the Faraday and Riemann tensors.  Therefore, $\phi^{ab}(x)$ is not simply a covariant derivative of $\phi^a(x)$, as it would be the case in the linearized theory (see e.g.,  \cite{Bekaert:2004qos}), but a covariant derivative of $\phi^a(x)$ plus nonlinear terms composed of other fields.

The natural question is how does spin locality of $\Upsilon(\omega, C, C)$ manifest itself in \eqref{spin-0:example2}? \textit{A priori} the rhs of \eqref{spin-0:example2} may contain infinitely many contractions $m>0$ of the form
\begin{equation}
    F^{a b(m) , c}F_{b(m) f, c} \, e^f, \quad R^{ab(m), cd}R_{cb(m), df} \, e^f \, .
\end{equation}
In the lower spin case these would lead to space-time nonlocalities, because
\begin{equation}
    F^{ b(m) a, c} \thicksim  \,  \partial^{\mu_1} \dots \partial^{\mu_m} e^{(b_1}_{\mu_1} \dots e^{b_m}_{\mu_m}F^{a), c} + \dots  \,,
\end{equation}
where the missing terms contain less derivatives.
However, spin locality of $\Upsilon(\omega, C, C)$ ensures that only finite number of such combinations contribute to \eqref{spin-0:example2}.

The equations for the spin one field and its lowest descendants are
\begin{equation}
     \dr_x F^{a,b} + \big(\omega^{a}\,_{c} F^{cb} - \omega^{b}\,_{c} F^{ca}\big) - \frac{3}{4} \, e_c \big( F^{ca,b} - F^{cb,a} \big)  = 0 \, ,
\end{equation}
\begin{equation}
    \label{3.73}
    \begin{split}
    \dr_x F^{ab,c} + \big( \omega^{a}\,_{d} F^{db,c} &+ \omega^{b}\,_{d} F^{da,c} + \omega^{c}\,_{d} F^{ab,d} \big) - 3 e_d\big(F^{dab,c} + \frac{1}{3} F^{abc,d} \big) - \frac{4}{3} \, (e^a F^{b,c} + e^b F^{a,c}) = \\
    &= \frac{8}{3}  \, e_f \big(F^a\,_{d} R^{fd, bc} +  F^b\,_{d} R^{fd, ac} - 2F^c\,_{d} R^{fd, ab}\big) + 4 \, e_f F^f\,_{d} R^{ab,cd} \, .
    \end{split}
\end{equation}
{\it A priori} the rhs of \eqref{3.73} could have contained infinitely many terms of the following structure\footnote{The proper index symmetrization is assumed to match the Young symmetry of $F^{ab,c}$.}:
\begin{equation}
    F^{a g(n), d}R_{df g(n)}\, ^{bc} e^{f}, \quad e_f F^{d(n),f}  R_{d(n)c}\,^{ab} \,,
\end{equation}
but due to spin locality of $\Upsilon(\omega, C, C)$, there is only a finite number of such contributions remains.

\section{Vertex dualities}
\label{section:DualRel}
The important feature of the obtained 0-form vertices is
\begin{equation}\label{Ver0}
    \Upsilon(\omega, C^n) \Big|_{y_\alpha = 0} = 0\,.
\end{equation}
The analysis of this section relies on the property \eqref{Ver0} of vertices \eqref{verC} only and, in this regard, is more general, than just exploring properties of the generating system \eqref{generating}.

Suppose condition \eqref{Ver0} is imposed on the vertices from the 0-form sector. Then a consequence of the compatibility $\dr_x^2=0$ of the system \eqref{nonlinear}
and \eqref{Ver0} amounts to the following relation that holds at
$y_\alpha=0$ only:
\begin{equation}
    \label{DualityMap}
    \mathcal{V}(\omega, \omega, C^n) \ast C - \omega \ast \Upsilon(\omega, C^{n+1}) - \Upsilon(\omega, C^{n+1}) \ast \pi\big(\omega\big) - C \ast \pi\bigg( \mathcal{V}(\omega, \omega, C^n) \bigg) \overset{y_\alpha=0}{=}
    0\,.
\end{equation}
We call the relation \eqref{DualityMap} the \textit{vertex duality} because it actually leads to a rather simple connection between the sources $\Psi$ and $\Phi$, \eqref{VertexSourcesDef}.
Condition \eqref{Ver0} appears to be quite natural because the minimal projectively compact spin-local vertices \cite{Vasiliev:2022med}, in fact, are forced to obey it.

Before proceeding with the general case, we illustrate with a simple example how sources $\Psi$ and $\Phi$ are related by \eqref{Ver0} at the lowest orders of perturbation theory. For $n=1$, there are three independent vertex orderings in each sector: 
\begin{equation}
    \begin{split}
       &\mathcal{V}(\omega, \omega, C) = \mathcal{V}_{\omega \omega C} + \mathcal{V}_{\omega C \omega } + \mathcal{V}_{C \omega \omega} \,,  \\
    &\Upsilon(\omega, C, C) = \Upsilon_{\omega C C} + \Upsilon_{C \omega  C} + \Upsilon_{C C \omega} \,. 
    \end{split}
\end{equation}
Inserting the expressions above into \eqref{DualityMap} and taking into account, that the rows of the form $C_1 \star \dots \star \omega_{1} \star \dots \star \omega_{2} \star \dots \star C_{n+1}$ are independent\footnote{To make this statement rigorous one can make use of the 'matrix trick' by replacing fields $\omega$ and $C$ with their colored counterparts $\omega_{I}^{J}$ and $C_{I}^{J}$, correspondingly.} of each other, one notes that the following identities must hold:
\begin{equation}
    \label{DualExm}
    \begin{split}
    \mathcal{V}_{\omega \omega C} \ast C - \omega \ast \Upsilon_{\omega C C} \overset{y=0}{=} 0\,,& \quad \quad \mathcal{V}_{\omega C \omega} \ast C - \omega \ast \Upsilon_{C\omega C} \overset{y=0}{=} 0\,, \\
    \mathcal{V}_{C \omega \omega} \ast C - C \ast \pi(\mathcal{V}_{\omega \omega C}) \overset{y=0}{=} 0\,,& \quad \quad -\omega \ast \Upsilon_{C C \omega} - \Upsilon_{\omega C C} \ast \pi(\omega) \overset{y=0}{=} 0\,,  \\
    -\Upsilon_{C\omega C} \ast \pi(\omega) - C \ast \pi(\mathcal{V}_{\omega C \omega}) \overset{y=0}{=} 0\,,& \quad \quad -\Upsilon_{C C \omega} \ast \pi(\omega) - C \ast \pi(\mathcal{V}_{C \omega \omega}) \overset{y=0}{=} 0\,. \\
    \end{split}
\end{equation}
Now, it is natural to calculate these expressions in terms of sources $\Psi$ and $\Phi$. For example, the first equation in \eqref{DualExm} is
\begin{equation}\label{dual:wwCC.cond}
    \big(\Psi^{[0,0]}_1(-p_2|t_1, t_2; p_1) - \Phi^{[0]}_2(t_1|t_2; p_1, p_2) \big)\big(\omega \star \omega \star C \star C\big) = 0\,.
\end{equation}
Now, one can strip off $\omega \star \omega \star C \star C$ that leads to the following relation between $\Psi$ and $\Phi$: 
\begin{equation}\label{dual:wwCC.0}
    \Psi^{[0,0]}_1(-p_2|t_1, t_2; p_1) = \Phi^{[0]}_2(t_1|t_2; p_1, p_2)\,.
\end{equation}
One may argue that the consequence \eqref{dual:wwCC.0} of the original condition \eqref{dual:wwCC.cond} is a far-fetched one. Indeed, as $p_{1,2}$ and $t_{1,2}$ are operators acting on the string $\omega \star \omega \star C \star C$, it is not obvious that this implies \eqref{dual:wwCC.0}. We note in this regard that \eqref{dual:wwCC.cond} contains no space-time differentiation, which, if present, require taking equations \eqref{nonlinear} into account, but instead, Eq. \eqref{dual:wwCC.cond} is valid at any space-time point. So, the functions $\go$ and $C$ can be considered as an arbitrary analytic HS initial data. Equation \eqref{dual:wwCC.cond} then means that the operator in the parenthesis acting on an arbitrary analytic function that admits $\star$ factorized form gives zero. This entails that the operator itself is zero, \eqref{dual:wwCC.0}. Relabeling $sp(2)$ variables, $p_2 \longrightarrow -y, p_1 \longrightarrow p$, we finally obtain
\begin{equation}\label{dual:wwCC}
    \Psi^{[0,0]}_1(y|t_1, t_2; p) = \Phi^{[0]}_2(t_1|t_2; p, -y)\,.
\end{equation}

It can be noted from \eqref{DualExm}, that the vertex duality splits vertices of different orderings into different ``families,'' which can be represented by the diagram in Fig. 5.
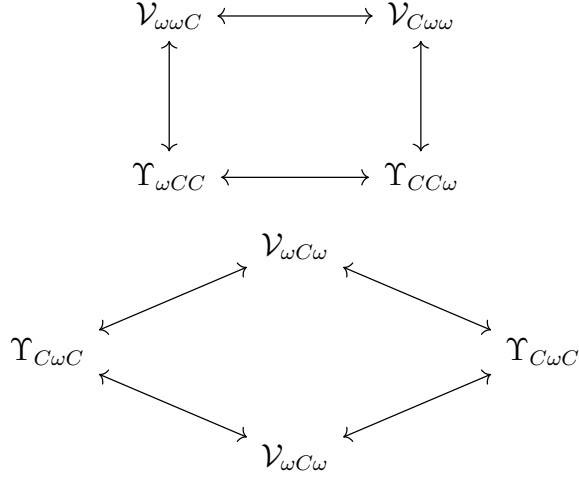
\begin{figure}[h!]
    \[\begin{tikzcd}
	{\mathcal{V}_{\omega \omega C}} && {\mathcal{V}_{C\omega\omega}} \\
	\\
	{\Upsilon_{\omega CC}} && {\Upsilon_{CC\omega }}
	\arrow[tail reversed, from=3-1, to=3-3]
	\arrow[tail reversed, from=3-1, to=1-1]
	\arrow[tail reversed, from=3-3, to=1-3]
	\arrow[tail reversed, from=1-1, to=1-3]
\end{tikzcd}\]
    \[\begin{tikzcd}
	&& {\mathcal{V}_{\omega C \omega}} \\
	{\Upsilon_{C\omega C}} &&&& {\Upsilon_{C\omega C}} \\
	&& {\mathcal{V}_{\omega C \omega}}
	\arrow[tail reversed, from=2-1, to=1-3]
	\arrow[tail reversed, from=2-1, to=3-3]
	\arrow[tail reversed, from=3-3, to=2-5]
	\arrow[tail reversed, from=1-3, to=2-5]
    \end{tikzcd}\]
    \caption{Diagrams showing which orderings of $\Upsilon(\omega, C, C)$ and $\mathcal{V}(\omega, \omega, C)$ are related to each other via duality.}
    \label{n=1}
\end{figure}

In the general case, Eq. \eqref{DualityMap} gives the following relations between sources for vertices of different orderings:
\begin{subequations}
    \label{DualRel}
    \begin{align}
        \Psi^{[0, k]}_n(-p_{n+1}|t_1, t_2; p_1, ..., p_n) -& \Phi^{[k]}_{n+1}(t_1|t_2; p_1, ..., p_{n+1}) = 0\\
        \Psi^{[k_1, k_2]}_n(-p_{n+1}|t_1, t_2; p_1, ..., p_n) - \Psi^{[k_1-1, k_2 -1]}_n & (-p_1|t_1, t_2; p_2, ..., p_{n+1}) = 0, \quad k_1 > 0\\
        -\Psi^{[k,n]}_n(-p_1|t_1, t_2; p_2, ..., p_{n+1}) -& \Phi^{[k+1]}_{n+1}(t_2|t_1; p_1, ..., p_{n+1}) = 0\\
        -\Phi^{[n+1]}_{n+1}(t_1|t_2; p_1, ..., p_{n+1}) -& \Phi^{[0]}_{n+1}(t_2|t_1; p_1, ..., p_{n+1}) = 0\,.
    \end{align}
\end{subequations}
The relations above allow us expressing every source $\Psi$ in terms of $\Phi$ and also give us a few relations between sources from the same sector. Namely, from the first two equations in \eqref{DualRel}, it follows that
\begin{equation}
    \label{Psi-Phi}
    \Psi^{[k_1, k_2]}_n(y|t_1, t_2; p_1, ..., p_n) = \, \,\Phi^{[k_2-k_1]}_{n+1}(t_1|t_2; p_{k_1+1}, ..., p_n, -y, p_1, ...)\, .
\end{equation}
Substituting this into the third equation of \eqref{DualRel}, one obtains the relations between different $\Phi$'s: 
\begin{equation}
    \label{Phi-Phi}
    \Phi^{[k]}_n(y|t; p_1, ..., p_n) = -\Phi^{[n-k]}_n(t|y; p_{k+1}, ..., p_n, p_1, ...) \, .
\end{equation}
It is important to stress that the relation \eqref{Phi-Phi} does not provide connection between vertices $\Upsilon(\omega, C^n)$ of mirrored orderings; instead the connection between vertices of mirrored ordering comes from the reality conditions for $\omega$ and $C$. The properties of $\star$ product and the reality conditions \eqref{RealCondW}, \eqref{RealCondC} give
\begin{equation}
    \Big[ \Upsilon\big( \overbrace{C, \dots, C}^{k} , \,
         \omega, \,\overbrace{C, \dots, C}^{n-k}\big) \Big]^\dag = - \pi\Big( \Upsilon\big( \overbrace{C, \dots, C}^{n-k} , \,
         \omega, \,\overbrace{C, \dots, C}^{k}\big)  \Big) \,,
\end{equation}
which in terms of sources amounts to 
\begin{equation}
    \label{6.12}
    \big[\Phi^{[k]}_n(y|t; -p_n, -p_{n-1}, \dots , -p_1) \big]^\dag = (-)^{n} \, \Phi^{[n-k]}_n(-y|t; p_1, \dots , p_n) \,. 
\end{equation}
Notice, that  \eqref{Phi-Phi} combined with \eqref{6.12} result in a constraint relating $\Phi_n^{[k]}$ with its Hermitian conjugate of the properly mixed variables $y$, $t$ and $p_i$.

On the other hand, \eqref{Psi-Phi} says that all sources corresponding to the vertices with $k$ pieces of $C$'s between two $\omega$'s are related to the source for the vertex in the sector of 0-forms, which has $k$ pieces of $C$'s before $\omega$. Summarizing, this can be represented by the diagram, as shown in Fig. 6.
\begin{figure}[h!]
\[\begin{tikzcd}
	&& {\mathcal{V}^{[0, k]}_{n}} && {\mathcal{V}^{[1, k+1]}_{n}} && \dots && {\mathcal{V}^{[n-k,n]}_{n}} \\
	{\Upsilon_{n+1}^{[k]}} &&&&&&&&& {\Upsilon_{n+1}^{[n+1-k]}} \\
	&& {\mathcal{V}^{[k-1,n]}_{n}} && {\mathcal{V}_n^{[k-2,n-1]}} && \dots && {\mathcal{V}^{[0,n-k+1]}_n}
	\arrow[tail reversed, from=2-1, to=1-3]
	\arrow[tail reversed, from=1-3, to=1-5]
	\arrow[tail reversed, from=1-5, to=1-7]
	\arrow[tail reversed, from=1-7, to=1-9]
	\arrow[tail reversed, from=1-9, to=2-10]
	\arrow[tail reversed, from=2-1, to=3-3]
	\arrow[tail reversed, from=3-3, to=3-5]
	\arrow[tail reversed, from=3-5, to=3-7]
	\arrow[tail reversed, from=3-7, to=3-9]
	\arrow[tail reversed, from=3-9, to=2-10]
\end{tikzcd}\]
    \caption{Vertices related by dualities. The lower index denotes perturbation order in $C$, while the upper index convention is the same as for sources. For $k=0$ the lower branch degenerates to a  straightforward duality between $\Upsilon$'s. Something similar happens to the upper branch for $k=n+1$.}
    \label{DualDiagram}
\end{figure}
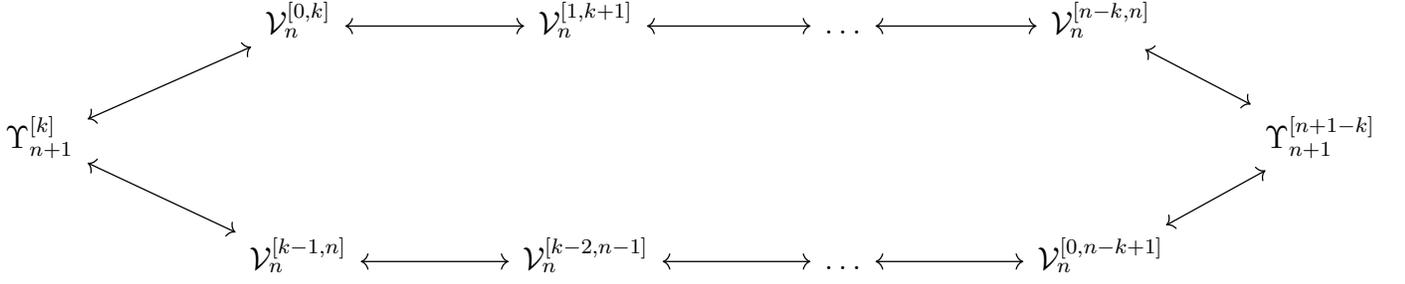

An important comment is now in order. In obtaining the above expressions, like \eqref{Psi-Phi}, we used the fact that the variables $y$, $p$, and $t$ that enter sources are independent. Thus, if two sources equal each other, they remain so for arbitrary values of
these variables. This allows us to set, say, $p_i=y$ for a particularly convenient number $i$. This is how $y$ reappears in
\eqref{Psi-Phi}, \eqref{Phi-Phi}. As an example, for $k_1 = 0$ we have
\begin{equation*}
    \Psi^{[0, k]}_{n}(y|t_1, t_2; p_1, ..., p_n) = \Phi^{[k]}_{n+1}(t_1|t_2; p_1, ..., p_n,
    -y)\,.
\end{equation*}
We refer to identities \eqref{Psi-Phi}, \eqref{Phi-Phi} relating various vertex sources with each other also as to {\it vertex dualities}. The more detailed proof is straightforward and we leave it to the  Appendix \ref{sec:Appendix4}. It may look surprising, however, that the compatibility condition \eqref{DualityMap} at $y=0$ gives access to the vertex $\Psi$ at an
arbitrary point $y$. This happens thanks to the manifest star
product in \eqref{DualityMap} that acts nonlocally. In other
words, the star product relates a given point in $y$ space with any
other.

Identity \eqref{Psi-Phi} not only allows recovering any
source $\Psi$ from $\Phi$, but it results in some consequences for
their locality in addition. Namely, if $\Phi$ is spin local, i.e., it
carries  no $(p_i p_j)$ contractions, then there is no $(y p)$
contractions in $\Psi$, which is the strongest version of spin
ultralocality. The opposite is also true: If $\Psi$ is ultralocal, then $\Phi$ is spin local. So, one proves the important statement: \\
\\
\textit{If all vertices $\Upsilon(\omega, C^n)$ are subject of condition \eqref{Ver0}, then the vertex $\mathcal{V}(\omega, \omega, C^{n})$ is ultralocal if and only if the vertex $\Upsilon(\omega, C^{n+1})$ is spin local.}\\
\\

Let us point out, that condition \eqref{Ver0} is automatically fulfilled for projectively compact spin-local  vertices introduced in \cite{Vasiliev:2022med}, which play important role in the context of space-time locality.

Another consequence of the vertex duality has to do
with the recently observed \textit{shift symmetry}, \cite{Didenko:2022qga, Didenko:2022eso, Didenko:2023vna}. Suppose $\Phi$ satisfies the
following property:
\begin{equation}
    \Phi^{[k]}_n(y|t; p_i+b) = e^{ib(t+y)} \Phi^{[k]}_n(y|t;
    p_i)\,,
\end{equation}
where $b_{\al}$ is an arbitrary $sp(2)$-parameter, then  it follows
from \eqref{Psi-Phi} that
\begin{equation}
    \Psi_n^{[k_1, k_2]}(y-b|t_1, t_2; p_i + b) = e^{ib(t_1 + t_2)}\Psi_n^{[k_1, k_2]}(y|t_1, t_2; p_i)
\end{equation}
i.e., shift symmetry of $\Phi$ implies shift symmetry of $\Psi$.
The opposite is also true. Therefore, if the vertices $\Upsilon(\omega, C^n)$ and $\mathcal{V}(\omega, \omega, C^n)$ are
subject to the duality identities \eqref{DualityMap} and
vertices of a given sector (say $\Phi$) feature shift-symmetry,
then and only then, vertices of another sector ($\Psi$) are
shift symmetric. Shift symmetry plays an interesting role in the HS locality problem. In particular, as was shown in \cite{Didenko:2022eso}, it gives one a class of field redefinitions that respects spin locality. In this regard, let us note that the shift-symmetry transformation applied to the vertices \eqref{eq:0-VertexSource} and \eqref{eq:1-VertexSource}  $p_i \longrightarrow p_i + a$ results in 
\begin{equation}
    P_n(\zeta) \longrightarrow P_n(\zeta) + \Big( \sum_{k=1}^n \zeta_k \Big) a\, ,
\end{equation}
which in combination with specific properties of the integration domains $\mathcal{D}^{k}_n$ and $\mathcal{D}^{[k_1, k_2]}_n$, leads to shift symmetry of the obtained vertices.

\section{Implication for the holomorphic sector in $4d$}\label{Sec:holomor}

The generating system \eqref{generating} is a straightforward generalization of the earlier obtained equations for the
(anti)holomorphic sector of the four-dimensional HS theory
\cite{Didenko:2022qga}, which describes the on-shell propagation. Thus, our findings on the HS vertices, shift symmetry, duality relations, and all their
consequences can be literally applied to propagation of the holomorphic HS fields in $4d$.

In four dimensions, the HS algebra is the associative algebra of functions of the following oscillators \cite{Vasiliev:1992av}:
\be
[y_{\al}, y_{\gb}]_*=2i\gep_{\al\gb}\,,\qquad [\bar{y}_{\dal},
\bar{y}_{\dgb}]_*=2i\gep_{\dal\dgb}\,.
\ee
The product is the usual Moyal star product
\be
f(y,\bar y)*g(y, \bar y)=\int f(y+u, \bar y+\bar u)g(y+v, \bar
y+\bar v)e^{iuv+i\bar u\bar v}\,.
\ee
Compared to the $d$-dimensional off-shell HS algebra \eqref{moyal}, the analog of variable $\vec\y$ is $\bar y$ in four dimensions. Correspondingly, the star product \eqref{starvec} is
replaced with the analogous one
\be\label{barstar}
f(y, \bar y)\bar{*}g(y, \bar y)=\int f(y, \bar y+\bar u)g(y, \bar
y+\bar v)\,e^{i\bar u\bar v}\,.
\ee
The unfolded form of the dynamical equations is \eqref{nonlinear},
which can be generated using the master fields $W(z; y, \bar y)$,
$\Lambda(z; y, \bar y)$ and $C(y, \bar y)$. These are reproduced
from
\begin{subequations}
    \label{generating:hol}
    \begin{align}
        \label{generating:1-sectorhol}& \dr_x W + W \ast W = 0\,, \\
        \label{generating:Whol}& \dr_z W + \{W, \Lambda\}_{\ast} + \dr_x \Lambda = 0\,, \\
        \label{generating:0-sectorhol}& \dr_x C + \big(W(z'; y, \bar y) \ast C - C \ast W(z'; -y, \bar y) \big)\Big|_{z'=-y} =
        0\,,
    \end{align}
\end{subequations}
where this time the large star product $*$ in
\eqref{generating:hol} is given by \eqref{limst} with $\bar{*}$ in
place of $\star$. So, one can use expressions for the HS vertices
\eqref{eq:0-VertexSource}, \eqref{eq:1-VertexSource} in the
holomorphic case too with the prescription
\eqref{VertexSourcesDef:0-form}, \eqref{VertexSourcesDef:1-form}
being modified by the star product \eqref{barstar}. Similarly, the
vertex dualities \eqref{Psi-Phi} and \eqref{Phi-Phi} remain intact.

The holomorhic HS sector was
also investigated in \cite{Sharapov:2022awp, Sharapov:2022nps} using the limiting star product \eqref{limst}. Some vertices extracted in \cite{Sharapov:2022awp, Sharapov:2022nps} were divergent, however. To overcome their regularization, a certain duality map was postulated, which look reminiscent of the more general relations \eqref{DualRel}.

\section{Comparison with Vasiliev's system}\label{Sec:Vasiliev}
As the generating equations \eqref{generating} differ from the original Vasiliev ones \cite{Vasiliev:2003ev}, it is instructive to compare the HS vertices \eqref{nonlinear} of the two systems. Naturally, one expects the results should agree.  However, this might not be easy to see due to freedom in field redefinition, which should be properly adjusted for the matching. Since vertices \eqref{nonlinear} obtained from \eqref{generating} are already fixed,  the problem boils down to an appropriate resolution for the field $z$ dependence in the Vasiliev case. We will demonstrate the agreement of the two approaches at least to the lowest interaction order.  To this end, let us proceed with Vasiliev's equations, which can be written down in the following form\footnote{The original equations \cite{Vasiliev:2003ev} are presented in the AdS covariant fashion by means of introducing a constant compensator field $V^A$. Since the AdS symmetry is in any way broken by the automorphism $\pi$, we gauge fix it by setting $V^A=(1, \vec{0})$. As a result, one is left with only two $z$ oscillators $z_\al:=V^AZ_{A\al}$ that properly encode HS interactions, as opposed to the $2(d+1)$ original oscillators $Z_{\al}^A$.}: 
\begin{subequations}
    \label{generating:Vasiliev}
    \begin{align}
        \label{Vasiliev:1-sector}& \dr_x W + W \ast W = 0\,, \\
         \label{Vasiliev:0-sector}& \dr_x B + W \ast B-B*\pi(W) = 0\,, \\
        \label{Vasiliev:W}& \dr_x S + \{W, S\}_{\ast} = 0\,, \\
         \label{Vasiliev:S}& S*S=i\,\dr z^{\al}\dr z_{\al}(1+B*\gk)\,, \\
          \label{Vasiliev:B}& S*B-B*\pi(S) = 0\,.
    \end{align}
\end{subequations}
Here a set of master fields depend on the generating variables $z$ and $Y$ of \eqref{Y}
\begin{align}
    W=\dr x^{\mu}W_{\mu}(z, Y|x)\,,\quad B=B(z, Y|x)\,,\quad S=\dr z^{\al}S_{\al}(z, Y|x)\,.
\end{align}
The automorphism $\pi$ is defined as
\begin{equation}
\pi f(z,y, \dr z)=f(-z,-y,-\dr z)
\end{equation}
and the standard Klein operator is
\begin{equation}
    \gk=e^{iz_{\al}y^{\al}}\,,\qquad \gk*f(z,y)=f(-z,-y)*\gk\,,\quad \gk*\gk=1\,.
\end{equation}
The original Vasiliev star product differs from \eqref{limst} and is given by
\be\label{Vasiliev:star}
(f*g)(z; Y)=\int f(z+u, y+u)\star g(z-v, y+v)e^{i u_{\al}v^{\al}}
\ee
with $\star$ being the same as \eqref{starvec}. The star product  \eqref{Vasiliev:star} amounts to the following actions (cf. \eqref{star:gen})
\begin{subequations}\label{star:gen1}
\begin{align}
&y* =y+i\ff{\p}{\p y}-i\ff{\p}{\p z}\,,\qquad z* =z+i\ff{\p}{\p
y}-i\ff{\p}{\p z}\,,\\
&* y=y-i\ff{\p}{\p y}-i\ff{\p}{\p z}\,,\qquad * z=z+i\ff{\p}{\p
y}+i\ff{\p}{\p z}\,.
\end{align}
\end{subequations}
While star products \eqref{limst} and \eqref{Vasiliev:star} are different, they prove to be the same for the following products:
\begin{equation}\label{star:equiv}
    f(y)*g(z,y)\quad\textnormal{and}\quad g(z,y)*f(y)\,.
\end{equation}
It is important to notice that the equations \eqref{generating:Vasiliev} become ill defined once the star product \eqref{Vasiliev:star} is replaced with \eqref{limst}. The opposite is also true; Eqs. \eqref{generating} are not consistent for the star product \eqref{Vasiliev:star}, since the projective identity \eqref{iden} fails for this choice. This makes comparison of the two systems not obvious. A possible way out is to use the one-parametric $\gb$-star product introduced in \cite{Didenko:2019xzz} for the reordering ambiguity of \eqref{generating:Vasiliev} that interpolates the two star products. The contraction $\gb\to-\infty$ yields \eqref{limst}, while $\gb=0$ reproduces \eqref{Vasiliev:star}. The limit $\gb\to-\infty$ can be studied at the level of the Vasiliev generating equations. What makes it highly nontrivial is the control over associativity of the large algebra. The associativity appears to be lost in the limit, in general, while it may survive on a certain classes of functions. We refer to \cite{Didenko:2022qga} for more details on that matter. Nevertheless, one may attempt to compare the results of the two systems at the level of vertices when available. 

\subsection{Perturbations}
Perturbation theory for the Vasiliev equations is well elaborated (see, e.g., \cite{Sezgin:2002ru, Didenko:2015cwv}). It starts with the proper vacuum 
\begin{subequations}
\begin{align}
    &W^{(0)}=\go(Y|x)\,,\\
    &S^{(0)}=\dr z^{\al}z_{\al}\,,\\
    &B^{(0)}=0\,.
\end{align}
\paragraph{Order $O(C)$.} At the first order we have from \eqref{Vasiliev:B}
\begin{equation}
[S^{(0)}, B^{(1)}]_*=0\quad\to\quad [z_\al, B^{(1)}]_*=0\quad\Rightarrow\quad B^{(1)}=C(Y|x)\,.    
\end{equation}
   \end{subequations}
and 
\begin{equation}
    \{S^{(0)}, S^{(1)}\}_*=i\,\dr z^{\al}\dr z_{\al}\,C*\gk\quad\Rightarrow\quad \dr_z S^{(1)}=\ff12 C*\gk\,\dr z^{\al}\dr z_{\al}
\end{equation}
with the solution being
\begin{equation}\label{S1}
S^{(1)}=\dr z^\alpha \, z_\alpha \int_0^1 d\tau \, \tau C(-\tau z, \Vec{\bf{y}}) \, e^{i \tau z_\alpha
    y^\alpha}\,.
\end{equation}
Notice that $S^{(1)}\equiv\Lambda$, \eqref{LambdaDef}. This fact is not accidental, but rather follows from the exact equivalence of the star products \eqref{limst} and \eqref{Vasiliev:star} for products with a single $z$-independent function \eqref{star:equiv}. The field $W^{(1)}$ is determined from \eqref{Vasiliev:W}
\begin{equation}
\dr_xS^{(1)}+\{\go,S^{(1)}\}_*+\{W^{(1)}, S^{(0)}\}_*=0\,,    
\end{equation}
which entails
\begin{equation}\label{eq:W1}
\dr_xS^{(1)}+\{\go,S^{(1)}\}_*-2i\,\dr_z W^{(1)}=0\,.
\end{equation}
The solution for $W^{(1)}$ can be found using the standard homotopy prescription
\begin{equation}
W^{(1)}=\ff{1}{2i}\Delta\left(\{\go,S^{(1)}\}_*\right)\,,  
\end{equation}
where the contracting homotopy is defined in \eqref{homotopy}.
We note again that up to field normalization,  Eq. \eqref{eq:W1} is exactly the same as Eq. \eqref{generating:W} at the linear order $O(C)$ due to the argument \eqref{star:equiv}. The lowest vertex is then found from \eqref{Vasiliev:1-sector} 
\begin{equation}\label{Vasiliev:wwC}
\mathcal{V}(\go, \go, C)=-(\go*W^{(1)}+W^{(1)}*\go)\big|_{z=0}\,.
\end{equation}
It coincides with the one earlier calculated in \eqref{wwC} up to the factor of $\ff{1}{2i}$ (which can be redefined away by rescaling $C\to 2i C$) simply because \eqref{Vasiliev:wwC} is the same (up to a number) as \eqref{V1} for $n=1$. 
Drawing a line here, the agreement of the vertices $\mathcal{V}(\go, \go, C)$ for the both generating systems can be reached without having calculated them. This happens because the two systems coincide at the order $O(C)$.

\paragraph{Order $O(C^2)$.} The analysis gets slightly more involved at the second order, where the Vasiliev module $B$ comes to differ from $C$. Still, one can show that the $C^2$ vertices are the same through a direct calculation. To this end, we proceed with the convenient source prescription \eqref{source} and ignore field dependence on $\vec y$, that enters the final expressions via $\star$ products \eqref{starvec}, which can be easily restored,    
\begin{equation}\label{wC:exp}
\go\to e^{-iyt}\,,\quad C\to e^{-iyp}\,.
\end{equation}
So,
\begin{equation}\label{S1:exp}
S^{(1)}\to \dr z^{\al} z_{\al}\int_{0}^{1}d\tau \tau\,e^{i\tau z(y+p)}\,.
\end{equation}
Up to a factor the source for the field $W^{(1)}$ is the same as in \eqref{W_k|n} for $n=1$
\begin{equation}\label{W1:exp}
W^{(1)}\to \ff{i}{2}\mathcal{W}^{(0|1)}+\ff{i}{2}\mathcal{W}^{(1|1)}\,,
\end{equation}
where $\mathcal{W}^{(0|1)}$ corresponds to $W^{(1)}\big|_{\go C}$, while  $\mathcal{W}^{(1|1)}$ to $W^{(1)}\big|_{C \go}$, respectively. In what follows we stick to the $\oast$-factorized form, \eqref{oast:Appendix} for the fields under consideration. In these terms these are given by
\begin{subequations}
\begin{align} 
&\mathcal{W}^{(0|1)}=\int_0^1 d\gs e^{i(1-\gs)\,ty-i\gs\,
tp}\oast \int_{0}^{1}d\tau (1-\tau)z t\,e^{i\tau z(y+p+t)}\,,\label{W1:wC}\\
&\mathcal{W}^{(1|1)}=-\int_0^1 d\gs e^{i(1-\gs)\,ty-i\gs\,
tp}\oast \int_{0}^{1}d\tau (1-\tau)z t\,e^{i\tau z(y+p-t)}\,,\label{W1:Cw}
\end{align}
\end{subequations}
The evolution of $B^{(2)}$ along $z$ is governed by \eqref{Vasiliev:B}
\begin{equation}
    [S^{(0)}, B^{(2)}]_*=C*\pi(S^{(1)})-S^{(1)}*C
\end{equation}
or, equivalently, 
\begin{equation}\label{eq:B2}
    \dr_z B^{(2)}=\ff i2\left(C*\pi(S^{(1)})-S^{(1)}*C\right)\,.
\end{equation}
Its solution via the standard contracting homotopy is known to be inconsistent with locality \cite{Giombi:2009wh}. Therefore, one should solve for \eqref{eq:B2} differently. The shifted homotopy calculation does reproduce the local vertex $\Upsilon(\go, C, C)$ along the lines of \cite{Didenko:2018fgx}, which, however, is not equal to the one found in \eqref{verC} for $n=2$. This is precisely the difficulty with the proper frame choice that one has to deal with, as we have already discussed above. Our strategy in what follows is to seek for the solution of \eqref{eq:B2} in the $\oast$-factorized form. The rationale behind this choice is to keep the same functional class for the Vasiliev master fields $B$ and $W$ as in the case of Eqs. \eqref{generating} with the field $W$\footnote{This makes sense because the functional class in question turns out to be $\gb$-reordering invariant, \cite{Didenko:2022qga} and, therefore, remains the same for $\gb=0$ and $\gb=-\infty$.}. The $\oast$-factorized solution does exist. Indeed, using \eqref{wC:exp}, \eqref{S1:exp} and \eqref{oast:Appendix}, one easily finds
\begin{subequations}
\begin{align}
     &C*\pi(S^{(1)})=e^{-iyp_1}*\dr z^{\al} z_{\al}\int_{0}^{1}d\tau \tau\,e^{i\tau z(y-p_2)}=\dr z^{\al}\,e^{-iyp_1}\oast  z_{\al}\int_{0}^{1}d\tau \tau\,e^{i\tau z(y+p_1-p_2)}\,,\\
     &S^{(1)}*C=\dr z^{\al} z_{\al}\int_{0}^{1}d\tau \tau\,e^{i\tau z(y+p_1)}*e^{-iyp_2}=\dr z^{\al}\,e^{-iyp_2}\oast  z_{\al}\int_{0}^{1}d\tau \tau\,e^{i\tau z(y+p_1-p_2)}\,.
\end{align}
\end{subequations}
Using the identity
\be\label{iden:int}
f(x)-f(y)=\int_{0}^{1}d\gs\,\ff{\p}{\p\gs}f\left(\gs x+(1-\gs)y\right)
\ee
we further rewrite \eqref{eq:B2} as
\begin{align}\label{eq:dzB2}
\dr_z B^{(2)}=&\ff i2 \dr z^{\al}\int_{0}^{1}d\gs\,\ff{\p}{\p \gs} e^{-i\gs\,yp_1-i(1-\gs)\,yp_2}\oast  z_{\al}\int_{0}^{1}d\tau \tau\,e^{i\tau z(y+p_1-p_2)}=\\
=&\ff12  \dr z^{\al}\,y(p_1-p_2)\int_{0}^{1}d\gs\, e^{-i\gs\,yp_1-i(1-\gs)\,yp_2}\oast  z_{\al}\int_{0}^{1}d\tau \tau\,e^{i\tau z(y+p_1-p_2)}\,.
\end{align}
The following identity proved in \cite{Didenko:2022eso}
\begin{equation}   
\ff{\p}{\p z^{\al}}\int_{0}^{1}d\tau\, (1-\tau)f^{\gb}(y)\oast z_{\gb}e^{i\tau z(y+q)}=-i\int_{0}^{1}d\tau\,\tau f^{\gb}(y)(y+q)_{\gb}\oast z_{\al}e^{i\tau z(y+q)}\,,
\end{equation}
which is valid for any function $f(y)$ and parameter $q_\al$ greatly simplifies further analysis. Using it, \eqref{eq:dzB2} casts into
\begin{equation}    
\dr_z B^{(2)}=\ff i2  \dr_z\left(y^{\al}\int_{0}^{1}d\gs\, e^{-i\gs\,yp_1-i(1-\gs)\,yp_2}\oast  z_{\al}\int_{0}^{1}d\tau (1-\tau)\,e^{i\tau z(y+p_1-p_2)}\right)\,.
\end{equation}
Thus, we can take the solution in the form 
\begin{equation}\label{B2}
  B^{(2)}=\ff i2  y^{\al}\int_{0}^{1}d\gs\, e^{-iy(\gs\,p_1+(1-\gs)\,p_2)}\oast  z_{\al}\int_{0}^{1}d\tau (1-\tau)\,e^{i\tau z(y+p_1-p_2)}\,.  
\end{equation}
We note that \eqref{B2} is the $d$-dimensional counterpart of the four-dimensional solution originally found in \cite{Vasiliev:2017cae} by the proper field redefinition carried out by hand. The solution cannot be reproduced using shifted homotopies of \cite{Didenko:2018fgx}, although it is accessible via the recent technique of \cite{Vasiliev:2023yzx}. In our case this solution arises from the requirement to belong to the functional class (called $\mathbf{C}^0$) that evolves on the generating equations \eqref{generating}. This class is conveniently generated by the $\oast$ product; see \cite{Didenko:2022qga} for details.

Naturally, there is freedom in any $z$-independent function, which one can add to \eqref{B2}. However, it is this particular $B^{(2)}$ given by \eqref{B2} that eventually leads to \eqref{wCC:expl}. Let us show that this is indeed so. 

At this order the 0-form vertex comes from \eqref{Vasiliev:0-sector} as
\begin{equation}
\Upsilon(\go, C, C)=-\dr_x B^{(2)}-W^{(1)}*C+C*\pi(W^{(1)})-\go*B^{(2)}+B^{(2)}*\pi(\go)\,.    
\end{equation}
It contains three different orderings 
\begin{align}
&\Upsilon(\go, C, C)=\\
&=\Phi_2^{[0]}(y|t; p_1, p_2)(\go\star C\star C)+\Phi_2^{[1]}(y|t; p_1, p_2)(C\star\go\star  C)+ \Phi_2^{[2]}(y|t; p_1, p_2)(C\star C\star \go)\,.\nonumber   
\end{align}
We focus here on $\Phi_2^{[0]}$, for simplicity, which collects
\begin{equation}\label{Vas:wCC}
    \Upsilon_{\go C C}=-(\dr_x B^{(2)})\big|_{\go C C}-W^{(1)}\big|_{\go C}*C-\go*B^{(2)}\,.
\end{equation}
The contribution from $\dr_x B^{(2)}$ brings $\dr_x C$ that should be expressed via the equations of motion to the first order:
\begin{equation}
(\dr_x C)\big|_{\go C}=-\go*C=-e^{-iy t}*e^{-iyp_1}=-e^{-iy(t+p_1)+ip_1t}\,,    \end{equation}
implying that one should replace $p_1\to p_1+t$ in \eqref{B2} and add up the factor $e^{ip_1 t}$,
\begin{equation}
(\dr_x B^{(2)})\big|_{\go C C}=-\ff i2  y^{\al}\int_{0}^{1}d\gs\, e^{-i\gs\,y(p_1+t)-i(1-\gs)\,yp_2+ip_1 t}\oast  z_{\al}\int_{0}^{1}d\tau (1-\tau)\,e^{i\tau z(y+p_1+t-p_2)}\,.
\end{equation}
The contribution from the second term in \eqref{Vas:wCC} is easy to calculate using \eqref{W1:wC}
\begin{align}
 &W^{(1)}\big|_{\go C}*C=\ff{i}{2}\mathcal{W}^{(0|1)}*e^{-iyp_2}=\\ &\ff{i}{2}\,t^{\al}\int_0^1 d\gs e^{i((1-\gs)\,t-p_2)y-i\gs\,
tp_1+i(1-\gs)\,p_2 t}\oast z_{\al}\int_{0}^{1}d\tau (1-\tau)\,e^{i\tau z(y+p_1-p_2+t)}\nonumber\,.  
\end{align}
Lastly, the third contribution is found to be 
\begin{align}
    &\go*B^{(2)}=e^{-iyt}*B^{(2)}=\\
    &=\ff i2  (y+t)^{\al}\int_{0}^{1}d\gs\, e^{-i(y+t)(\gs\,p_1+(1-\gs)\,p_2)-i\,yt}\oast  z_{\al}\int_{0}^{1}d\tau (1-\tau)\,e^{i\tau z(y+t+p_1-p_2)}\,.\nonumber
\end{align}
Notice, that each of the three contributions have the same factor on the right of the $\oast$ product so that \eqref{Vas:wCC} acquires the following form:
\begin{equation}\label{ver:gen}
    \Phi_2^{[0]}=F^{\al}(y|t, p_1, p_2)\oast z_{\al}\int_{0}^{1}d\tau (1-\tau)\,e^{i\tau z(y+q)}\,,
\end{equation}
where 
    \begin{align}\label{F_alpha}
        F_{\al}=\ff i2\int_{0}^{1}d\gs&\Big(y_{\al}e^{-i\gs\,y(p_1+t)-i(1-\gs)\,yp_2+ip_1 t}- t_{\al}e^{i((1-\gs)\,t-p_2)y-i\gs\,
tp_1+i(1-\gs)\,p_2 t}-\\
-&(y+t)_{\al}e^{-i(y+t)(\gs\,p_1+(1-\gs)\,p_2)-i\,yt}\Big)\nonumber
    \end{align}
and 
\begin{equation}\label{q}
    q=t+p_1-p_2\,.
\end{equation}
The vertex \eqref{ver:gen} is $z$ independent by construction, which is not manifest, given the apparent presence of $z$ in its expression. The condition 
\begin{equation}
    \dr_z\Upsilon=0
\end{equation}
for the expression like \eqref{ver:gen} was analyzed in \cite{Didenko:2022eso} and shown to be equivalent to 
\begin{equation}
    (y+q)^{\al}F_{\al}(y|t; p_1, p_2)=0\,.
\end{equation}
Therefore, the latter condition is resolved by
\begin{equation}\label{verfunc}
    F_{\al}(y|t; p_1, p_2)=(y+q)_{\al}F(y|t; p_1, p_2)
\end{equation}
with some analytic function $F$, due to the two-component range of the $sp(2)$ indices $\al=1,2$. Plugging \eqref{verfunc} into \eqref{ver:gen} and performing the $\oast$ integration, which reduces the result down to the total $\tau$ derivative (see \cite{Didenko:2022eso} for more details), one arrives at
\begin{equation}\label{ver:F}
     \Phi_2^{[0]}=i F(y|t; p_1, p_2)\,.
\end{equation}
So, in order to obtain the quadratic vertex one needs to read off the function $F$ from \eqref{F_alpha} using \eqref{verfunc}. This can be arranged by contracting \eqref{verfunc} with, e.g., $t^{\al}$ and using the definition \eqref{q}
\begin{equation}\label{contr:tF}
    t^\al F_{\al}=(y-p_1+p_2)t\,F\,.
\end{equation}
From \eqref{F_alpha}, it follows that
\begin{equation}
    t^{\al}F_{\al}=\ff{i}{2}yt\int_{0}^{1}d\gs\,e^{-i\gs\,yp_1-i(1-\gs)\,yp_2}\left(e^{ip_1t}f(y|t; p_1+t, p_2)-e^{-iyt}f(y+t|t; p_1, p_2)\right)\,,
\end{equation}
where we introduced 
\begin{equation}
    f(y|t; p_1, p_2):=e^{-\gs\,yp_1-i(1-\gs)\,yp_2}\,.
\end{equation}
Making use of \eqref{iden:int}, we have
\begin{align}
    &t^{\al}F_{\al}=\\
    &\ff{i}{2}yt\int_{0}^{1}d\gs\,e^{-i\gs\,yp_1-i(1-\gs)\,yp_2}\int_0^1d\rho\ff{\p}{\p\rho}\left(e^{i\rho\,p_1t-i(1-\rho)\,yt}f(y+(1-\rho)t|t; p_1+\rho t, p_2)\right)=\nonumber\\
    &=-\ff{1}{2}(yt)\,(y+p_1-p_2)t\times\nonumber\\
    &\times\int_{[0,1]^2}d\gs d\rho\, (1-\gs)e^{-i y(\gs p_1+(1-\gs)p_2)-i t((\rho+\gs(1-\rho))p_1+(1-\gs)(1-\rho)p_2)+i(\gs\rho+1-\rho)\,ty}\,.\nonumber
\end{align}
Comparing the above expression with \eqref{contr:tF} one notes that the factor $(y-p_1+p_2)t$ exactly cancels out, while from \eqref{ver:F} it follows that
\begin{align}
    &\Upsilon_{\go C C}\to\Phi_2^{[0]}=\\
    &=-\frac{i}{2} (yt)\int_{[0,1]^2}d\gs d\rho\, (1-\gs)e^{-i y(\gs p_1+(1-\gs)p_2)-i t((\rho+\gs(1-\rho))p_1+(1-\gs)(1-\rho)p_2)+i(\gs\rho+1-\rho)\,ty}\nonumber\,,
\end{align}
which can be compared with $\Phi_2^{[0]}$ from \eqref{eq:0-VertexSource}. The two expressions match up to the prefactor of $\frac{i}{2}$ attributed to the specific normalization of Vasiliev's $z$-commutation relations. This can be seen after the following change of the integration variables is made:
\begin{align}
    &\xi_1=\gs\,,\qquad \xi_2=1-\gs\,,\\
    &\eta_1=\rho+\gs(1-\rho)\,,\qquad \eta_2=(1-\gs)(1-\rho)\,. 
\end{align}
The other orderings $\Upsilon_{C \go C}$ and $\Upsilon_{C C \go}$ of the $C^2$ vertex can be checked to agree with \eqref{eq:0-VertexSource} for $n=2$,  $k=1$ and $k=2$, correspondingly, too.

The vertex $\mathcal{V}(\go, \go, C, C)$ is harder to calculate in the Vasiliev case. While we do not intend to perform this calculation here, one may argue that the result should match \eqref{eq:1-VertexSource} for $n=2$. In this case the matching relies on the identity (see Sec. 6.2 from \cite{Didenko:2019xzz}):
\begin{equation}\label{struc:eq}
    S^{(1)}*S^{(1)}-i \dr z^{\al}\dr z_{\al}B^{(2)}*\gk\approx 0\,,
\end{equation}
where the sign $\approx$ means equality up to terms that do not contribute to the final vertex. The identity is valid for $B^{(2)}$ from \eqref{B2} only, allowing one to discard contribution to the vertex from $S^{(2)}$. Indeed, from \eqref{Vasiliev:S}, one concludes using \eqref{struc:eq} that at the order $C^2$:
\begin{equation}
    \dr_z S^{(2)}\approx 0\,.
\end{equation}

\paragraph{Higher orders.} Extracting vertices gets much more involved within the Vasiliev theory at the order $O(C^3)$ and higher. The only available in the literature results so far is the contribution to the $4d$ holomorphic sector\footnote{The central purpose of the paper \cite{Gelfond:2021two} was to demonstrate manifest locality of the holomorphic $O(C^3)$ vertex. The obtained vertex contains various pieces that might be related via partial integration. Its form makes comparison with the results of this paper a highly nontrivial task. As shown in \cite{Didenko:2022eso}, the vertex is shift symmetric, however, which suggests that  the $C^3$ contributions may also coincide.} \cite{Gelfond:2021two} and the recent result of Gelfond for the $4d$ mixed sector, \cite{Gelfond:2023fwe}. Given that vertex $\mathcal{V}(\go, \go, C, C)$ matches \eqref{eq:1-VertexSource} for $n=2$ and given there is the duality between the 1-form $C^2$ vertices and the 0-form $C^3$ vertices, that provides evidence that the latter should match too.

\section{Conclusion}\label{Sec:Conclusion}
In this paper, we completed the analysis of the nonlinear vertices from \eqref{nonlinear} corresponding to interaction of symmetric gauge fields in arbitrary dimensions at the unconstrained level. The unconstrained system \eqref{nonlinear} is not dynamical but describes a set of the HS Bianchi identities along with conditions relating the auxiliary fields with the derivatives of the primary ones. The importance of the proposed analysis is due to the fact that the vertices in question are governed by the so-called off-shell HS algebra \cite{Vasiliev:2003ev} yielding the mechanism of setting the equations \eqref{nonlinear} on shell by modding out the trace ideal. In other words, in order to arrive at the actual dynamics one has to solve first for the vertex problem at the off-shell level and then subtract extra degrees of freedom associated with the traceful components. Here we give the complete answer to the off-shell problem, while leaving the on-shell reduction for the future along the lines of \cite{Vasiliev:2003ev, Didenko:2023vna}.

Finding manifest HS vertices is a highly nontrivial task even at the unconstrained level. In principle, one can use Vasiliev's equations \cite{Vasiliev:2003ev}, which are designed for this purpose. In practice it is very hard to do so beyond order $C^2$ so far (see, however, \cite{Vasiliev:2023yzx} for a progress in this direction). Our strategy was to use the generating equations \eqref{generating} of \cite{Didenko:2023vna} instead, which are expected to be a certain reduction  of the original Vasiliev system,\footnote{Practically speaking, the class of functions evolving on the equations \eqref{generating} (identified in \cite{Didenko:2022qga}) is a subclass of the one from \eqref{generating:Vasiliev}.} where locality becomes manifest. The great advantage of Eqs. \eqref{generating} is that they give straightforward access to all-order vertices in their minimal (local) form. That this is indeed the case has already been shown for the $4d$ holomorphic on-shell higher spins in \cite{Didenko:2022qga} and also in \cite{Didenko:2023vna}, where a particular all-order vertex was manifestly found for the unconstrained case in any $d$. The rest vertices are calculated in this paper as we also observe their remarkable structure, which we now briefly summarize and comment on.
\begin{itemize}
    \item The manifest expressions for the right-hand sides of \eqref{nonlinear} are given by the concise formula \eqref{eq:0-VertexSource} or  \eqref{verC} for the 0-form sector and by \eqref{eq:1-VertexSource} or \eqref{verw} for the 1-form sector, respectively. The 1-form result is ultralocal (see also \cite{Didenko:2023vna}), while the 0-form one is projectively compact spin local. Therefore, these vertices are space-time spin local and in addition contain a minimal number of derivatives in accordance with \cite{Vasiliev:2022med, Vasiliev:2023yzx}. Moreover, the vertices manifest proliferated nonlinearities: The degree of the nonlinearity for a particular descendant $C^{a(s+k), b(s)}$ is bounded by its depth $k$ for all perturbation orders.    

    \item The obtained HS vertices feature interesting geometric structure. Specifically, they are organized in the form of integrals, the phase space of which is given by a set of polygons with the number of vortexes growing with the order of perturbation $O(C^n)$. The polygons in addition are characterized by the place of the impurity $\omega$ in the vertex line. So, in the case of a 0-form vertex $\Upsilon_{C..C\omega C..C}$ the integration polygon is a junction of the concave part associated with $C$'s standing before $\omega$  and the convex one associated with the rest of $C$'s after $\omega$. Specific functions entering these integrals acquire interpretation in terms of polygon areas.    
    
    \item Curiously, the ``coupling constants'' coefficients in \eqref{nonlinear} that result from integration over the space of polygons in \eqref{eq:0-VertexSource} are given by rational numbers, as we checked up to order $O(C^3)$ in Appendix \ref{App:Integrals} for a particular ordering. It would be interesting to see whether this is also true at higher orders. As a side remark, it would be interesting to check whether integral \eqref{Integral}, which generates the coupling constants is expressible in terms of the generalized hypergeometric functions ${}_nF_{n-1}$ for $n>3$, as these functions are known to have remarkable monodromy properties \cite{Beukers}.

    \item The standard Moyal star product, which has the well-known geometric representation, see, e.g., \cite{Zachos:1999mp}, is a particular $n=1$ example of the obtained vertex expression \eqref{verC}, even though it was not meant to be so by default. The Moyal star product comes as the underlying algebra within the HS deformation problem \cite{Vasiliev:1992av, Vasiliev:2003ev} of the more general product that enjoys the $A_{\infty}$ relations in place of the usual associativity (for a review and references see \cite{Stasheff}). We thus  found manifest expressions for such $A_\infty$ structures realized minimally, that is satisfying locality requirement. In this regard, the observed polygonal geometry of the interaction vertices might admit further interesting generalizations. 

    \item An intriguing feature of the calculated 0-form vertex
    \begin{equation}\label{Ups:zero}
        \Upsilon(\omega, C\dots C)\Big|_{y=0}=0\,,
    \end{equation}
which is a manifestation of its being projectively compact spin local, results in a chain of vertex dualities that hold universally; i.e., they rely on no particular generating system, such as \eqref{generating}. Being universal, they have much broader applicability. The dualities (i) tie the vertices $\Upsilon$ of various orderings ($\omega$ impurities) within a given perturbation order, (ii)  tie the $n$th order vertices $\mathcal{V}_{n}$ and the vertices $\Upsilon_{n+1}$. Therefore, having $\Upsilon$ supplemented with the condition \eqref{Ups:zero} suffices to restore $\mathcal{V}(\omega, \omega, C\dots C)$ unambiguously. Notice that generally the $n$th vertex does not define the $(n+1)$ one uniquely, rather up to a remaining freedom in solution of homogeneous equation. That there is a one-to-one map in our case might be attributed to the fact that our vertices are minimal. It is also of interest that the vertex dualities relate projectively compact spin locality of the $\Upsilon$ sector with ultralocality of the $\mathcal{V}$ sector and vice versa, as well as, the closely related shift symmetry \cite{Didenko:2022eso} of the two sectors.  

\item As a byproduct of our analysis, the on-shell vertices of the $4d$ holomorphic HS theory immediately follow. The result is obtained by simply replacing the $\star$ product in the final expressions by the antiholomorphic star product of the $4d$ model.  
\end{itemize}
As it was argued, we expect the generating HS system \eqref{generating} follows from the original Vasiliev equations \eqref{generating:Vasiliev}. On that occasion, our results for vertices should agree with those that could be potentially extracted from the Vasiliev equations. Since we do not have proof of that claim, we carried out the detailed calculation of the few lower-order vertices using the original Vasiliev equations. As expected, the agreement was established, although proceeding this way to higher orders gets impressively laborious. Bridging the two approaches would be highly desirable.  

In conclusion, let us specify the problem related to the proposed research for the future. It will be interesting to trace to which extent the observed geometric (convex/concave polygon phase space) and algebraic (vertex dualities, shift symmetries) structures of the off-shell vertices survive upon on-shell reduction. For example, they do survive in the $4d$ holomorphic sector \cite{Didenko:2022qga}, yet certainly do not in the whole $4d$ theory, while imposing strong constraints on the mixed sector.    In particular, once the off-shell interactions are proved to be maximally local, the effect caused by the ideal factorization becomes crucial for the locality issue. The natural starting point in the analysis of this problem would be the $O(C^2)$ on-shell cubic interaction known to be spin local. The challenging technical issue on this way is factorization of traces, which might be resolved using the effective approach of \cite{Joung:2014qya}.      

\section*{Acknowledgments}
We would like to thank Anatoly Korybut and Mikhail Vasiliev for useful discussions. We are thankful to Ilgam Falyahov and Kirill Ushakov for valuable comments on the draft. The financial support from the Foundation for the Advancement of Theoretical Physics and Mathematics ``BASIS'' is kindly acknowledged.     
    \appendix

\section{Useful formulae}
It is often convenient to use the $\oast$ product defined as\footnote{Similar to our definitions of star products, the integration measure is chosen in such a way that $1 \oast 1 = 1$.}
\begin{equation}
    \label{oast:Appendix}
    f(y) \oast g(z, y) = \int f(y+u) g(z-v, y) e^{iuv} \,.
\end{equation}
In particular, it is helpful for analysis of functional classes along the lines \cite{Didenko:2022qga, Didenko:2022eso, Didenko:2023vna}. It enjoys plenty of useful properties for exponentials. The simplest one is
\begin{equation}
    \label{oast:Decompose}
    e^{i \, yA} \oast e^{i\tau z(y+B)} = e^{i\tau z(y + B) + i(1-\tau) yA - i\tau BA} \, ,
\end{equation}
and the two most important in the following computations are
\begin{equation}
    \label{useful:1}
    \begin{split}
      \Bigg(\smashoperator{ \int_{0}^{1} } &d\tau \frac{1 - \tau}{\tau}  f(-i\partial_B, -i \partial_A) e^{iyA} \oast e^{i\tau z(y+B)} \Bigg) \ast \Bigg(\smashoperator{ \int_{0}^{1} } d\sigma e^{i\sigma z(y+p)} \Bigg) =  \\
      &=\int\displaylimits_{[0,1]^2} \frac{d\tau d\sigma }{1-\sigma} f(-i\partial_B, -i \partial_A) e^{iyA} \oast e^{i \tau z(y + \sigma(p-A)+(1-\sigma)B)} \, ,  
    \end{split}
\end{equation}
\begin{equation}
    \label{useful:2}
    \begin{split}
       \Bigg(\smashoperator{ \int_{0}^{1} } &d\sigma e^{i\sigma z(y+p)} \Bigg) \ast \Bigg(\smashoperator{ \int_{0}^{1} } d\tau \frac{1 - \tau}{\tau} f(-i\partial_B, -i \partial_A) e^{iyA} \oast e^{i\tau z(y+B)} \Bigg)  = \\
       &=\int\displaylimits_{[0,1]^2}  \frac{d\tau d\sigma}{1-\sigma} f(-i\partial_B, -i \partial_A) e^{iyA} \oast e^{i \tau z(y + \sigma(p+A)+(1-\sigma)B)} \, .
    \end{split}
\end{equation}
Here, the function $f\big( -i\frac{\partial}{\partial B}, -i\frac{\partial}{\partial A} \big)$ is such a polynomial of its arguments, which ensures integrals on the lhs of \eqref{useful:1}, \eqref{useful:2} are not divergent at $\tau = 0$. Such a requirement automatically makes integrals on the rhs of these formulas convergent. 

\section{$\mathcal{W}^{(k|n)}$ derivation}
\label{W-derivation}

\subsection*{Recurrent system}
We want to solve the following recurrent system:
\begin{equation}
    \label{reccurentAppendix}
    \begin{cases}
        \mathcal{W}^{(k|n)} = \Delta \big[\mathcal{W}^{(k|n-1)} \ast \Lambda' - \Lambda' \ast \mathcal{W}^{(k-1|n-1)}  \big], \, \, k \in [0, n] \\
        \mathcal{W}^{(0|0)} = e^{-iyt}\,,
    \end{cases}
\end{equation}
where $\Lambda'$ is the source term of $\Lambda$,
\begin{equation}
    \Lambda'(z; y | p) = \dr z^\alpha \, z_\alpha \, \smashoperator{\int_{0}^{1}} d\sigma \, \sigma \, e^{i\sigma z(y+p)} \equiv \dr z^\alpha \, \Big(-i\frac{\partial}{\partial p^\alpha}\Big) \, \smashoperator{\int_{0}^{1}} d\sigma \, e^{i\sigma z(y+p)} \, . 
\end{equation}
What we will prove by induction is 
\begin{equation}
    \begin{split}
    &\mathcal{W}^{(k|n)}(z; y| t, p) =\\
    &=(-)^{k}    \int\displaylimits_{\Delta(n-1)} \frac{d\lambda}{\lambda_1 \, ... \, \lambda_{n}} \quad \smashoperator{\int_{\Delta^{*}_{k,n}(\lambda)} }d\nu
    \smashoperator{\int_{0}^{1}} d\tau \, \frac{1 - \tau}{\tau} \, \prod_{s=1}^{n}\bigg(-i t^\alpha \frac{\partial}{\partial p^\alpha_s} \bigg) \, e^{i y A_n + i C_n} \oast e^{i\tau z(y+B_{k,n})} \, .
    \end{split}
\end{equation}
The base case for $n=1, k = 0, 1$ was proven in \cite{Didenko:2023vna}, so we proceed with the induction step and will derive $\mathcal{W}^{(k|n)}$ from \eqref{reccurentAppendix}.\\

\subsection*{The first term derivation}
For now, let us concentrate our attention on the derivation of $\Delta \big[\mathcal{W}^{(k|n-1)} \ast
\Lambda' \big]$. Using
\eqref{useful:1}, we get
\begin{align*}
     &\mathcal{W}^{(k|n-1)} \ast \Lambda' =\\
     &=(-)^{k} \int\displaylimits_{[0,1]^2} \frac{d\sigma d\tau}{1 - \sigma}\int\displaylimits_{\Delta(n-2)} \frac{d\lambda}{\lambda_1 \lambda_2 \, ... \, \lambda_{n-1}} \int\displaylimits_{\Delta^{*}_{k,n-1}(\lambda)}d\nu \, \bigg(-i\dr z^\alpha \frac{\partial}{\partial p^\alpha_{n}} \bigg)\prod_{s=1}^{n-1}\bigg(-i t^\alpha \frac{\partial}{\partial p^\alpha_s} \bigg) \, e^{iyA_{n-1} + iC_{n-1}} \oast e^{i\tau z(y+B')} \, ,
\end{align*}
where
\begin{equation*}
    B' = (1 - \sigma) B_{k,n-1} + \sigma(p_{n} - A_{n-1}) =
\end{equation*}
\begin{equation*}
    = (1 - \sigma) \sum_{s=1}^{n-1} \lambda_s p_s + \sigma p_n +  (1-\sigma)\bigg[ - \sum_{s=1}^{k} \lambda_s + \sum_{s=k+1}^{n-1} \lambda_s + \sum_{i<j}^{n-1} (\lambda_i \nu_j - \lambda_j  \nu_i) \bigg]t + \sigma(1 - \sum_{s=1}^{n-1} \nu_s) t \, .
\end{equation*}
Now we perform the first change of integration variables:
\begin{subequations}
    \begin{align}
        \Big\{\lambda'_1 = (1-\sigma) \lambda_1, \quad ..., \quad \lambda'_{n-1} =(1-\sigma) \lambda_{n-1}, &  \quad \lambda_n = \sigma \Big\} \longrightarrow J^{-1} = (1-\sigma)^{n-1} \, ,\\
        \smashoperator{\int_{0}^{1}} \frac{d\sigma}{1 - \sigma} \int\displaylimits_{\Delta_{n-2}} \frac{d\lambda}{\lambda_1 \, ... \, \lambda_{n-1}} = &  \int\displaylimits_{\Delta_{n-1}} \frac{d\lambda'}{\lambda'_1 \, ... \, \lambda'_{n-1}} \, .
    \end{align}
\end{subequations}
This transformation does not affect $A_{n-1}, C_{n-1}$, but changes $B'$ into
\begin{equation*}
    B' = \sum_{s=1}^{n} \lambda'_s p_s + \Bigg[ - \sum_{s=1}^{k} \lambda'_s + \sum_{s=k+1}^{n} \lambda'_s + \sum_{i<j}^{n-1}(\lambda'_i \nu_j - \lambda'_j \nu_i) - \lambda'_{n} \sum_{s=1}^{n-1} \nu_s \Bigg]t \,,
\end{equation*}
after that, one should use the decomposition \eqref{oast:Decompose} in
order to calculate the action of homotopy operator \eqref{homotopy} on
$\mathcal{W}^{(k|n-1)} \ast \Lambda'$. Once it is done, we are left
with
\begin{equation*}
    \begin{split}
        &\Delta \bigg[ \mathcal{W}^{(k|n-1)} \ast \Lambda' \bigg] = (-)^{k} \Bigg[ \smashoperator{\int_{0}^{1}} d\rho \smashoperator{\int_{0}^{1}} d\tau \int\displaylimits_{\Delta(n-1)} \frac{d\lambda}{\lambda_1 \lambda_2 \, ... \, \lambda_{n-1}}  \\
        \int\displaylimits_{\Delta^{*}_{k,n-1}(\lambda)} d&\nu  \bigg(-i z^\alpha \frac{\partial}{\partial p^\alpha_{n}} \bigg)\prod_{s=1}^{n-1}\bigg(-i t^\alpha \frac{\partial}{\partial p^\alpha_s} \bigg) \, e^{i(1-\tau)yA_{n-1} + i C_{n-1} + i\tau\rho z(y+B') - i\tau B' A_{n-1} }  \Bigg] \, .
    \end{split}
\end{equation*}
For the expression above there is the following identical transformation:
\begin{equation*}
    z^\alpha \frac{\partial}{\partial p^\alpha_{n}} = \frac{1}{\rho} \, \big(1 - \sum_{s=1}^{n-1} \nu_s \big) t^\alpha \frac{\partial}{\partial p^\alpha_{n}} \, ,
\end{equation*}
which applied gives us
\begin{equation*}
    \begin{split}
        &\Delta \bigg[ \mathcal{W}^{(k|n-1)} \ast \Lambda' \bigg] = (-)^k  \Bigg[ \smashoperator{\int_{0}^{1}} d\rho \smashoperator{\int_{0}^{1}} d\tau \int\displaylimits_{\Delta(n-1)} \frac{d\lambda}{\lambda_1 \lambda_2 \, ... \, \lambda_{n-1}} \\
        \int\displaylimits_{\Delta^{*}_{k,n-1}(\lambda)}d\nu \, \frac{1}{\rho} & \, \big(1 - \sum_{s=1}^{n-1} \nu_s \big) \, \prod_{s=1}^{n}\bigg(-i t^\alpha \frac{\partial}{\partial p^\alpha_s} \bigg) \, e^{i(1-\tau)yA_{n-1} + i C_{n-1} + i\tau\rho z(y+B') - i\tau B' A_{n-1} }  \Bigg] \, .
    \end{split}
\end{equation*}
And now we perform the second change of integration variables:
\begin{subequations}
    \begin{align}
        & \Big\{ \tau' = \tau \rho, \quad \rho' = 1 - \tau \Big\} \longrightarrow J^{-1} = \tau \, ,\\
        & \smashoperator{\int_{0}^{1}} d\tau \smashoperator{\int_{0}^{1}} d\rho \,  \frac{1}{\rho} = \smashoperator{\int_{0}^{1}} d\tau' \smashoperator{ \int_{0}^{1-\tau'} } d\rho' \, \frac{1}{\tau'}  \, ,
    \end{align}
\end{subequations}
which results into
\begin{align*}
    &\Delta \bigg[ \mathcal{W}^{(k|n-1)} \ast \Lambda' \bigg] =(-)^k \Bigg[ \smashoperator{\int_{0}^{1}} \frac{d\tau}{\tau} \int\displaylimits_{\Delta(n-1)} \frac{d\lambda}{\lambda_1 \lambda_2 \, ... \, \lambda_{n-1}}\\
    \smashoperator{\int_{\Delta^{*}_{k,n-1}(\lambda)} }d\nu \int_{0}^{1-\tau} &d\rho \, \big(1 - \sum_{s=1}^{n-1} \nu_s \big) \, \prod_{s=1}^{n}\bigg(-i t^\alpha \frac{\partial}{\partial p^\alpha_s} \bigg) \, e^{i\rho yA_{n-1} + i C_{n-1} + i\tau z(y+B') - i (1-\rho) B'A_{n-1}}  \Bigg] \, .
\end{align*}
Now we proceed with the third change of integration variables:
\begin{subequations}
    \begin{align}
        \Big\{\nu'_i = \nu_i, \, \nu'_{n} = \Big( 1 - \frac{\rho}{1-\tau} \Big)  & \Big( 1 - \sum_{s=1}^{n-1} \nu_s \Big) \Big\} \longrightarrow J^{-1} = \frac{1}{1 - \tau} \,  \Big( 1 - \sum_{s=1}^{n-1} \nu_s \Big) \, ,\\
        \smashoperator{\int_{0}^{1-\tau}} d\rho \quad \smashoperator{ \int_{\Delta^{*}_{k,n-1}(\lambda)} } \, d\nu \, \big(1 - \sum_{s=1}^{n-1} \nu_s \big) =& (1-\tau) \smashoperator{\int_{\Delta^{*}_{k,n-1}(\lambda)} }d\nu'_1 \, ... \, d\nu'_{n-1} \quad \smashoperator{ \int_{0}^{1-\nu'_1 - ... - \nu'_{n-1}} } \quad d\nu'_{n} \, ,
    \end{align}
\end{subequations}
which yields 
\begin{align*}
    \Delta \bigg[  \mathcal{W}^{(k|n-1)} & \ast \Lambda' \bigg] = (-)^k  \Bigg[\smashoperator{\int_{0}^{1}} d\tau \, \frac{1 - \tau}{\tau} \int\displaylimits_{\Delta(n-1)} \frac{d\lambda}{\lambda_1 \lambda_2 \, ... \, \lambda_{n-1}} \smashoperator{ \int_{\Delta^{*}_{k,n-1}(\lambda)} } \,d\nu_1 \, ... \, d\nu_{n-1}  \\
    & \smashoperator{ \int_{0}^{1-\nu_1 - ... - \nu_{n-1}} } \quad d\nu_{n} \, \prod_{s=1}^{n}\bigg(-i t^\alpha \frac{\partial}{\partial p^\alpha_s} \bigg) \, e^{i(1-\tau) yA' + i C' + i\tau z(y+B') - i \tau B'A'}  \Bigg] \, ,
\end{align*}
where
\begin{align*}
    &A' = -\Big(1 - \sum_{s=1}^{n} \nu_s\Big) t \, , \\
    &B' = \sum_{s=1}^{n} \lambda_s p_s + \Bigg[ - \sum_{s=1}^{k} \lambda_s + \sum_{s=k+1}^{n} \lambda_s + \sum_{i<j}^{n-1}(\lambda_i \nu_j - \lambda_j \nu_i ) -\lambda_{n} \sum_{s=1}^{n-1} \nu_s \Bigg]t \, ,\\
    &C' = C_{n-1} - B' A_{n-1} + B' A' = \sum_{s=1}^{n-1} (\nu_s + \lambda_s \nu_{n})(p_s t) + (\lambda_{n}\nu_{n})(p_{n} t) \, .
\end{align*}
The final and the most complicated fourth change of variables
is
\begin{equation}
    \Big\{ \nu'_1 = \nu_1 + \lambda_1 \nu_{n}, \quad ..., \quad \nu'_{n-1} = \nu_{n-1} + \lambda_{n-1} \nu_{n}, \quad \nu'_{n} = \lambda_{n}\nu_{n} \Big\} \longrightarrow J^{-1} = \lambda_{n} \, .
\end{equation}
Provided $\lambda_1 + ... + \lambda_n = 1$, the following
properties arise:
\begin{subequations}
    \begin{align}
        \label{eq:sum} \sum_{s=1}^{n}\nu_s &= \sum_{s=1}^{n}\nu'_s \, ,  \\
        \label{eq:NuLambda}i<j<n-1: \quad \nu_i &\lambda_j - \nu_j \lambda_i = \nu'_i \lambda_j - \nu'_j \lambda_i \, ,\\
        \sum_{i<j}^{n-1}(\lambda_i \nu_j  - \lambda_j\nu_i) - \lambda_{n} & \sum_{s=1}^{n-1} \nu_s = \sum_{i<j}^{n}(\lambda_i \nu'_j  - \lambda_j\nu'_i ) \, ,
    \end{align}
\end{subequations}
and therefore,
\begin{subequations}
    \begin{align}
       A' = -\Big(1 - \sum_{s=1}^{n} & \nu'_s \Big) t; \quad C' = \Big(\sum_{s=1}^{n} \nu'_s p_s \Big) t \, . \\
       B' = \sum_{s=1}^{n} \lambda_s p_s + \Bigg[  - \sum_{s=1}^{k} & \lambda_s + \sum_{s=k+1}^{n}\lambda_s + \sum_{i<j}^{n}(\lambda_i \nu'_j  - \lambda_j\nu'_i )\Bigg]t \, .
    \end{align}
\end{subequations}
As we can see, $A', B', C'$ coincide with $A_n, B_{k,n}, C_n$, correspondingly. However, we still need to find the change of integration domain under such transformation, which is somewhat tricky.

Due to $\nu_i \geqslant 0, \lambda_i \geqslant 0$ and
(\ref{eq:sum}), the following restrictions on $\nu'$ arise:
\begin{equation}
    \nu'_s \geqslant 0; \quad \sum_{s=1}^{n}\nu'_s  \leqslant 1 \, .
\end{equation}
From \eqref{eq:NuLambda} it follows that
\begin{subequations}
    \label{ConstraintNums_1st}
    \begin{align}
        \label{ConstraintNums_1st:1}\nu'_i \lambda_{i+1} &- \nu'_{i+1}  \lambda_i \leqslant 0, \quad i \in [1, k-1] \, ,\\
        \label{ConstraintNums_1st:2}\nu'_i \lambda_{i+1} - &\nu'_{i+1}  \lambda_i \geqslant 0, \quad i \in [k+1, n-2] \, .
    \end{align}
\end{subequations}
From $\nu_s \geqslant 0$ there are the following conditions:
\begin{equation}
    \label{ConstraintLetters_1st}
    \begin{split}
    \nu'_1 \lambda_n - \nu'_n & \lambda_1 \geqslant 0\,,\\
    ... \quad & \, ,\\
    \nu'_{n-1} \lambda_n  - \nu'_n & \lambda_{n-1}  \geqslant 0 \, .
    \end{split}
\end{equation}
However, not all of the conditions \eqref{ConstraintNums_1st},
\eqref{ConstraintLetters_1st} are independent from each over. One notes
that the linear combination of
\eqref{ConstraintNums_1st:2} and \eqref{ConstraintLetters_1st}
leads to
\begin{equation}
    \nu'_i \lambda_n - \nu'_n \lambda_i \geqslant 0, \quad i \in [k+1, n-2] \, ,
\end{equation}
and a consequence of \eqref{ConstraintNums_1st:1} and
\eqref{ConstraintLetters_1st} gives
\begin{equation}
    \nu'_i \lambda_n - \nu'_n \lambda_i \geqslant 0, \, \, i \in [2, k] \, .
\end{equation}
Therefore, the set of independent constraints is
\begin{align*}
    \nu'_1 + ... + \nu'_n  \leqslant 0,& \quad \nu'_i \geq 0 \, , \\
    \nu'_i \lambda_{i+1} - \nu'_{i+1} \lambda_i \geqslant 0&, \quad i \in [1, k-1] \, ,\\
    \nu'_i \lambda_{i+1} - \nu'_{i+1} \lambda_i \geqslant 0, &\quad i \in [k+1, n-1]  \, , \\
    \nu'_1 \lambda_n - &\nu'_n \lambda_1 \geqslant 0  \, .
\end{align*}
So, we finally obtain 
\begin{equation}
    \label{eq:1-term}
    \begin{split}
        &\Delta \bigg[  \mathcal{W}^{(k|n-1)} \ast \Lambda' \bigg] = \\
        &=(-)^k \int\displaylimits_{\Delta(n-1)} \frac{d\lambda}{\lambda_1 \lambda_2 \, ... \, \lambda_n} \quad \smashoperator{ \int_{\Delta^{'}_{k,n}(\lambda)} } \, d\nu \smashoperator{\int_{0}^{1}} d\tau \, \frac{1-\tau}{\tau} \, \prod_{s=1}^{n}\bigg(-i t^\alpha \frac{\partial}{\partial p^\alpha_s} \bigg)  \, e^{iyA_n + iC_n} \oast e^{i\tau z(y+B_{k,n})} \, ,
    \end{split}
\end{equation}
with the integration domain $\Delta^{'}_{k,n}(\lambda)$ defined as
\begin{equation}
    \label{eq:DOMAIN1}
    \Delta^{'}_{k,n}(\lambda) = \begin{cases}
       \nu_1 + ... + \nu_n \leqslant 1, \quad \nu_i \geq 0 \, , \\
        \nu_1 \lambda_n - \nu_n \lambda_1 \geqslant 0 \, , \\
        \nu_i \lambda_{i+1} - \nu_{i+1} \lambda_{i} \leqslant 0, \quad i \in [1,k-1] \, , \\
        \nu_i \lambda_{i+1} - \nu_{i+1} \lambda_{i} \geqslant 0, \quad i \in [k+1 , n-1] \, ,
    \end{cases}
\end{equation}
\subsection*{The second term derivation}
Now we want to calculate $\Delta \bigg[ \Lambda'  \ast
\mathcal{W}^{(k-1|n-1)} \bigg]$. The derivation is pretty much
the same. The only difference is numeration of $p$ being
shifted. $p$ in $\Lambda'$ becomes $p_1$, while $p$ from
$\mathcal{W}^{(k-1|n-1)}$ are $p_2, ..., p_n$. Given this,
from \eqref{useful:2} it follows that
\begin{align*}
    &\Lambda' \ast \mathcal{W}^{(k-1|n-1)} = \\
    &=(-)^{k-1} \int\displaylimits_{[0,1]^2} \frac{d\sigma d\tau}{1 - \sigma} \int\displaylimits_{\Delta(n-2)} \frac{d\lambda}{\lambda_1 \lambda_2 \, ... \, \lambda_{n-1}} \quad \smashoperator{\int_{\Delta^{*}_{k-1,n-1}(\lambda)} } \,  d\nu \, \bigg(-i\dr z^\alpha \frac{\partial}{\partial p^\alpha_{1}} \bigg)\prod_{s=2}^{n}\bigg(-i t^\alpha \frac{\partial}{\partial p^\alpha_s} \bigg) \, e^{iyA_{n-1} + iC_{n-1}} \oast e^{i\tau z(y+B')} \, .
\end{align*}
Due to shifts in $p$ we have
\begin{align*}
    \quad C_{n-1} = \Big(&\sum_{s=1}^{n-1} \nu_s p_{s+1} \Big)t \, ,\\
    B' = \sigma(p_1 + A_{n-1})& + (1-\sigma)B_{k-1,n-1} =  \\
    = \sigma p_1 -\sigma\Big( 1 - \sum_{s=1}^{n-1} \nu_s \Big)t + (1-\sigma)\bigg( \sum_{s=1}^{n-1} &\lambda_s p_{s+1} + \Big[ - \sum_{s=1}^{k-1} \lambda_s + \sum_{s=k}^{n-1} \lambda_s + \sum_{i<j}^{n-1}(\lambda_i \nu_j -  \lambda_j \nu_i) \Big] t \bigg) \, .
\end{align*}
The change of variables of integration gets modified due to the shift in
$p$. For this term the first change of integration variables is
\begin{subequations}
    \begin{align}
        \Big\{ \lambda'_{1} = \sigma, \quad \lambda'_{2} = (1 - \sigma) \lambda_{1}, \quad ..., \quad  \lambda'_{n}=(1-&\sigma)\lambda_{n-1}  \Big\}  \longrightarrow J^{-1} = (1-\sigma)^{n-1} \, , \\
        \smashoperator{\int_{0}^{1}} \frac{d\sigma}{1 - \sigma} \int\displaylimits_{\Delta(n-2)} \frac{d\lambda}{\lambda_1 \, ... \, \lambda_{n-1}} = &\int\displaylimits_{\Delta(n-1)} \frac{d\lambda'}{\lambda'_2 \, ... \, \lambda'_{n}} \, .
    \end{align}
\end{subequations}
This transformation does not affect $A_{n-1}, C_{n-1}$, but
\begin{equation*}
    B' = \sum_{s=1}^{n} \lambda'_s p_s + \bigg[-\sum_{s=1}^{k} \lambda'_s + \sum_{s=k+1}^{n} \lambda'_s + \sum_{i<j}^{n-1}(\lambda'_{i+1} \nu_j  - \lambda'_{j+1}\nu_i) + \lambda'_1 \sum_{s=1}^{n-1} \nu_s  \bigg]t \, .
\end{equation*}
With the help of \eqref{oast:Decompose} we calculate the homotopy
operator action on this product
\begin{equation*}
    \begin{split}
        &\Delta \bigg[\Lambda' \ast \mathcal{W}^{(k-1|n-1)}\bigg] = (-)^{k-1}  \Bigg[ \smashoperator{\int_{0}^{1}} d\rho \smashoperator{\int_{0}^{1}} d\tau \int\displaylimits_{\Delta(n-1)} \frac{d\lambda}{\lambda_2 \lambda_3 \, ... \, \lambda_{n}} \\
        \smashoperator{ \int_{\Delta^{*}_{k-1,n-1}(\lambda)} } \, d\nu & \, \bigg(-i z^\alpha \frac{\partial}{\partial p^\alpha_{1}} \bigg)\prod_{s=2}^{n}\bigg(-i t^\alpha \frac{\partial}{\partial p^\alpha_s} \bigg) \,  e^{i(1-\tau)yA_{n-1} + i C_{n-1} + i\tau\rho z(y+B') - i\tau B' A_{n-1} }  \Bigg] \, .
    \end{split}
\end{equation*}
For the integrand the following identity holds
\begin{equation*}
    z^\alpha \frac{\partial}{\partial p^\alpha_1} = \frac{1}{\rho} \, \big(1 - \sum_{s=1}^{n-1} \nu_s \big) t^\alpha \frac{\partial}{\partial p^\alpha_{1}}\,.
\end{equation*}
Now we can perform the second change of integration variables, which is the same as in the previous calculations
\begin{subequations}
    \begin{align}
        & \Big\{ \tau' = \tau \rho, \quad \rho' = 1 - \tau \Big\} \longrightarrow J^{-1} = \tau \, ,\\
        & \smashoperator{\int_{0}^{1}} d\tau \smashoperator{\int_{0}^{1}} d\rho \,  \frac{1}{\rho} = \smashoperator{\int_{0}^{1}} d\tau' \smashoperator{ \int_{0}^{1-\tau'} } d\rho' \, \frac{1}{\tau'} \, ,
    \end{align}
\end{subequations}
which gives
\begin{equation*}
    \begin{split}
      &\Delta \bigg[\Lambda \ast \mathcal{W}^{(k-1|n-1)}\bigg] = (-)^{k-1}   \Bigg[ \smashoperator{\int_{0}^{1}} \frac{d\tau}{\tau} \int\displaylimits_{\Delta(n-1)} \frac{d\lambda}{\lambda_2 \lambda_3 \, ... \, \lambda_{n}}
       \quad \smashoperator{ \int_{\Delta^{*}_{k-1,n-1}(\lambda)} } \, d\nu \\
      \smashoperator{\int_{0}^{1-\tau}} \, d\rho & \, \big(1 - \sum_{s=1}^{n-1} \nu_s \big) \, \prod_{s=1}^{n}\bigg(-i t^\alpha \frac{\partial}{\partial p^\alpha_s} \bigg) \, e^{i\rho yA_{n-1} + i C_{n-1} + i\tau z(y+B') - i (1-\rho) B'A_{n-1}} \Bigg] \, .
    \end{split}
\end{equation*}
Proceeding with the third change of variables,
\begin{subequations}
    \begin{align}
        \Big\{ \nu'_1 = \Big( 1 - \frac{\rho}{1-\tau} \Big) \Big( 1 - \sum_{s=1}^{n-1} \nu_s \Big), \, \nu'_{i+1} = \nu'_i & \Big\} \longrightarrow J^{-1} = \frac{1}{1 - \tau} \,  \Big( 1 - \sum_{s=1}^{n-1} \nu_s \Big) \, ,\\
        \smashoperator{ \int_{0}^{1-\tau} } \, \, d\rho \quad   \smashoperator{ \int_{\Delta^{*}_{k-1,n-1}(\lambda)} } \, d\nu \big(1 - \sum_{s=1}^{n-1} \nu_s \big) = (1-\tau) & \smashoperator{ \int_{\Tilde{\Delta}_{k-1,n-1}(\lambda)} } \, d\nu'_2 \, ... \, d\nu'_{n} \quad \smashoperator{ \int_{0}^{1-\nu'_2 - ... - \nu'_{n}} } \,  d\nu'_{1} \, .
    \end{align}
\end{subequations}
Due to the shift in numeration, the integration domain for $\nu'_2, \,
..., \, \nu'_n$ is
\begin{equation}
    \Tilde{\Delta}_{k-1,n-1}(\lambda) = \begin{cases}
        \nu_i \lambda_{i+1} - \nu_{i+1} \lambda_{i} \leqslant 0, \, \, \, i \in [2,k-1] \, , \\
        \nu_i \lambda_{i+1} - \nu_{i+1} \lambda_{i} \geqslant 0, \, \, \, i \in [k+1 , n-1] \, .
    \end{cases}
\end{equation}
So, we have
\begin{equation*}
    \begin{split}
        \Delta \bigg[\Lambda' \ast & \mathcal{W}^{(k-1|n-1)}\bigg] = (-)^{k-1}   \Bigg[ \smashoperator{\int_{0}^{1}} d\tau \,  \frac{1-\tau}{\tau}  \int\displaylimits_{\Delta(n-1)} \frac{d\lambda}{\lambda_2 \lambda_3 \, ... \, \lambda_{n}} \smashoperator{ \int_{\Tilde{\Delta}_{k-1,n-1}(\lambda)} }d\nu_2 \, ... \, d\nu_{n} \\
        &\smashoperator{ \int_{0}^{1-\nu_2 - ... - \nu_{n}} } \, d\nu_{1} \, \prod_{s=1}^{n}\bigg(-i t^\alpha \frac{\partial}{\partial p^\alpha_s} \bigg) \,  e^{i(1-\tau) yA' + i C' + i\tau z(y+B') - i \tau B'A'}  \Bigg] \, ,
    \end{split}
\end{equation*}
where
\begin{align*}
    &A' = -\Big( 1 - \sum_{s=1}^{n} \nu_s \Big)t \, ,\\
    &C' = C_{n-1} - B'A_{n-1} + B'A' = \lambda_1 \nu_1 \, (p_1 t) + \sum_{s=2}^{n}(\nu_s + \lambda_s \nu_1)(p_s t) \, , \\
    &B' = \sum_{s=1}^{n} \lambda_s p_s + \bigg[-\sum_{s=1}^{k} \lambda_s + \sum_{s=k+1}^{n} \lambda_s + \sum_{2\leqslant i < j}^{n}(\lambda_{i} \nu_j  - \lambda_{j} \nu_i ) + \lambda_1 \sum_{s=2}^{n} \nu_s  \bigg]t \, ,
\end{align*}
and the final fourth change of variables is
\begin{equation}
    \Big\{ \nu'_1 = \lambda_1 \nu_1, \quad \nu'_2 = \nu_2 + \lambda_2 \nu_1, \quad ..., \quad \nu'_n = \nu_n + \lambda_n \nu_1  \Big\} \longrightarrow J^{-1} = \lambda_1 \,.
\end{equation}
For $\lambda_1 + ... +\lambda_n = 1$, we have
\begin{subequations}
    \begin{align}
        \sum_{s=1}^{n} &\nu_s = \sum_{s=1}^{n} \nu'_s \, , \\
        \label{PROP1} 2\leqslant i < j: \quad \nu_i & \lambda_j - \nu_j\lambda_i =   \nu'_i \lambda_j - \nu'_j \lambda_i \, , \\
        \sum_{2\leqslant i < j}^{n}(\lambda_{i} \nu_j  - \lambda_{j}\nu_i ) +  &\lambda_1 \sum_{s=2}^{n}  \nu_s = \sum_{1\leqslant i < j}^{n}(\lambda'_{i} \nu_j  - \lambda'_{j} \nu_i )  \, .
    \end{align}
\end{subequations}
So,
\begin{align*}
    &A' = -\Big(1 - \sum_{s=1}^{n} \nu'_s \Big) t \, , \\
    &C' = \Big(\sum_{s=1}^{n} \nu'_s p_s \Big) t \, , \\
    &B' = \sum_{s=1}^{n} \lambda_s p_s + \bigg[-\sum_{s=1}^{k} \lambda_s + \sum_{s=k+1}^{n} \lambda_s - \sum_{i < j}^{n}(\nu'_i \lambda_{j} - \nu'_j \lambda_{i}) \bigg]t \, ,
\end{align*}
and we see that $A', B', C'$ coincide with $A_n, B_{k,n}, C_n$ correspondingly. Now we only have to determine the integration domain. Obviously,
\begin{equation*}
    \nu'_i \geqslant 0; \quad \sum_{s=1}^{n}\nu'_s  \leqslant 1 \, ,
\end{equation*}
and due to \eqref{PROP1} there are constraints coming from
$\Tilde{\Delta}_{k-1,n-1}(\lambda)$,
\begin{subequations}
    \label{ConstraintNums_2nd}
    \begin{align}
        \label{ConstraintNums_2nd:1}\nu'_i \lambda_{i+1} - \nu'_{i+1} \lambda_{i} \leqslant 0, \quad i \in [2, k-1] \, , \\
        \label{ConstraintNums_2nd:2}\nu'_i \lambda_{i+1} - \nu'_{i+1} \lambda_{i} \geqslant 0, \quad i \in [k+1, n-1] \, ,
    \end{align}
\end{subequations}
but from $\nu_i \geqslant 0$ the following constraints arise
\begin{equation}
    \label{ConstraintLetters_2nd}
        \begin{split}
            \lambda_1 \nu'_2 -  \lambda_2 &  \nu'_1 \leqslant 0\\
            ... \quad & \, ,\\
            \lambda_1 \nu'_{n}  -  \lambda_{n} & \nu'_1  \leqslant 0 \,.
        \end{split}
\end{equation}
Linear combination of \eqref{ConstraintNums_2nd:1} and
\eqref{ConstraintLetters_2nd} generates
\begin{equation}
    \nu'_i \lambda_1 - \nu'_1 \lambda_i \geqslant 0, \quad i \in [3, k] \, ,
\end{equation}
while the consequence of \eqref{ConstraintNums_2nd:2} and
\eqref{ConstraintLetters_2nd} is
\begin{equation}
    \nu'_i \lambda_1 - \nu'_1 \lambda_i \geqslant 0, \quad i \in [k+1, n-1] \, ,
\end{equation}
and so we have
\begin{equation}
    \label{eq:2-term}
    \begin{split}
        &\Delta \bigg[\Lambda' \ast \mathcal{W}^{(k-1|n-1)}\bigg] =  \\
        &=(-)^{k-1} \int\displaylimits_{\Delta(n-1)} \frac{d\lambda}{\lambda_1 \lambda_2 \, ... \, \lambda_n} \quad \smashoperator{ \int_{\Delta^{''}_{k,n}(\lambda)} } d\nu \smashoperator{ \int_{0}^{1} } d\tau \, \frac{1-\tau}{\tau}  \prod_{s=1}^{n}\bigg(-i t^\alpha \frac{\partial}{\partial p^\alpha_s} \bigg)  \, e^{iyA_n + iC_n} \oast e^{i\tau z(y+B_{k,n})} \, ,
    \end{split}
\end{equation}
with the integration domain $\Delta^{''}_{k,n}(\lambda)$ being
\begin{equation}
    \label{eq:DOMAIN2}
    \Delta^{''}_{k,n}(\lambda) = \begin{cases}
        \nu_1 + ... + \nu_n \leqslant 1, \quad \nu_i \geq 0 \, , \\
        \nu_1 \lambda_n - \nu_n \lambda_1 \leqslant 0 \, , \\
        \nu_i \lambda_{i+1} - \nu_{i+1} \lambda_{i} \leqslant 0, \quad i \in [1,k-1] \, , \\
        \nu_i \lambda_{i+1} - \nu_{i+1} \lambda_{i} \geqslant 0, \quad i \in [k+1 , n-1] \, .
    \end{cases}
\end{equation}\\

\subsection*{Summing up the contributions}
Since
\begin{equation*}
    \mathcal{W}^{(k|n)} = \Delta \bigg[ \mathcal{W}^{(k-1|n-1)} \ast \Lambda' \bigg]-\Delta \bigg[\Lambda' \ast \mathcal{W}^{(k-1|n-1)}\bigg] \, ,
\end{equation*}
we now only need to sum these terms up. From \eqref{eq:1-term} and
\eqref{eq:2-term}, it is evident that both terms has the same sign
$(-)^k$ and the same integrand, while different integration domains.
However, it is easy to see from \eqref{eq:DOMAIN1} and
\eqref{eq:DOMAIN2} that
\begin{equation}
    \Delta^{'}_{k,n}(\lambda) \cap \Delta^{''}_{k,n}(\lambda) = \varnothing; \quad \Delta^{'}_{k,n}(\lambda) \cup \Delta^{''}_{k,n}(\lambda) = \Delta^{*}_{k,n}(\lambda) \, ,
\end{equation}
and, therefore, we proved that
\begin{equation*}
    \begin{split}
        &\mathcal{W}^{(k|n)}(z; y| t, p) =\\
        &=(-)^{k}   \int\displaylimits_{\Delta(n-1)} \frac{d\lambda}{\lambda_1 \, ... \, \lambda_{n}} \quad \smashoperator{\int_{\Delta^{*}_{k,n}(\lambda)} }d\nu \smashoperator{\int_{0}^{1}} d\tau \, \frac{1 - \tau}{\tau} \, \prod_{s=1}^{n}\bigg(-i t^\alpha \frac{\partial}{\partial p^\alpha_s} \bigg)  \, e^{i y A_n + i C_n} \oast e^{i\tau z(y+B_{k,n})}\,.
    \end{split}
\end{equation*}
Now, using the decomposition rule \eqref{oast:Decompose} for
$\oast$ product, we can get rid of the $p$ derivatives and obtain
\begin{equation}
    \label{eq:W-useful1}
    \begin{split}
       &\mathcal{W}^{(k|n)}(z; y| t; p) = \\
       &=(-)^{k} \, (zt)^n \, \, \smashoperator{\int_{\Delta(n-1)}} \dr \lambda \, \, \,
        \smashoperator{ \int_{\Delta^{*}_{k,n}(\lambda) } } \dr\nu\smashoperator{ \int_{0}^{1} } \dr\tau \, (1-\tau)\tau^{n-1}\, e^{i\tau z(y+B_{k,n}) + i (1-\tau) y A_n - i\tau B_{k,n}A_n + i C_n} \, .
    \end{split}
\end{equation}

\section{Derivation of $\Phi^{[k]}_n$}
Recalling \eqref{eq:0-VertexFromSources}, we need to calculate two
terms
\begin{equation}
    \mathcal{W}^{(k-1|n-1)}(-y; -y - p_1 | t, p_2, ..., p_n)e^{-iyp_1} \, ,
\end{equation}
and 
\begin{equation}
    \mathcal{W}^{(k|n-1)}(-y; y - p_n | t, p_1, ..., p_{n-1})e^{-iyp_n}  \, .
\end{equation}
These calculations are pretty much the same as for $\mathcal{W}^{(k|n)}$, so we provide them only for the second term.

Using \eqref{eq:W-useful1} and \eqref{oast:Decompose}, we obtain
\begin{equation*}
    \mathcal{W}^{(k|n)}(z; y| t, p) =(-)^{k} \, (zt)^n \, \smashoperator{\int_{\Delta(n-1)} }\, d\lambda \, \, \smashoperator{\int_{\Delta^{*}_{k,n}(\lambda)} }d\nu \smashoperator{\int_{0}^{1}} d\tau \, \tau^{n-1}\,(1-\tau) \, \, e^{i(1-\tau)yA_n + i \tau z(y+ B_{k,n}) - i \tau B_{k,n}A_n + i C_n} \, .
\end{equation*}
Therefore,
\begin{align*}
    \mathcal{W}&^{(k|n-1)}(-y; y - p_n | t, p_1, ..., p_{n-1})e^{-iyp_n} =\\
    &=(-)^k \, (ty)^{n-1} \,  \smashoperator{\int_{\Delta(n-2)} }\, d\lambda \quad \smashoperator{\int_{\Delta^{*}_{k,n-1}(\lambda)} }d\nu \, \smashoperator{\int_{0}^{1}} d\tau \, \tau^{n-2}\,(1-\tau) \,  e^{i(1-\tau)(y-p_n)A_{n-1}-i\tau y(-p_n + B_{k, n-1}) - i\tau B_{k,n-1}A_{n-1} + i C_{n-1} - iy p_n} \, .
\end{align*}
There are three types of terms in the exponential: the contractions $(py)$, $(ty)$, and
$(tp)$,
\begin{subequations}
    \begin{align}
        (py) \longrightarrow i(1-\tau)(p_n y)& + i \tau \sum_{s=1}^{n-1} \lambda_s (p_s y) \, , \\
        (pt) \longrightarrow i(1-\tau) \Big( 1 - \sum_{s=1}^{n-1}\nu_s \Big)(p_n t) + i \sum_{s=1}^{n-1} &\nu_s (p_s t) + i \tau \Big( \sum_{s=1}^{n-1} \lambda_s p_s \Big) \Big( 1 - \sum_{s=1}^{n-1}\nu_s\Big) t \, ,  \\
        (ty) \longrightarrow i\tau \bigg[-\sum_{s=1}^{k}\lambda_s + \sum_{s=k+1}^{n-1}\lambda_s + \sum_{i<j}^{n}(\lambda_i \nu_j &- \lambda_j \nu_i) \bigg](ty) + i(1-\tau) \Big( 1- \sum_{s=1}^{n-1} \nu_s \Big) (ty) \, . 
    \end{align}
\end{subequations}
Now it is convenient to perform a series of changes of integration variables, which are basically the same as for the derivation of $\Delta_{0}\Big[ \mathcal{W}^{(k|n-1)} \ast \Lambda' \Big]$.

The first change is
\begin{subequations}
    \begin{align}
        \Big\{ \xi_1 = \tau \lambda_1, \quad ..., \quad \xi_{n-1} = \tau\lambda_{n-1}, \quad \xi_{n} &= \lambda_{n}  \Big\} \longrightarrow J^{-1} = \tau^{n-1} \, , \\
        \smashoperator{ \int_0^1 d\tau} \, \tau^{n-2} \, (1-\tau) \,  \smashoperator{\int_{\Delta(n-2)} }\, d\lambda \quad =& \quad  \smashoperator{\int_{\Delta(n-1)} }\, d\xi \, \xi_n \, .
    \end{align}
\end{subequations}
The terms in the exponential transform to
\begin{subequations}
    \begin{align}
        (py) \longrightarrow i \Big( \sum_{s=1}^{n} \xi_s &p_s \Big) y = -iy P_n(\xi) \, , \\
        (pt) \longrightarrow i \xi_n \Big( 1 - \sum_{s=1}^{n-1}\nu_s \Big)(p_n t) + i &\sum_{s=1}^{n-1} \nu_s (p_s t) + i \Big( \sum_{s=1}^{n-1} \xi_s p_s \Big) \Big( 1 - \sum_{s=1}^{n-1}\nu_s\Big) t  \, , \\
        (ty) \longrightarrow i \bigg[ -\sum_{s=1}^{k}\xi_s + \sum_{s=k+1}^{n} \xi_s &+  \sum_{i<j}^{n-1}(\xi_i\nu_j -\xi_j \nu_i) - \xi_{n+1} \sum_{s=1}^{n-1} \nu_s \bigg](ty) \, .
    \end{align}
\end{subequations}
Now we perform the second change, which is just the introduction of the new variable:
\begin{subequations}
    \begin{align}
        \Big\{ \eta_1 = \nu_1, \quad ..., \quad \eta_{n-1} =& \, \, \nu_{n-1}, \quad  \eta_n = 1 - \sum_{s=1}^{n-1}\nu_s \Big\} \, , \\
        \smashoperator{\int_{\Delta^{*}_{k,n-1}(\lambda)} } \, d\nu \quad =& \quad \smashoperator{\int_{D^{*}_{k,n-1}(\lambda)} } \, d\eta \, ,
    \end{align}
\end{subequations}
where we defined
\begin{equation*}
    D^{*}_{k,n-1}(\lambda) = \begin{cases}
        \eta_1 + ... + \eta_n = 1, \quad \eta_i \geq 0  \, ,\\
        \nu_i \lambda_{i+1} - \nu_{i+1} \lambda_{i} \leqslant 0, \quad i \in [1,k-1] \, , \\
        \nu_i \lambda_{i+1} - \nu_{i+1} \lambda_{i} \geqslant 0, \quad i \in [k+1 , n-2]  \, .
    \end{cases}
\end{equation*}
This transformation does not affect the term $(py)$, while the other two
become
\begin{subequations}
    \begin{align}
        (pt) \longrightarrow i \sum_{s=1}^{n-1} \big\{ \eta_s + &\eta_n \xi_s \big\} (p_s t) + i \xi_n \eta_n (p_n t) \, ,\\
        (ty) \longrightarrow i \bigg[ -\sum_{s=1}^{k}\xi_s + \sum_{s=k+1}^{n} \xi_s &+  \sum_{i<j}^{n-1}(\xi_i\eta_j -\xi_j \eta_i) - \xi_{n} \sum_{s=1}^{n-1} \eta_s \bigg](ty) \, .
    \end{align}
\end{subequations}
The last change is basically the fourth change from the derivation of the first term in $\mathcal{W}^{(k|n)}$:
\begin{equation}
    \Big\{\eta'_1 = \eta_1 + \eta_n \xi_1, \quad ..., \quad \eta'_{n-1} = \eta_{n-1} + \eta_n \xi_{n-1}, \quad \eta'_{n} = \xi_n \eta_n  \Big\} \longrightarrow J^{-1} = \xi_n \, .
\end{equation}
This transformation does not affect the term $(py)$, while the other two
become
\begin{subequations}
    \begin{align}
        (pt) \longrightarrow i \Big( \sum_{s=1}^{n} \eta'_s p_s \Big) &t = - i t P_n(\eta') \, , \\
        (ty) \longrightarrow i \bigg[ -\sum_{s=1}^{k}\xi_s + \sum_{s=k+1}^{n} \xi_s &+  \sum_{i<j}^{n}(\xi_i\eta'_j -\xi_j \eta'_i) \bigg](ty) \, .
    \end{align}
\end{subequations}
The analysis of the integration domain constraints is the same as for
$\Delta\Big[\mathcal{W}^{(k|n-1)} \ast \Lambda' \Big]$, and so
we provide the final result:
\begin{equation}
    \mathcal{W}^{(k|n-1)}(-y; y - p_n | t, p_1, ..., p_{n-1})e^{-iyp_n} = (-)^k \, (ty)^n \, , \, \smashoperator{\int_{\bar{D}^{[k]}_n } } d\eta d\xi \, e^{-iyP_n(\xi) - i t P_n(\eta) + i (ty) \cdot S^{[k]}_n} \, ,
\end{equation}
where
\begin{equation}
    \bar{D}^{[k]}_n =
    \begin{cases}
        \eta_1 + ... + \eta_{n} = 1, \quad \eta_i \geq 0 \, , \\
        \xi_1 + ... + \xi_{n} = 1, \quad \xi_i \geq 0 \, , \\
        \eta_1 \xi_n - \eta_n \xi_1 \geqslant 0 \, ,  \\
        \eta_i \xi_{i+1} - \eta_{i+1} \xi_{i} \leqslant 0, \, \, \, i \in [1, k-1] \\
        \eta_i \xi_{i+1} - \eta_{i+1} \xi_{i} \geqslant 0, \, \, \, i \in [k+1 , n-1]\,.
    \end{cases}
\end{equation}
Similar\footnote{The transformation of the integration variables are
almost the same as in the derivation of $\Delta\Big[\Lambda' \ast
\mathcal{W}^{(k-1|n-1)} \Big]$} calculations give
\begin{equation}
    \mathcal{W}^{(k-1|n-1)}(-y; -y - p_1 | t, p_2, ..., p_n)e^{-iyp_1} = (-)^{k-1} \, (ty)^n \, \smashoperator{\int_{\tilde{D}^{[k]}_n } } d\eta d\xi \, e^{-iyP_n(\xi) - i t P_n(\eta) + i (ty) \cdot S^{[k]}_n} \, ,
\end{equation}
where
\begin{equation}
    \tilde{D}^{[k]}_n =
    \begin{cases}
        \eta_1 + ... + \eta_{n} = 1, \quad \eta_i \geq 0 \, , \\
        \xi_1 + ... + \xi_{n} = 1, \quad \xi_i \geq 0 \, , \\
        \eta_1 \xi_n - \eta_n \xi_1 \leqslant 0 \\
        \eta_i \xi_{i+1} - \eta_{i+1} \xi_{i} \leqslant 0, \, \, \, i \in [1, k-1] \\
        \eta_i \xi_{i+1} - \eta_{i+1} \xi_{i} \geqslant 0, \, \, \, i \in [k+1 , n-1]\,.
    \end{cases}
\end{equation}
For the $\Phi^{[k]}_n$ those terms have the same sign $(-)^{k-1} =
(-)^{k+1}$, the same integrand and
\begin{equation}
   \tilde{D}^{[k]}_n \cap \bar{D}^{[k]}_n = \varnothing; \quad  \tilde{D}^{[k]}_n \cup \bar{D}^{[k]}_n = \mathcal{D}^{[k]}_n \,,
\end{equation}
which completes the derivation of \eqref{eq:0-VertexSource}.

\section{Vertex dualities} \label{sec:Appendix4}
\subsection*{Generalities}
A peculiar feature of the obtained sources for the $0$-form vertices is
\begin{equation*}
    \Phi^{[k]}_n(0|t; p) = 0 \, .
\end{equation*}

Let us consider the nonlinear off-shell equations
\begin{subequations}
    \begin{align}
        & \dr_x \omega + \omega \ast \omega = \mathcal{V}(\omega, \omega, C) + \mathcal{V}(\omega, \omega, C, C) + ... \, ,\\
        & \dr_x C + \omega \ast C - C \ast \pi(\omega) = \Upsilon(\omega, C, C) + \Upsilon(\omega, C, C, C) + ... \, .
\end{align}
\end{subequations}

Acting with $\dr_x$ on the second equation and  substituting\footnote{Substitution of $\dr_x \omega$ and $\dr_x C$
instead of $\omega$ and $C$ in the rhs should be understood somewhat
symbolically, as we do not keep track of signs.} equations themselves gives
\begin{align*}
    \big(\mathcal{V}(\omega, \omega, C) + ...\big) &\ast C - \omega \ast \big(\Upsilon(\omega, C, C) + ...\big) - \big( \Upsilon(\omega, C, C) + ... \big) \ast \pi(\omega) - \\
     - C \ast \pi\big(\mathcal{V}(\omega,& \omega, C) + ... \big) = \Upsilon(\dr_x \omega, C, C) + \Upsilon(\omega, \dr_x C, C) + \Upsilon(\omega, C, \dr_x C) + ... \,.
\end{align*}
Due to $\Phi^{[k]}_n(0|t; p) = 0$, the rhs is zero for $y = 0$;
therefore, we have
\begin{equation*}
    \big(\mathcal{V}(\omega, \omega, C) + ...\big) \ast C - \omega \ast \big(\Upsilon(\omega, C, C) + ...\big) - \big( \Upsilon(\omega, C, C) + ... \big) \ast \pi(\omega) - C \ast \pi\big(\mathcal{V}(\omega, \omega, C) + ... \big) \overset{y=0}{=} 0 \, .
\end{equation*}
Since it must hold for each power of $C$, we obtain 
\begin{equation}
    \label{DualityMap:App}
    \mathcal{V}(\omega, \omega, C^n) \ast C - \omega \ast \Upsilon(\omega, C^{n+1}) - \Upsilon(\omega, C^{n+1}) \ast \pi\big(\omega\big) - C \ast \pi\bigg( \mathcal{V}(\omega, \omega, C^n) \bigg) \overset{y_\alpha=0}{=}
    0\,.
\end{equation}
To extract \eqref{DualRel}, one should consider each line formed of two $\omega$'s and $n+1$ $C$'s in \eqref{DualityMap:App}. Since all rows are independent, each ordering must be equal to zero. 
 
Let us first consider the ordering $\big( \omega \, \overbrace{C...C}^{k} \,  \omega \,  C ... C  \big)$ which ends with $C$. Obviously, only the first two terms of \eqref{DualityMap:App} contribute to this ordering and result into 
\begin{equation}
    \big( \Psi^{[0,k]}_n(-p_{n+1} | t_1, t_2 ; p_1, ... , p_n) - \Phi^{[k]}_{n+1}(t_1 | t_2; p_1, ... , p_{n+1}) \big)  \big( \omega \, \overbrace{C...C}^{k} \,  \omega \, C ... C  \big) = 0 \, .
\end{equation}
Stripping off the line $\big( \omega \, \overbrace{C...C}^{k} \,  \omega \, C ... C  \big)$ from the expression above leads to 
\begin{equation}
    \Psi^{[0,k]}_n(-p_{n+1} | t_1, t_2 ; p_1, ... , p_n) - \Phi^{[k]}_{n+1}(t_1 | t_2; p_1, ... , p_{n+1}) = 0 \, .
\end{equation}
Considering all other possible orderings gives \eqref{DualRel}.
\subsection*{1-form vertices}
Using \eqref{Psi-Phi}, straightforwardly we have
\begin{equation*}
    \Psi^{[k_1, k_2]}_n(y|t_1, t_2; p_1, ..., p_n) = (-)^{k_2 - k_1 +1}(t_2t_1)^{n} \smashoperator{\int_{\mathcal{D}^{[k_2 - k_1]}_{n+1} } } d\xi d\eta \, e^{-it_1 \tilde{P}_{n+1}(\xi) - it_2 \tilde{P}_{n+1}(\eta) +  i (t_2t_1) \cdot S^{[k_2 - k_1]}_{n+1} } \, ,
\end{equation*}
where
\begin{equation*}
    \Tilde{P}_{n+1}(\zeta) = \bigg(\sum_{s=1}^{n-k_1} \zeta_s p_{k_1 + s} \bigg) - \zeta_{n-k_1 + 1} y + \bigg( \sum_{s=n-k_1 + 2}^{n+1} \zeta_s p_{s - (n-k_1 + 1)} \bigg) \, .
\end{equation*}
It is natural to do the following renaming of variables
\begin{subequations}
    \label{transformation}
    \begin{align}
        s \in [1, n-k_1 + 1]:& \quad \xi_s \longrightarrow \xi_{s+k_1}; \, \, \eta_s \longrightarrow \eta_{s+k_1} \\
        s \in [n - k_1 + 2, n+1]: \quad \xi_s &\longrightarrow \xi_{s-(n-k_1 +1)}; \, \, \eta_s \longrightarrow \eta_{s-(n-k_1 +1)} \, ,
    \end{align}
\end{subequations}
as this transformation gives
\begin{equation}
    \Psi_n^{[k_1, k_2]}(y|t_1, t_2; p) = (-)^{k_2 - k_1 + 1} (t_2 t_1)^{n} \smashoperator{\int_{\mathcal{D}^{[k_1, k_2]}_{n+1} }  } d\xi d\eta  \, e^{-iy (\xi_{n+1}t_1 + \eta_{n+1}t_2) - it_1 P_n(\xi) - it_2 P_n(\eta) +  i (t_2 t_1) \cdot S^{[k_1, k_2]}_{n}} \, ,
\end{equation}
with
\begin{equation}
    S^{[k_1, k_2]}_n = 1 - \sum_{s=1}^{k_2} \xi_s + \sum_{s=k_2+1}^{n}\xi_s + \sum_{s=1}^{k_1}\eta_s - \sum_{s=k_1 + 1}^{n}\eta_s + \sum_{i<j}^{n}(\xi_i \eta_j - \xi_j \eta_i) \, .
\end{equation}
Let us note that in the last formula, there is no $\xi_{n+1}$ and
$\eta_{n+1}$. In order to get rid of them, one should use two
identities that hold on the integration domain:
\begin{subequations}
    \begin{align}
        \xi_{n+1} = 1 - (\xi_1 + ... + \xi_n) \, , \\
        \eta_{n+1} = 1 - (\eta_1 + ... + \eta_n) \, .
    \end{align}
\end{subequations}
One last thing we need to do is to specify the integration domain
$\mathcal{D}^{[k_1, k_2]}_{n+1}$. As the constraints of
$\mathcal{D}^{[k_2 - k_1]}_{n+1}$ get shuffled under \eqref{transformation}, we must take that into account. So,
originally we have\footnote{We discard the positivity condition for
$\xi$ and $\eta$, as well as the condition for their sum being equal to 1,
because under the renaming of $\xi$ and $\eta$, they remain not affected.}
\begin{equation*}
    \begin{cases}
        \eta_i \xi_{i+1} - \eta_{i+1} \xi_{i} \leqslant 0, \, \, \, i \in [1, k_2-k_1-1] \, , \\
        \eta_i \xi_{i+1} - \eta_{i+1} \xi_{i} \geqslant 0, \, \, \, i \in [k_2-k_1+1 , n] \, .
    \end{cases}
\end{equation*}
Now we have to consider the two separate cases $k_1 = 0$ and $k_1 > 0$.

For $k_1 = 0$ the renaming \eqref{transformation} is the identity
transformation, and so the constraints are just
\begin{equation*}
    \begin{cases}
        \eta_i \xi_{i+1} - \eta_{i+1} \xi_{i} \leqslant 0, \, \, \, i \in [1, k_2-1] \, , \\
        \eta_i \xi_{i+1} - \eta_{i+1} \xi_{i} \geqslant 0, \, \, \, i \in [k_2+1 , n] \, .
    \end{cases}
\end{equation*}

The $k_1 > 0$ case is a bit tricky. First of all, considering $k_1 \in [1, k_2]$ and $k_2 \in [1, n]$, we have
the following inequalities:
\begin{equation*}
    k_2 - k_1 + 1 \leqslant n - k_1 + 1  \leqslant n \, . \\
\end{equation*}
Therefore, the original constraints from vertex dualities can be split into
\begin{equation*}
    \begin{cases}
        \eta_i \xi_{i+1} - \eta_{i+1} \xi_{i} \leqslant 0, \, \, \, i \in [1, k_2-k_1-1] \, , \\
        \eta_i \xi_{i+1} - \eta_{i+1} \xi_{i} \geqslant 0, \, \, \, i \in [k_2-k_1+1 , n - k_1] \, , \\
        \eta_{n-k_1 + 1} \xi_{n-k_1 + 2} - \eta_{n-k_1 + 2} \xi_{n-k_1 + 1} \geqslant 0 \, , \\
        \eta_i \xi_{i+1} - \eta_{i+1} \xi_{i} \geqslant 0, \, \, \, i \in [n - k_1 + 2 , n] \, .
    \end{cases}
\end{equation*}
Here, we stress that $n-k_1 + 2 \leqslant n+1$ due to $k_1 \in [1, k_2]$, making perfect sense as an index.\footnote{For $k_1=0$, we have $n+2\leq n+1$, which makes no sense, given there are only $(n+1)$ $\eta$'s.}

Upon the transformation \eqref{transformation}, the following bunch of constraints arise:
\begin{equation*}
    \begin{cases}
        \eta_i \xi_{i+1} - \eta_{i+1} \xi_{i} \leqslant 0, \, \, \, i \in [k_1+1, k_2-1] \, , \\
        \eta_i \xi_{i+1} - \eta_{i+1} \xi_{i} \geqslant 0, \, \, \, i \in [k_2+1 , n] \, , \\
        \eta_{n-k_1 + 1} \xi_{1} - \eta_{1} \xi_{n-k_1 + 1} \geqslant 0 \, , \\
        \eta_i \xi_{i+1} - \eta_{i+1} \xi_{i} \geqslant 0, \, \, \, i \in [1, k_1 - 1] \, .
    \end{cases}
\end{equation*}
So, all constraints can be summed up to
\begin{equation*}
    \mathcal{D}^{[k_1, k_2]}_{n+1} =
    \begin{cases}
        \eta_1 + ... + \eta_{n+1} = 1, \quad \eta_i \geq 0 \, , \\
        \xi_1 + ... + \xi_{n+1} = 1, \quad \xi_i \geq 0 \, , \\
        \eta_i \xi_{i+1} - \eta_{i+1}\xi_i \geqslant 0, \quad i \in [1, k_1-1] \, , \\
        \eta_i \xi_{i+1} - \eta_{i+1}\xi_i \leqslant 0, \quad i \in [k_1 + 1, k_2-1] \, , \\
        \eta_i \xi_{i+1} - \eta_{i+1}\xi_i \geqslant 0, \quad i \in [k_2 + 1, n] \, , \\
        \text{if} \, \, \, k_1 > 0: \quad \eta_{n+1}\xi_1 - \eta_1 \xi_{n+1} \geqslant 0 \, .
    \end{cases}
\end{equation*}
The last condition can be merged with the conditions for $i \in [1, k_1-1]$ through the identification:
\begin{equation*}
    \eta_{n+1} \equiv \eta_0, \quad \xi_{n+1} \equiv \xi_0 \, ,
\end{equation*}
which sets $\mathcal{D}^{[k_1, k_2]}_{n+1}$ into the form \eqref{eq:1-VertexDomain}.

\section{Gauge transformations}
\label{App:gauge}
Let us remind that the gauge transformation for $W$ is 
\begin{equation}
    \delta_\epsilon W = \dr_x \epsilon + W \ast \epsilon - \epsilon \ast W \, ,
\end{equation}
and the $z$ dependence for $W(z; Y)$ and $\epsilon(z; Y)$ is resolved via
\begin{subequations}
    \begin{align}
        W(z; Y) = \sum_{n=0}^{+\infty}\sum_{k=0}^{n} \mathcal{W}^{(k|n)}(z; y | t; p)\big(\overbrace{C...C}^{k} \, \omega \,\overbrace{C...C}^{n-k}\big) \,, \\
        \epsilon(z; Y) = \sum_{n=0}^{+\infty}\sum_{k=0}^{n} \mathcal{W}^{(k|n)}(z; y | t; p)\big(\overbrace{C...C}^{k} \, \varepsilon\,\overbrace{C...C}^{n-k}\big) \,,
    \end{align}
\end{subequations}
where $\mathcal{W}^{(k|n)}$ is given by \eqref{eq:1-VertexSource}. Since $W(z; Y)$ satisfies the \textit{canonical embedding}, the correspondence between $\delta_\varepsilon \omega$ and $\delta_\epsilon W$ is
\begin{equation}
    \delta_\varepsilon \omega = (\delta_\epsilon W)\big|_{z=0} \, .
\end{equation}
However, the expansion of $\epsilon(z; Y)$ in powers of $C$ is also canonically embedded. Therefore,
\begin{equation}
    ( \dr_x \epsilon )\big|_{z=0} = \dr_x \varepsilon \, ,
\end{equation}
\begin{equation}
    \begin{split}
        &(W \ast \epsilon)\big|_{z=0} = \sum_{n=0}^{+\infty} \Big( \sum_{m_1 + m_2 = n} \sum_{k_1 = 0}^{m_1} \sum_{k_2 - m_1 = 0}^{m_2} \mathcal{W}^{(k_1|m_1)}(z; y| t; p_1, ..., p_{m_1}) \ast  \\
        &\ast \mathcal{W}^{(k_2 - m_1|m_2)}(z; y|q; p_{m_1+1}, ..., p_{n}) \Big)\Big|_{z=0} \big(\overbrace{C...C}^{k_1} \,
    \omega \, \overbrace{C...C}^{k_2-k_1} \, \varepsilon \, \overbrace{C...C}^{n-k_2}\big)
    \end{split}
\end{equation}
\begin{equation}
    \begin{split}
        &(\epsilon \ast W)\big|_{z=0} = \sum_{n=0}^{+\infty} \Big( \sum_{m_1 + m_2 = n} \sum_{k_1 = 0}^{m_1} \sum_{k_2 - m_1 = 0}^{m_2} \mathcal{W}^{(k_1|m_1)}(z; y| q; p_1, ..., p_{m_1}) \ast  \\
        &\ast \mathcal{W}^{(k_2 - m_1|m_2)}(z; y|t; p_{m_1+1}, ..., p_{n}) \Big)\Big|_{z=0} \big(\overbrace{C...C}^{k_1} \,
    \varepsilon\, \overbrace{C...C}^{k_2-k_1} \, \omega \, \overbrace{C...C}^{n-k_2}\big)
    \end{split}
\end{equation}
However, due to relation between $\mathcal{W}$ and the sources $\Psi$ for the vertices $V(\omega, \omega, C^n)$, which is represented in \eqref{PsiFromGen}, the expression simplifies
\begin{equation}
    (W \ast \epsilon)\big|_{z=0} = \omega \ast \varepsilon - \mathcal{V}(\omega, \varepsilon, C) - \mathcal{V}(\omega, \varepsilon, C, C) - ... \, ,
\end{equation}
\begin{equation}
    (\epsilon \ast W)\big|_{z=0} =  \varepsilon \ast \omega - \mathcal{V}(\varepsilon, \omega,  C) - \mathcal{V}(\varepsilon, \omega, C, C) - ... \, ,
\end{equation}
which finally leads to 
\begin{equation}
    \delta_\varepsilon \omega = \dr_x \varepsilon + \omega \ast \varepsilon - \varepsilon \ast \omega - \mathcal{V}(\omega, \varepsilon, C) + \mathcal{V}(\varepsilon, \omega,  C) - \mathcal{V}(\omega, \varepsilon, C, C) +  \mathcal{V}(\varepsilon, \omega, C, C) + \dots \, .
\end{equation}

\section{Matching the lowest order}
\label{App:Match}
The vertex $\mathcal{V}(\omega, \omega, C)$ for the $d$-dimensional higher-spin theory was first derived in \cite{Didenko:2023vna}, though it was given in the form that differs from \eqref{eq:1-VertexSource}. Here, we show that actually it is in a perfect agreement with \eqref{eq:1-VertexSource}. 

The sources for $\mathcal{V}(\omega, \omega, C)$ derived in \cite{Didenko:2023vna} are
\begin{equation}
    \label{wwC:APP}
    \Psi^{[0,0]}_1(y| t_1, t_2; p) = -(t_2 t_1) \smashoperator{ \int_{\begin{subarray}{c}
    (\tau_1, \tau_2)\in [0,1]^2,\\
    \tau_1 + \tau_2 \, \leqslant \, 1
    \end{subarray} } } d\tau_1 d\tau_2 \, e^{i(1-\tau_1)(t_1 y) + i \tau_2 (t_2 y) - i (\tau_1 t_1 + (1-\tau_2) t_2)p + i(\tau_1 + \tau_2)(t_2t_1)}\,,
\end{equation}
\begin{equation}
    \label{Cww:APP}
    \Psi^{[1,1]}_1(y| t_1, t_2; p) = -(t_2 t_1) \smashoperator{ \int_{\begin{subarray}{c}
    (\tau_1, \tau_2)\in [0,1]^2,\\
    \tau_1 + \tau_2 \, \leqslant \, 1
    \end{subarray} } } d\tau_1 d\tau_2 \, e^{i(1-\tau_1)(t_2 y) + i \tau_2 (t_1 y) - i (\tau_1 t_2 + (1-\tau_2) t_1)p + i(\tau_1 + \tau_2)(t_2t_1)}\,,
\end{equation}
\begin{equation}
    \label{wCw:0}
    \begin{split}
    \Psi^{[0,1]}_1(y|t_1, t_2; p) = &(t_2 t_1) \smashoperator{ \int_{\begin{subarray}{c}
    (\tau_1, \tau_2)\in [0,1]^2,\\
    \tau_1 + \tau_2 \, \leqslant \, 1
    \end{subarray} } } d\tau_1 d\tau_2 \, e^{i(1-\tau_1)(t_2 y) + i \tau_2 (t_1 y) - i (\tau_1 t_2 +(1-\tau_2)t_1)p + i (\tau_2 - \tau_1)(t_2 t_1)} + \\
    +&(t_2 t_1) \smashoperator{ \int_{\begin{subarray}{c}
    (\tau_1, \tau_2)\in [0,1]^2,\\
    \tau_1 + \tau_2 \, \leqslant \, 1
    \end{subarray} } } d\tau_1 d\tau_2 \, e^{i(1-\tau_1)(t_1 y) + i \tau_2 (t_2 y) - i (\tau_1 t_1 +(1-\tau_2)t_2)p + i (\tau_2 - \tau_1)(t_2 t_1)}\,. 
    \end{split}
\end{equation}
One may notice, that the two terms \eqref{wCw:0} can actually be merged into one. To do so, one should make the following change of integration variables in the second term of \eqref{wCw:0}: 
\begin{equation*}
    \begin{matrix}
        \tau_1 \longrightarrow 1 - \tau_2 \\
        \tau_2 \longrightarrow 1 - \tau_1 
    \end{matrix} \implies
    \begin{matrix}
        (\tau_1, \tau_2) \in [0,1]^2 \longrightarrow (\tau_1, \tau_2) \in [0,1]^2 \,  \\
        \tau_1 + \tau_2 \leqslant 1 \longrightarrow \tau_1 + \tau_2 \geqslant 1 \, .
    \end{matrix}
\end{equation*}
Such a change of variables makes integrand of the second term in \eqref{wCw:0} the same as in the first, but the integration goes over the two non intersecting halves of the unit square. Therefore, $\Psi^{[0,1]}(y|t_1, t_2; p)$ can be written as
\begin{equation}
    \label{wCw:APP}
    \Psi^{[0,1]}_1(y|t_1, t_2; p) = (t_2 t_1) \smashoperator{ \int_{(\tau_1, \tau_2) \in [0,1]^2} } d\tau_1 d\tau_2 \, e^{i(1-\tau_1)(t_2 y) + i \tau_2 (t_1 y) - i (\tau_1 t_2 +(1-\tau_2)t_1)p + i (\tau_2 - \tau_1)(t_2 t_1)} \, .
\end{equation}
The next step is to introduce new variables $\xi$ and $\eta$. For \eqref{wwC:APP} these are 
\begin{equation}
    \begin{split}
        \xi_1 = \tau_1, \quad \xi_2 = 1-\tau_1 \, ,\\
        \eta_1 = 1 - \tau_2, \quad \eta_2 = \tau_2 \, ,
    \end{split}
\end{equation}
while for \eqref{Cww:APP} and \eqref{wCw:APP}, they are
\begin{equation}
    \begin{split}
        \xi_1 = 1 - \tau_2, \quad \xi_2 = \tau_2 \, , \\ 
        \eta_1 = \tau_1, \quad \eta_2 = 1 - \tau_1  \, .
    \end{split}
\end{equation}
Having done this, the sources for $\mathcal{V}(\omega, \omega, C)$ read
\begin{subequations}
    \label{V-Low}
    \begin{align}
        &\Psi^{[0,0]}_1(y|t_1, t_2; p) = -(t_2 t_1) \smashoperator{ \int_{\mathcal{D}^{[0,0]}_2}} d\xi d\eta \, e^{-iy(\xi_2 t_1 + \eta_2 t_2) - i \xi_1 (t_1 p) - i \eta_1 (t_2 p) + i (t_2 t_1) (1+\xi_1 - \eta_1)} \, , \\
        &\Psi^{[0,1]}_1(y|t_1, t_2; p) = +(t_2 t_1) \smashoperator{ \int_{\mathcal{D}^{[0,1]}_2}} d\xi d\eta \, e^{-iy(\xi_2 t_1 + \eta_2 t_2) - i \xi_1 (t_1 p) - i \eta_1 (t_2 p) + i (t_2 t_1) (1-\xi_1 - \eta_1)} \, , \\
        &\Psi^{[1,1]}_1(y|t_1, t_2; p) = -(t_2 t_1) \smashoperator{ \int_{\mathcal{D}^{[1,1]}_2}} d\xi d\eta \, e^{-iy(\xi_2 t_1 + \eta_2 t_2) - i \xi_1 (t_1 p) - i \eta_1 (t_2 p) + i (t_2 t_1) (1-\xi_1 + \eta_1)} \, ,
    \end{align}
\end{subequations}
with the integration domains being 
\begin{subequations}
    \begin{align}
        &\mathcal{D}^{[0,0]}_2 = \begin{cases}
            \eta_1 + \eta_2 = 1, \quad \eta_{1,2} \geq 0 \, , \\
            \xi_1 + \xi_2 = 1, \quad \xi_{1,2} \geq 0 \, , \\
            \eta_1 \xi_2 - \eta_2 \xi_1 \geqslant 0 \, ,
        \end{cases}\\
        &\mathcal{D}^{[0,1]}_2 = \begin{cases}
            \eta_1 + \eta_2 = 1, \quad \eta_{1,2} \geq 0 \, , \\
            \xi_1 + \xi_2 = 1, \quad \xi_{1,2} \geq 0 \, ,
        \end{cases}\\
         &\mathcal{D}^{[1,1]}_2 = \begin{cases}
            \eta_1 + \eta_2 = 1, \quad \eta_{1,2} \geq 0 \, , \\
            \xi_1 + \xi_2 = 1, \quad \xi_{1,2} \geq 0 \, , \\
            \eta_2 \xi_1 - \eta_1 \xi_2 \geqslant 0 \, ,
        \end{cases}
    \end{align}
\end{subequations}
which is literally as in  \eqref{eq:1-VertexSource}.

\section{Polygon integrals and hypergeometric functions}
\label{App:Integrals}
A natural question to ask is how the numerical coefficients accompanying descendants on the rhs of \eqref{nonlinear} depend on spins upon integration over $\mathcal{D}_{n}^{[k]}$ in \eqref{eq:0-VertexSource}. While we do not provide the full answer to this question, we present calculations suggesting a deep connection of the generating integrals of monomials over $\mathcal{D}_{n}^{[k]}$ with the  generalized hypergeometric functions, which can be defined as 
\begin{equation}
    \label{Hypergeom:expansion}
    {}_{A}F_{B}\bigg(
    \begin{array}{c}
        {a_1, \dots, a_A} \\
        {b_1, \dots, b_B}\\
    \end{array}
        \! \bigg| z \bigg) = \sum_{n=0}^{+\infty} \frac{(a_1)_n \dots (a_A)_n}{(b_1)_n \dots (b_B)_n} \frac{z^n}{n!} \, ,
\end{equation}
with the Pochhammer symbol defined as 
\begin{equation}
    (a)_n = a(a+1)(a+2)\dots(a+n-1) = \frac{(a+n-1)!}{(a-1)!} \, .
\end{equation}
A useful property of the generalized hypergeometric functions is captured by Euler's integral transformation
\begin{equation}
    \label{Hypergeom}
    {}_{A+1}F_{B+1}\bigg(
    \begin{array}{c}
        {a_1, \dots, a_A, c} \\
        {b_1, \dots, b_B, d}\\
    \end{array}
        \! \bigg| z \bigg) = \frac{1}{B(c, d-c)} \int_0^1 dt \, t^{c-1} (1-t)^{d-c-1} \, {}_{A}F_{B}\bigg(
    \begin{array}{c}
        {a_1, \dots, a_A} \\
        {b_1, \dots, b_B}\\
    \end{array}
        \! \bigg| tz \bigg) \, ,
\end{equation}
where $B(x,y)$ is the well-known beta-function.

\subsection*{Leading order}
Consider the integrals that contribute to
$\Phi_2^{[0]}$ of \eqref{eq:0-VertexSource}. These can be
generated by 
\begin{equation}
    \int_{\mathcal{D}^{[0]}_2}  d\xi d\eta \, \xi_1^{a_1}\xi_2^{a_2}\,\eta_1^{b_1}\eta_2^{b_2}\,(\xi_1\eta_2-\xi_2\eta_1)^{\Delta}\,.
\end{equation}
The above generating integral is overdetermined, since $a_{1,2}, b_{1,2}$ and $\Delta$ are all non-negative integers. Using that $\xi_2 = 1-\xi_1, \eta_2 = 1-\eta_1$ this allows us to reduce the analysis to integrals of the form
\begin{equation}
    I_2(a; b) = \int_{\mathcal{D}^{[0]}_2} d\xi_1 d\xi_2 d\eta_1 d\eta_2 \, (\xi_1)^a (\eta_1)^b = \int_{\mathcal{A}_1} d\xi d\eta \, \xi^a \eta^b \, ,
\end{equation}
where 
\begin{equation}
    \mathcal{A}_1 = \begin{cases}
        (\xi, \eta) \in [0,1]^2 \, , \\
        \eta - \xi \geqslant 0 \, .
    \end{cases}
\end{equation}
Therefore, this integral is given by 
\begin{equation}
    I_2(a; b) = \int_0^1 d\eta \, \eta^b \int_0^\eta d\xi \, \xi^a = \frac{1}{(1+a)(2+a+b)} \, ,
\end{equation}
which is clearly a rational number. Due to spin locality of the vertices, it is evident that all coefficients are rational numbers at this order. 

\subsection*{Next to leading order}
Starting with $\Phi^{[0]}_3$ the calculations become more complicated. The corresponding integration domain is 
\begin{equation*}
    \mathcal{D}^{[0]}_3 = \begin{cases}
        \xi_1 + \xi_2 + \xi_3= \eta_1 + \eta_2 + \eta_3 = 1 \, , \\
        \eta_1 \xi_2 - \eta_2 \xi_1 \geqslant 0 \, , \\
        \eta_2 \xi_3 - \eta_3 \xi_2 \geqslant 0 \, .
    \end{cases}
\end{equation*}
So, the analysis boils down to 
\begin{equation}
    I_3(a_1, a_2; b_1, b_2) = \int_{\mathcal{A}_2} d\xi_1 d\xi_2 d\eta_1 d\eta_2 \, (\xi_1)^{a_1}(\xi_2)^{a_2} (\eta_1)^{b_1}(\eta_2)^{b_2} \, ,
\end{equation}
with the integration domain 
\begin{equation}
    \mathcal{A}_2 = \begin{cases}
        \xi_{1,2}, \eta_{1,2} \in [0,1]^4 \, , \\
        \xi_1 + \xi_2 \leqslant 1  \, ,\\
        \eta_1 + \eta_2 \leqslant 1 \, , \\
        \eta_1 \xi_2 - \eta_2 \xi_1 \geqslant 0 \, , \\
        \eta_2(1-\xi_1) - (1-\eta_1)\xi_2 \geqslant 0 \, .
    \end{cases}
\end{equation}
It is convenient to perform the change of integration variables 
\begin{subequations}
    \begin{align}
        \xi_1 = \xi, \quad \xi_2 = \sigma(1 - \xi) \longrightarrow \xi, \sigma \in [0,1]^2 \, ,\\
        \eta_1 = \eta, \quad \eta_2 = \mu(1-\eta) \longrightarrow \eta, \mu \in [0,1]^2 \, ,
    \end{align}
\end{subequations}
with the following Jacobi determinants 
\begin{subequations}
    \begin{align}
        \frac{\partial(\xi_1, \xi_2)}{\partial(\xi, \sigma)} = 1 - \xi \, , \\
        \frac{\partial(\eta_1, \eta_2)}{\partial(\eta, \mu)} = 1 - \eta \, .
    \end{align}
\end{subequations}
This gives us
\begin{equation}
    I_3(a_1, a_2; b_1, b_2) = \int_{\tilde{\mathcal{A}}} d\xi d\eta d\mu d\sigma \, \xi^{a_1}\sigma^{a_2} (1-\xi)^{1+a_2} \eta^{b_1} \mu^{b_2}(1-\eta)^{1+b_2} \, ,
\end{equation}
with the domain $\tilde{\mathcal{A}}$ being
\begin{equation} 
    \tilde{\mathcal{A}} = \begin{cases}
        \xi, \mu, \sigma, \eta \in [0,1]^4 \, , \\
        \eta \sigma (1-\xi) - \mu \xi (1-\eta) \geqslant 0 \, ,\\
        \mu - \sigma \geqslant 0 \, .
    \end{cases}
\end{equation}
To resolve the last inequality, we introduce 
\begin{equation}
    \sigma = \mu \tau \longrightarrow \tau \in [0,1] \, ,
\end{equation}
which gives 
\begin{equation*}
    I_3(a_1, a_2; b_1, b_2) = \int_0^1 d\mu \, \mu^{1+a_2 + b_2} \int_{[0,1]^3}d\xi d\eta d\tau \, \theta\big(\eta \tau (1-\xi) - \xi(1-\eta) \big) \, \tau^{a_2} \xi^{a_1}(1-\xi)^{1+a_2} \eta^{b_1} (1-\eta)^{1+b_2} \, .
\end{equation*}
The Heaviside step function provides the following inequality 
\begin{equation}
    \label{app:inequality}
    \eta \tau (1-\xi) - \xi(1-\eta) \geqslant 0 \, ,
\end{equation}
which can be resolved via the following parametrization: 
\begin{equation}
    \tau(\zeta) = \frac{1-\eta}{\eta} \cdot \frac{\zeta}{1-\zeta} \, .
\end{equation}
This parametrization is monotonous, provided $\eta < 1$, and $\tau \in [0,1]$. It results in $\zeta \in [0, \eta]$. The Jacobi determinant is given by
\begin{equation}
    d\tau = \frac{1-\eta}{\eta} \cdot \frac{1}{(1-\zeta)^2} \, ,
\end{equation}
and \eqref{app:inequality} transforms into 
\begin{equation}
    \frac{\zeta}{1-\zeta} \geqslant \frac{\xi}{1-\xi} \, ,
\end{equation}
which due to the map $\tau \rightarrow \zeta$ is monotonous, results in $\zeta \geqslant \xi$. Therefore, we have 
\begin{equation*}
     I_3(a_1, a_2; b_1, b_2) = \frac{1}{2+a_2 + b_2} \int_{[0,1]^2} d\xi d\eta \int_0^{\eta} d\zeta \, \theta(\zeta - \xi) \, \zeta^{a_2} (1-\zeta)^{-(2+a_2)} \xi^{a_1}(1-\xi)^{1+a_2}\eta^{b_1 - a_2 - 1} (1-\eta)^{2+b_2+a_2} \, ,
\end{equation*}
which is equivalent to 
\begin{equation*}
    I_3(a_1, a_2; b_1, b_2) = \frac{1}{2+a_2 + b_2} \int_{[0,1]^2} d\xi d\eta \, \theta(\eta - \xi) \, \xi^{a_1}(1-\xi)^{1+a_2}\eta^{b_1 - a_2 - 1} (1-\eta)^{2+b_2+a_2} \int_\xi^\eta d\zeta  \, \zeta^{a_2} (1-\zeta)^{-(2+a_2)} \, .
\end{equation*}
Surprisingly enough, the $\zeta$ integral has a very simple form 
\begin{equation}
    \int_\xi^\eta d\zeta  \, \zeta^{a_2} (1-\zeta)^{-(2+a_2)} = \frac{1}{1+a_2} \bigg[ \Big( \frac{\eta}{1-\eta} \Big)^{1+a_2} - \Big( \frac{\xi}{1-\xi} \Big)^{1+a_2} \bigg] \,,
\end{equation}
which leaves us with 
\begin{equation}
    I_3(a_1, a_2; b_1, b_2) = \frac{1}{(1+a_2)(2+a_2 + b_2)} \Big(\Tilde{I} - \Bar{I} \Big) \, ,
\end{equation}
where 
\begin{subequations}
    \begin{align}
    &\Tilde{I} = \int_{[0,1]^2} d\xi d\eta \, \theta(\eta - \xi) \, \xi^{a_1}(1-\xi)^{1+a_2}  \eta^{b_1} (1-\eta)^{1+b_2} \, , \\
    &\Bar{I} = \int_{[0,1]^2} d\xi d\eta \, \theta(\eta - \xi) \, \xi^{1+a_1 + a_2}  \eta^{b_1 - a_2 - 1} (1-\eta)^{2+b_2+a_2} \, .
    \end{align}
\end{subequations}
First we work out $\Tilde{I}$. In order to get rid of the Heaviside step function we introduce $\xi = \lambda \eta \longrightarrow \lambda \in [0,1]$. This yields 
\begin{align*}
    &\Tilde{I} = \int_{[0,1]^2} d\lambda d\eta \, \lambda^{a_1}(1-\lambda \eta)^{1+a_2} \eta^{1+a_1 + b_1}(1-\eta)^{1+b_2} = \\
    &= \int_0^1 d\lambda \, \lambda^{a_1} \int_0^1 d\eta \, \eta^{1+a_1 + b_1}(1-\eta)^{1+b_2}(1-\lambda \eta)^{1+a_2} = \\
    &=  B(2+a_1+b_1, 2+b_2) \cdot \int_0^1 d\lambda \, \lambda^{a_1} \, {}_2F_{1}\bigg(
    \begin{array}{c}
        {-1-a_2, 2+a_1 + b_1} \\
        {4+a_1+b_1+b_2}\\
    \end{array}
        \! \bigg| \lambda \bigg) \, .
\end{align*}
The remaining integral is easy to evaluate via \eqref{Hypergeom}, which just gives us
\begin{equation}
    \Tilde{I} = B(2+a_1+b_1, 2+b_2) \cdot B(a_1 + 1, 1) \cdot \, {}_3F_{2}\bigg(
    \begin{array}{c}
        {-1-a_2, 2+a_1 + b_1, 1+a_1} \\
        {4+a_1+b_1+b_2, 2+a_1}\\
    \end{array}
        \! \bigg| 1 \bigg) \, .
\end{equation}
Note that one of the upper coefficients of the resulting hypergeometric function is a negative integer. Due to \eqref{Hypergeom:expansion} this hypegeometric function has a finite number of terms and, therefore, $\Tilde{I}$ is some rational number.

In order to simplify $\bar{I}$ we make the same change of variables $\xi = \lambda \eta \longrightarrow \lambda \in [0,1]$:
\begin{equation}
    \Bar{I} = \smashoperator{\int_{[0,1]^2} } d\lambda d\eta \, \lambda^{1+a_1 + a_2} \, \eta^{1+a_1 + b_1} (1-\eta)^{2+b_2 + a_2} = \frac{\Gamma(2+a_1+b_1)\Gamma(3+a_2+b_2)}{(2+a_1+a_2)\Gamma(5+a_1+a_2+b_1+b_2)} \,,
\end{equation}
which is clearly also a rational number due to the fact that $a_{1,2}$ and $b_{1,2}$ are non-negative integers. Therefore, $I_3(a_1, a_2; b_1, b_2)$ itself is rational.

    \subsection*{Higher orders}
Higher-order analysis amounts to evaluation of the following generating integrals:
\begin{equation}\label{Integral}
    I_n^{[k]}(a_i; b_i):=\int_{\mathcal{D}_{n}^{[k]}}\prod_{i=1}^{n}\xi^{a_i}_{i}\eta^{b_i}_{i}\,.
\end{equation}
It is interesting whether the result is still a rational number for integer $a_i$ and $b_i$ and whether it is expressible in terms of ${}_{n}F_{n-1}$. This problem we leave unattended.   


\end{document}